\documentclass[sigconf]{acmart}
\AtBeginDocument{%
  }

\setcopyright{acmlicensed}
\copyrightyear{2018}
\acmYear{2018}
\acmDOI{XXXXXXX.XXXXXXX}
\acmConference[Conference acronym 'XX]{Make sure to enter the correct
  conference title from your rights confirmation email}{June 03--05,
  2018}{Woodstock, NY}
\acmISBN{978-1-4503-XXXX-X/2018/06}

\newcommand{\priIcdcsDelete}[1]{{\color{red}{#1}}}

\newcommand{\priCCSDelete}[1]{{\color{red}{#1}}}

\renewcommand{\priIcdcsDelete}[1]{}
\renewcommand{\priCCSDelete}[1]{}
\makeatletter
\newsavebox{\@brx}
\newcommand{\llangle}[1][]{\savebox{\@brx}{\(\m@th{#1\langle}\)}%
  \mathopen{\copy\@brx\kern-0.5\wd\@brx\usebox{\@brx}}}
\newcommand{\rrangle}[1][]{\savebox{\@brx}{\(\m@th{#1\rangle}\)}%
  \mathclose{\copy\@brx\kern-0.5\wd\@brx\usebox{\@brx}}}
\makeatother

\usepackage{subfig}
\usepackage{enumitem}
\usepackage{stmaryrd}
\usepackage{xurl}
\usepackage{graphicx}
\usepackage{booktabs} 
\usepackage{amsmath}
\usepackage{mathtools}
\usepackage{amsthm}
\usepackage{algorithm}
\usepackage{algorithmic}
\usepackage{wrapfig} 
\usepackage{makecell}
\usepackage{balance}
\usepackage{float}
\usepackage{stfloats}

\begin{document}

\title{CryptGNN: Enabling Secure Inference for Graph Neural Networks}

\author{Pritam Sen}
\orcid{0000-0003-2482-0057}
\email{ps37@njit.edu}
\affiliation{%
  \institution{New Jersey Institute of Technology }
  \city{Newark}
  \state{New Jersey}
  \country{US}
}

\author{Yao Ma}
\orcid{0000-0002-4985-8724}
\email{may13@rpi.edu}
\affiliation{%
  \institution{Rensselaer Polytechnic Institute}
  \city{Troy}
  \state{New York}
  \country{US}
}

\author{Cristian Borcea}
\orcid{0000-0003-0020-0910}
\email{borcea@njit.edu}
\affiliation{%
  \institution{New Jersey Institute of Technology }
  \city{Newark}
  \state{New Jersey}
  \country{US}
}


\begin{abstract}
  We present CryptGNN, a secure and effective inference solution for third-party graph neural network (GNN) models in the cloud, which are accessed by clients as ML as a service (MLaaS). The main novelty of CryptGNN is its secure message passing and feature transformation layers using distributed secure multi-party computation (SMPC) techniques. CryptGNN protects the client’s input data and graph structure from the cloud provider and the third-party model owner, and it protects the model parameters from the cloud provider and the clients. CryptGNN works with any number of SMPC parties, does not require a trusted server, and is provably secure even if $\mathcal{P}-1$ out of $\mathcal{P}$ parties in the cloud collude. Theoretical analysis and empirical experiments demonstrate the security and efficiency of CryptGNN.
\end{abstract}




\maketitle

\section{Introduction}
\label{sec:intro}

Unleashing the power of Graph Neural Networks (GNNs) ~\cite{kipf2016semi, Hamilton2017InductiveRL, xu2018how} requires large amounts of training data and computing resources, which are not available to small businesses or individuals. A promising way to democratize access to large-scale GNN models is to make them available in the form of MLaaS~\cite{7424435}.
In MLaaS, third-party owners can monetize their trained models, and clients can perform inference by uploading data to the ML service.

Many types of applications can benefit from GNN models made available as MLaaS, in domains such as pharmaceuticals, finance, software engineering, cybersecurity, and IoT.
In pharmaceuticals, a company may offer a cloud-based GNN model trained on a proprietary collection of organic compound data to help researchers and small start-ups screen out unqualified molecules in the drug discovery process. 
In finance, transaction networks can also be modeled as graphs, where user accounts are represented as nodes with features such as account age, transaction frequency, and reputation score, and edges represent financial transactions. GNNs trained on these graphs can be used to detect fraudulent activity.
In software engineering, GNNs trained on codebase represented as program dependence graphs (PDGs)~\cite{Liu2021} can support automated code analysis, allowing developers to upload code-derived graphs for inference. In cybersecurity, control flow graphs (CFGs) generated from binary executables can serve as inputs to GNNs for effective malware detection. In IoT networks, devices and their interconnections can be represented as graphs, where nodes correspond to devices and edges capture communication links or physical layout, and inference on these graphs enables system fault detection. 

However, operating GNN models in MLaaS settings faces two major privacy concerns. First, the input graph data (such as molecular graphs, financial transactions, source code as PDGs, binary executables as CFGs, or IoT topologies) submitted by clients often contains highly sensitive or proprietary information that must be protected from both the cloud provider and the model owner.
Second, the GNN model parameters trained on valuable proprietary datasets need to be protected from both the cloud provider and the clients to prevent model theft or leakage. This paper addresses these two privacy challenges for GNN inference in the cloud.
We do not address training privacy, as we assume the model is trained in the private infrastructure of its owner.

Existing protocols for privacy-preserving ML inference~\cite{217515, 10.1145/3196494.3196522,  10.5555/3045390.3045413, 9152660} use cryptographic techniques such as homomorphic encryption (HE), trusted execution environment (TEE), and secure multi-party computation (SMPC). Applying these solutions for secure inference over graph-structured data is difficult, because we need to protect not only the features of the graph nodes, but also the graph structure that contains the relationships between the nodes. Protecting the graph structure is especially challenging for GNNs that use message passing layers (MPL)~\cite{NIPS2015_f9be311e, kipf2016semi}
(i.e., the majority of GNNs) because the structure needs to be exploited to exchange messages through edges. Furthermore, it is challenging to design an efficient algorithm to secure the computations in the feature transformation layers (FTL) in GNN, which is required to protect the intermediate and final results, thereby safeguarding the model parameters and node features. 

We propose \textbf{CryptGNN}, a privacy-preserving inference system for GNN models in the cloud, which protects the privacy of the model parameters and the client data. Targeting privacy assurance and high efficiency, we develop distributed SMPC~\cite{TRAN2021245, ZHAO2019357, NEURIPS2021_27545182} protocols that enable a set of mutually distrusting cloud providers (parties) to compute a function on their secret inputs without disclosing each other's inputs and outputs.  
Practically, we outsource the encrypted GNN models across several honest-but-curious parties, assuming $\mathcal{P}-1$ out of $\mathcal{P}$ parties can collude with each other. The SMPC providers compute the forward pass of the model, while CryptGNN protects the model parameters in additive secret-shared format. To protect the client input graph, CryptGNN encrypts the node features and the graph structure in an additive secret-shared format before uploading the data to the SMPC parties.

CryptGNN consists of two novel distributed protocols to enable privacy-preserving inference of encrypted GNN models on encrypted input graph data in the cloud. \textbf{CryptMPL} executes the message-passing layer, while preserving the privacy of input data (i.e., node features and graph structure). It employs novel operations that rotate and shift the input data to securely perform the read and write steps in MPLs. These operations use a data preprocessing step at the client, which helps the SMPC parties mask the private data, thereby eliminating the need for any trusted servers.
\textbf{CryptMUL} executes the secure multiplication operations required for evaluating the linear and nonlinear FTLs in GNN models. In this protocol, the SMPC parties conduct offline preprocessing to generate auxiliary data that allows them to execute matrix and element-wise multiplications without expensive cryptographic operations or relying on a trusted server. CryptMPL and CryptMUL are invoked from the GNN models to guarantee the cloud providers do not learn partial results of functions executed over secure inputs or the final inference results.

Our theoretical analysis proves that CryptGNN is correct and secure, as models using CryptGNN achieve the same accuracy as plain-text models while protecting the input graph and the model parameters.
Our experiments demonstrate that CryptGNN and its protocols achieve lower latency and overhead than baselines based on CrypTen~\cite{NEURIPS2021_27545182}, SecGNN\cite{wang2023secgnn}, and adjacency matrices for graph representation.
The main contributions of this paper are as follows: 
\begin{itemize}[leftmargin=*,topsep=0pt,itemsep=0ex,partopsep=0ex,parsep=0ex]
\item CryptGNN is the first system to enable secure inference for GNNs, ensuring data privacy for both the model owner and the data owner under a strong threat model while maintaining efficiency.

\item We propose a novel algorithm, CryptMPL, for message-passing in GNNs (Section~\ref{sec:cryptmpl_design}) that eliminates the need for a trusted server by using client-side noise for data masking and server-side data transformations. This design works with an arbitrary number of SMPC servers, offering stronger security guarantees compared to prior approaches.

\item CryptMPL proposes a novel graph structure representation, using an edge list combined with SMPC and plaintext relative indexing. This design allows for batch processing of edges, significantly improving efficiency.

\item We design novel CryptMUL protocols (Section~\ref{sec:cryptmul_design}) that integrate existing SMPC techniques and leverage an input-independent preprocessing step to enable secure matrix multiplication and element-wise multiplication, which are essential for the linear and non-linear layers of a GNN. 

\item CryptMUL exploits the fact that model parameters remain fixed across all GNN inferences to optimize the protocol for linear layers. It also leverages the fixed number of multiplications in the GNN to generate auxiliary data offline, enabling efficient online execution of element-wise multiplications in non-linear layers.

\item We provide a security and overhead analysis (Section~\ref{sec:analysis}) and demonstrate the efficacy and security of our proposed system through experiments (Section~\ref{sec:evaluation}) on benchmark graph datasets using a well-known GNN architecture.

\end{itemize}

Overall, the novelty of our contributions lies in creating application-specific protocols for GNN models that can have significant impact in real-life. We utilize existing cryptographic primitives as foundational building blocks to develop new secure and efficient protocols for GNN inference.
\section{Background and Related Work}
\label{sec:background_related_works}

This section covers background information and related work. As a matter of notation: (i) $x \stackrel{\$}{\leftarrow} \mathbb{Z}_L$ denotes that $x$ is uniformly randomly sampled from $\mathbb{Z}_L$, where $L = 2^l$ represents $l$-bit values; (ii) regular and bold characters represent a scalar and matrix, respectively.

\setlength{\heavyrulewidth}{1.5pt}
\begin{table*}[t!]
\small
\centering
\caption{Comparison between CryptGNN and Related Work}
\label{table:comparison}
\addtolength{\tabcolsep}{0.6em}
\resizebox{0.79\textwidth}{!}{
\begin{tabular}{*8c}
\toprule
\thead{} & \thead{Number of \\SMPC \\Parties} & \thead{Does not \\require a \\trusted party} & \thead{Protects \\Model \\Parameters} & \thead{Protects \\Node \\Features} & \thead{Protects \\Graph \\Structure} & \thead{Supports \\heterogeneous \\graphs} & \thead{Supports \\weighted \\directed edges}\\

\midrule
CryptoGCN~\citep{NEURIPS2022_f5332c82}  &  $-$ & \checkmark  & $\times$    &   \checkmark      &   $\times$  & $\times$ & \checkmark\\
SecGNN~\citep{wang2023secgnn}      &   2  & $\times$    & \checkmark    &   \checkmark      &   \checkmark  & $\times$ & \checkmark\\
CryptGNN        &   Any      &   \checkmark  & \checkmark    &   \checkmark      &   \checkmark  & $\times$ & \checkmark\\
\bottomrule
\label{related_work_table}
\end{tabular}

} 

\end{table*}

\subsection{Background on GNN and Cryptographic Primitives}
\label{subsec:background}
\textbf{Message-passing layer (MPL) in GNN.}
\label{subsec-GNN} 
This key operation is executed on graph data $\mathcal{G}=(\mathbf{X}, \mathbf{S}, \mathbf{D})$. $\mathbf{X} \in \mathbb{R}^{N \times K}$ represents the node features as a matrix, where $N$ is the number of nodes in the graph and $K$ is the number of features for each node. The graph structure is often stored via edges, represented as source/destination indices $\mathbf{S}$ and $\mathbf{D}$, where $(\mathbf{S}[j], \mathbf{D}[j])$ represents the $j$-th edge. We consider the most common MPL, where the features of neighboring nodes are aggregated at each node. For the $i$-th node, the MPL processing is expressed as in Eq.~\ref{eqn:MPL}, where $\mathcal{N}(i)$ is the neighbor set of node $i$, ${\bf x'}_i$ is the aggregated feature vector, and ${\bf x}_j$ is the current feature vector of node $j$. A GNN model often consists of multiple MPLs, with FTLs in between. 

\begin{equation}
{\bf x'}_i= \sum_{j \in \mathcal{N}(i)} {\bf x}_j
\label{eqn:MPL}
\end{equation}



\textbf{Feature transformation layers (FTLs) in GNN.}
\label{subsec:background_gnn_layers}
A GNN model incorporates several FTLs, which can involve linear operations, non-linear activations, and other operations that modify the feature representations of the nodes.
Below, we describe some common types of FTLs in GNNs. We use $\times$ and $\otimes$ symbols for scalar and matrix multiplication, respectively. 

{\it Linear Layers.} 
A linear layer uses learned parameters (weight matrix $\bf{H}$ and bias matrix $\bf{B}$) to transform intermediate feature matrices during inference. Mathematically, it involves matrix multiplication and addition operations:
\begin{equation}
\bf{X'} = \bf{X} \otimes \bf{H} + \bf{B} 
\label{eqn:linear_layer}
\end{equation}

{\it Non-linear Layers.}
A non-linear layer modifies the input representation by applying a non-linear function to each element of the input. For instance, a sigmoid layer applied to the vector $\bf{X}$ computes the sigmoid function for each element in $\bf{X}$. Non-linear functions can be implemented using standard approximations~\cite{NEURIPS2021_27545182}.

\textit{Batch Normalization Layer.} During inference, a batch normalization layer utilizes learned parameters, including mean and variance calculated from the training data, along with model-specific parameters (e.g., $\epsilon$, $\gamma$, and $\beta$), to normalize an input value $x$ to the value $y$ as defined in Equation~\ref{eqn:batch_norm_layer}.

\begin{equation}
y = \frac{x - \mathrm{E}[x]}{\sqrt{\mathrm{Var}[x] + \epsilon}} * \gamma + \beta
\label{eqn:batch_norm_layer}
\end{equation}

\textbf{Cryptographic primitives.}
A secret sharing~\cite{10.1145/359168.359176} scheme shares a secret $x$ among $\mathcal{P}$ parties, s.t. the parties can collectively reconstruct the secret, while learning nothing about the secret. We use $\mathcal{P}$-out-of-$\mathcal{P}$ secret sharing schemes, which require the shares of all $\mathcal{P}$ parties to reconstruct the data. We denote the parties by $CP_i$, $i \in \{1, \ldots, \mathcal{P}\}$. 

\textit{Additive secret sharing (A-SS):} In our work, we primarily use A-SS approach. In A-SS, the secret value and its shares are defined over the ring $\mathbb{Z}_L$. A real value $x_R \in \mathbb{R}$ is represented using a fixed-point encoding with a scaling factor $B$ to obtain $x = \lfloor Bx_R \rfloor \in [-2^{l-1}, 2^{l-1})$, where $B = 2^f$ for a given precision of $f$ bits. $x$ can be decoded as $x_R \approx \frac{x}{B}$. We denote the shares of $x$ across the parties by $\llbracket x \rrbracket=\left\{\llbracket x \rrbracket _p\right\}_{p \in \mathcal{P}}$, where $\llbracket x \rrbracket _p$ indicates $CP_p$ 's share of $x$. In A-SS, $\mathcal{P}$ shares are chosen s.t. $\sum_{i=1}^\mathcal{P} x_i=x\bmod L$. The reconstruction algorithm simply adds all the shares as $x=(\sum_{p \in \mathcal{P}}[x]_p)\bmod L$.

\textit{Multiplicative secret sharing (M-SS):} 
We define M-SS over real field $\mathbb{R}$, where 
$\mathcal{P}$ values are chosen uniformly at random, such that $x=\prod_{i=1}^\mathcal{P} x_i$, $x_i \in \mathbb{R}$ and $x_i > 0$. We denote the M-SS of $x$ across the parties $p \in \mathcal{P}$ by $\llangle x \rrangle=\left\{\llangle x \rrangle _p\right\}_{p \in \mathcal{P}}$, where $\llangle x \rrangle _p$ indicates party $CP_p$ 's share of $x$. 

\textit{Beaver Triples:}
\label{sec:Beaver_triple}
Given the additive secret shares of values $X, Y \in \mathbb{Z}_L$, computing the shares of $X \times Y$ requires interaction between the parties. A commonly used approach for this secure multiplication is using a Beaver triple~\cite{beaver1992efficient}, which consists of three elements $(A, B, C)$ such that $C \leftarrow A \times B$, and $A, B \stackrel{\$}{\leftarrow} \mathbb{Z}_L$. The A-SS shares of a Beaver triple $(A, B, C)$ can be used to compute the shares of $Z \leftarrow X \times Y$ by following the protocol $ \mathcal{F}_{BeaverMul}(\llbracket X \rrbracket, \llbracket Y \rrbracket, \llbracket A \rrbracket, \llbracket B \rrbracket, \llbracket C \rrbracket)$ shown below:

\begin{itemize}[leftmargin=*,topsep=0pt,itemsep=0ex,partopsep=0ex,parsep=0ex]
\item Each party gets the share of triples as ($\llbracket A \rrbracket, \llbracket B \rrbracket, \llbracket C \rrbracket)$.
\item Each party computes $\llbracket U \rrbracket \leftarrow  \llbracket X \rrbracket - \llbracket A \rrbracket$, $\llbracket V \rrbracket \leftarrow  \llbracket Y \rrbracket - \llbracket B \rrbracket$.
\item All parties interact to reveal $U \leftarrow (X - A)$, $V \leftarrow (Y - B)$.
\item Each party computes the shares of $Z$ as $\llbracket Z \rrbracket \gets U \times \llbracket B \rrbracket + V \times \llbracket A \rrbracket + \llbracket C \rrbracket + U \times V$.
\end{itemize}

Matrix multiplication can also be performed using Beaver triples following the above steps, just by replacing $\times$ with $\otimes$ to represent the multiplication of different matrices.


\subsection{Related Work} 
\label{related_work}

SecGNN \cite{wang2023secgnn} is an SMPC-based solution for GNN models that processes graph data encrypted in A-SS. Unlike CryptGNN, it relies on a trusted server, which is a strong assumption in practice. Moreover, it works only for 2 parties, which cannot collude with each other. In contrast, CryptGNN works with an arbitrary number of parties, even when $\mathcal{P}-1$ out of the $\mathcal{P}$ parties collude, offering a stronger security guarantee.

One intuitive approach for secure MPL computation is to represent node features in an encrypted feature matrix and the graph structure in an encrypted adjacency matrix, and then employ state-of-the-art matrix multiplication methods~\cite{crypto-1991-1013, NEURIPS2021_27545182}.
However, this incurs high communication overhead, may require an additional trusted party, and results in unnecessary computations for sparse real-world graphs. CryptGNN represents the graph structure as source/destination arrays and achieves substantially lower overhead.
To reduce computation overhead, CryptoGCN~\cite{NEURIPS2022_f5332c82} proposed an efficient matrix multiplication using homomorphic encryption (HE). 
However, it does not protect the graph structure and assumes the GNN model parameters are in plain-text. In fact, protecting the input graph and model parameters with HE is challenging due to the involvement of two different entities (client and model owner) encrypting the data, which introduces additional overhead for bootstrapping~\cite{FHE_Bootstrapping} and key relinearization~\cite{10.1007/978-3-319-70503-3_20}.
CryptGNN uses a much cheaper A-SS approach, reducing the overhead. 

ORAM techniques~\cite{10.1145/3133956.3134090, 10.1145/3177872} could enable a client to access graph data without leaking the access patterns, and thus the graph structure; however, they require the client to download and decrypt data, making it more computationally expensive at resource-constrained clients. In contrast, CryptGNN's secure message-passing protocol offers a more efficient approach that reduces client involvement in GNN inference computations.

Several approaches exist for secure element-wise and matrix multiplication for FTLs in SMPC settings.
CrypTen~\cite{NEURIPS2021_27545182} requires the parties to communicate with a trusted third party. This is not required in CryptGNN. Protocols using HE~\cite{cryptoeprint:2020/451} and oblivious transfer~\cite{235489} tend to have high overhead, making them impractical, especially for numerous inference requests. Our CryptMUL employs HE or OT-based techniques only during preprocessing, enabling the design of FTLs with very low overhead for computing multiple inference requests.

Table~\ref{table:comparison} presents a comparison of our work with closely related approaches~\citep{wang2023secgnn, NEURIPS2022_f5332c82} that directly support GNN inference. As discussed so far in this section, the table emphasizes that CryptGNN is the only solution that works with any numbers of SMPC parties, does not require a trusted server, and is able to protect the model parameters, the node features, and the graph structure. Let us notice that all these works can be easily extended to support weighted directed edges, but they lack support for heterogeneous graphs, as discussed in Section~\ref{sec:discussion}. 
We also compare our work in Section~\ref{sec:evaluation} with~\cite{NEURIPS2021_27545182}, a generic framework that can be adapted to design secure protocols in SMPC settings; however, it relies on a trusted party and introduces higher overhead for GNN inference. Other works~\citep{TRAN2021245, ZHAO2019357} on SMPC cannot be directly extended to support the complex requirements of GNN inference tasks.

\section{Threat Model and System Overview}
\label{sec:overview}

\textbf{Threat Model.}
In our system, there are three key entities:
\begin{itemize}[leftmargin=*,topsep=0pt,itemsep=0ex,partopsep=0ex,parsep=0ex]
\item Model owner (MO): Its primary concern is safeguarding the parameters of the trained GNN model, while ensuring accurate results for each inference request.
\item Data owners (DO): The system can accommodate multiple DOs (clients), each making numerous inference requests and concerned about ensuring the privacy of input graph data.
\item Cloud servers (referred to as parties or CP): We consider an honest-but-curious adversary in the $\mathcal{P}$-party SMPC settings, where each of the $\mathcal{P}$ cloud servers honestly follows the protocols, but may attempt to learn the private data of the MO or DO individually or through collusion.
\end{itemize}

\begin{figure}
\centering
\subfloat[]
{%
\includegraphics[width=1.0\linewidth]{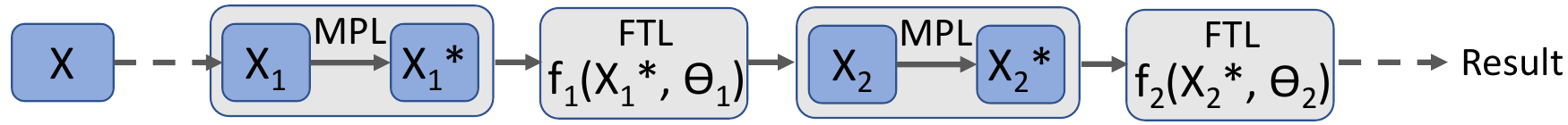}
}%
\quad
\subfloat[]
{%
\includegraphics[width=1.0\linewidth]{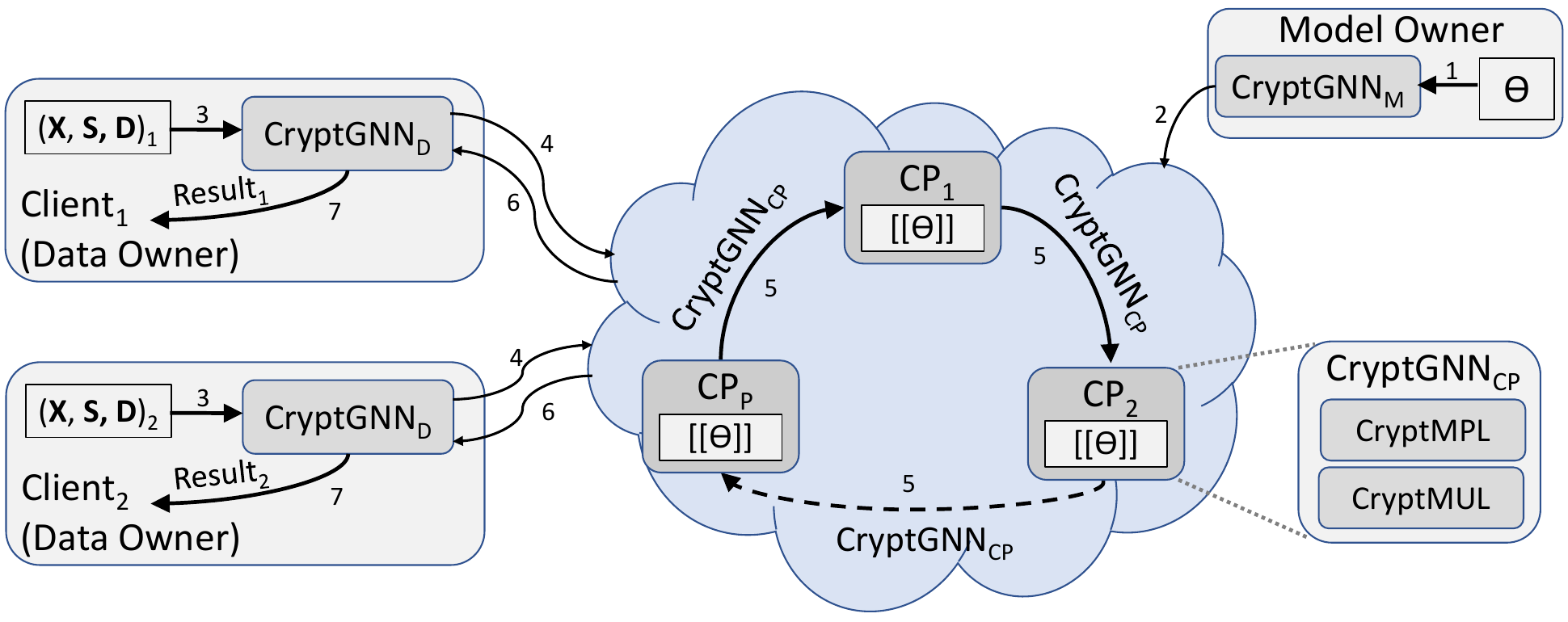}
}%
\caption{(a) A computational flow showing input graph features 
$X$ passed through several MPLs and FTLs of a GNN to generate an inference result (b) CryptGNN architecture illustrating the major components in a 
$p$-party SMPC setting}
\label{fig:system}
\end{figure}

Our threat model $TM$ assumes that at most $\mathcal{P}-1$ parties may collude to learn DO's input data or MO's model parameters. Within $TM$, we also consider the cases where $\mathcal{P}-1$ colluding parties may collude either with the MO to gain access to the DO's input data or with a DO that they control, $DO_{fake}$, to access the MO's model parameters or the input graph data of other DOs. We assume that parties communicate using a secure channel. As the colluding parties can monitor the computation’s control flow and analyze data access patterns, we must use oblivious operations to ensure the input, output, and intermediate results are secured.


\textbf{System Overview.} 
Fig.~\ref{fig:system}(a) shows the flow of a typical GNN, where the initial node features $\mathbf{X}$ are passed through GNN layers to get the intermediate node features $\mathbf{X_1}$. 
An MPL takes the current node features $\mathbf{X_1}$ as input and exchanges messages between the nodes through the edges to compute new node features $\mathbf{X_1^*}$. The FTL transforms $\mathbf{X_1^*}$ into $\mathbf{X_2} = f_1(\mathbf{X_1^*};\Theta_1)$, where $\Theta_1$ summarizes the parameters in an FTL. In GNN, after executing multiple MPLs and FTLs, the final node features are computed to generate the inference result.

The CryptGNN system architecture, shown in Fig.~\ref{fig:system}(b), has components at the SMPC parties, the MO, and the DOs. The components at the SMPC parties execute most of the secure inference protocols. The component at MO uploads the proprietary GNN model to the $\mathcal{P}$ SMPC parties in A-SS format, such that model parameters $\Theta$ are protected (Fig.~\ref{fig:system}(b) Steps 1 and 2). $\Theta$ comprises of the parameters $\{\Theta_1, \Theta_2, \cdots\}$, where $\Theta_i$ is associated with the $i$-th FTL of the model. Following prior work~\cite{wang2023secgnn}, we consider the GNN model architecture (i.e., type, sequence, and number of layers) to be shared by the MO with the parties and the clients, enabling the parties to invoke the secure versions of the insecure layers.
 
CryptGNN's client-side component allows DOs to upload graph data to the cloud as $( \llbracket \mathbf{X} \rrbracket, \llbracket \mathbf{S} \rrbracket, \llbracket \mathbf{D} \rrbracket )$, such that the node features $\mathbf{X}$ and the graph structure, i.e., the list of source indices $\mathbf{S}$ and destination indices $\mathbf{D}$, are protected (Fig.~\ref{fig:system}(b) Steps 3-4). We consider directed, unweighted graphs, although CryptGNN protocols can be extended in a straightforward way for weighted graphs. During each inference request, the parties execute the secure protocols of CryptGNN to compute the output of each layer of the GNN (Fig.~\ref{fig:system}(b) Step 5). Finally, the client receives (Fig.~\ref{fig:system}(b) Step 6) the shares of the final output from all parties to reconstruct the result (Fig.~\ref{fig:system}(b) Step 7). CryptGNN comprises of the following two novel protocols:

{\bf CryptMPL:} This protocol is used for secure message-passing in GNN in a $\mathcal{P}$-party A-SS setting. Executing MPL in the A-SS domain is difficult, as the encrypted features need to be passed through edges, while the source and destination nodes of each edge are encrypted. Our goal is to take the graph structure ($\mathbf{S}, \mathbf{D}$) and the feature matrix $\mathbf{A}$ (e.g., $\mathbf{X_1}$ in Fig.~\ref{fig:system}(a)) as input in A-SS format, and compute the feature matrix $ \llbracket \mathbf{A^*} \rrbracket  = MPL( \llbracket \mathbf{A} \rrbracket ,  \llbracket \mathbf{S} \rrbracket, \llbracket \mathbf{D} \rrbracket )$ after the execution of an MPL layer, while preserving the privacy of the input graph, intermediate results, and model parameters.

{\bf CryptMUL:} The FTLs are computed using additions, multiplications, and comparisons. In A-SS, addition is cheap and can be computed locally, and comparison can be implemented using state-of-the-art techniques~\cite{catrina2010improved}. 
To eliminate the need for a trusted server and costly online step in secure multiplications, CryptMUL employs a preprocessing step to generate auxiliary data for each client, which is used for multiple inference requests from the same client.

\section{CryptMPL}
\label{sec:cryptmpl_design}

\begin{figure}[!t]
\centering
  \subfloat[]
  {%
    \raisebox{6mm}{\includegraphics[width=0.28\linewidth]{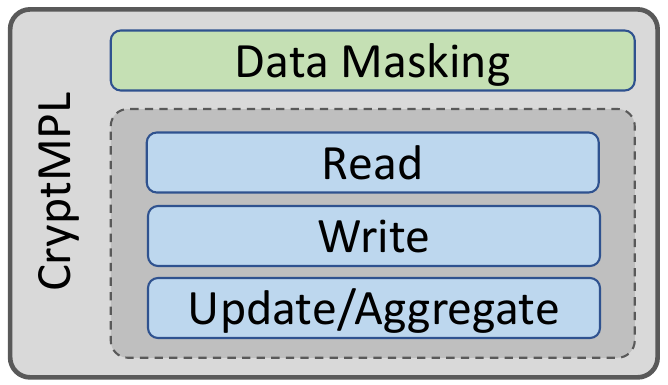}}
  }%
  \hspace{3mm}
  \subfloat[]
  {%
    \includegraphics[width=0.65\linewidth]{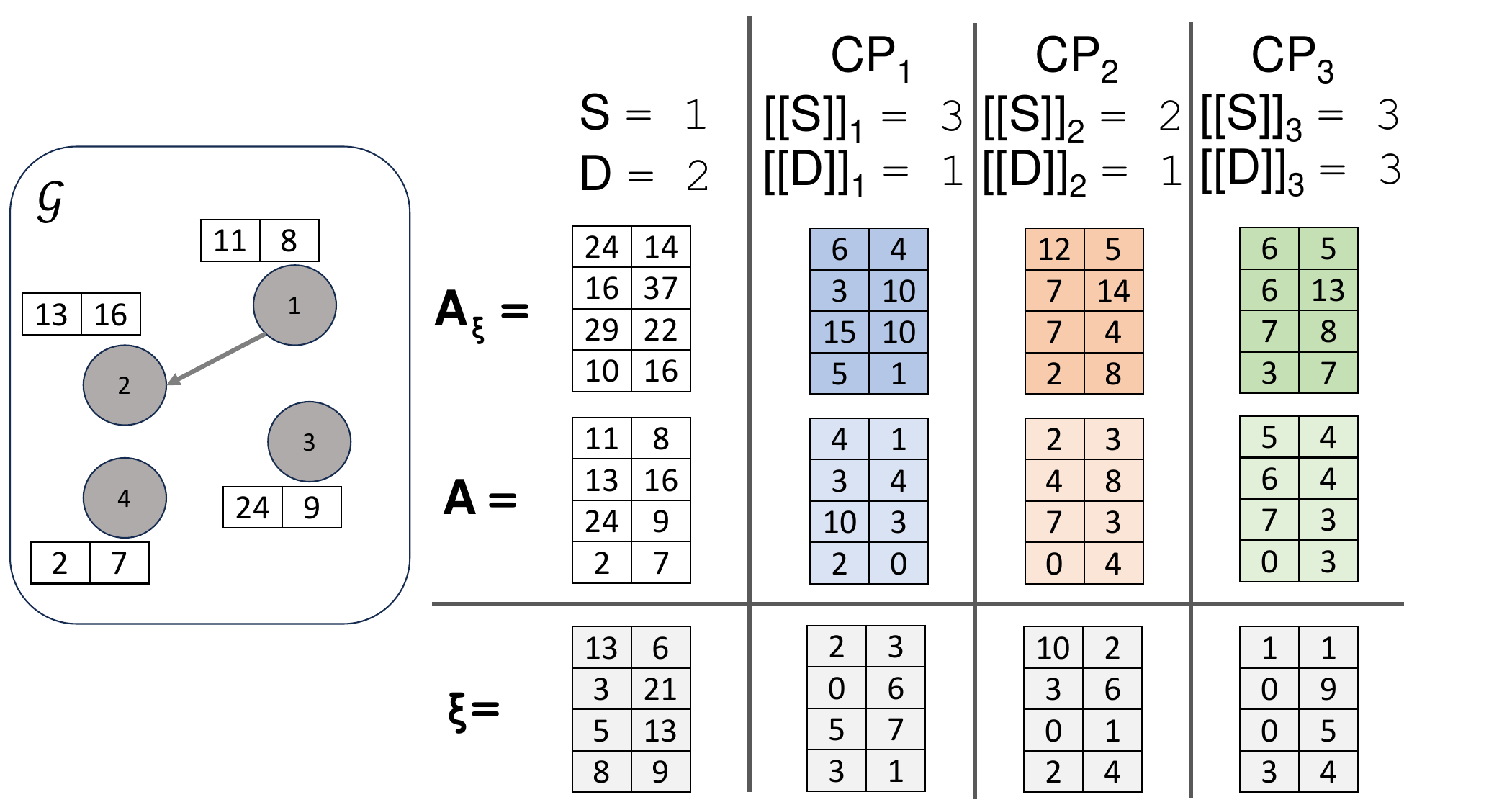}
  }%
\caption{(a) CryptMPL protocol stack (b) An example scenario showing the input feature matrices, source and destination edges of a graph $\mathcal{G}$ with $N=4$ nodes and $K=2$ features, in a 3-party SMPC setting}
\label{fig:protocol_example}
\end{figure}

This section presents the CryptMPL stack of protocols, shown in Fig.~\ref{fig:protocol_example}(a), for privacy-preserving message-passing in GNN using a $\mathcal{P}$-party SMPC setting. To privately exchange messages through edges of a graph, represented as source/destination arrays, we develop novel protocols enabling the SMPC parties to read the feature vector of a source node, write the vector at the destination node, and update the node features by aggregation of intermediate vectors. CryptMPL also uses a novel data masking technique, where the client collaborates with the parties to protect the data against $TM$. 
Fig.~\ref{fig:protocol_example}(b) shows a simple graph $\mathcal{G}$,  where CryptMPL needs to execute the MPL through an edge from node \textcircled{1} to node \textcircled{2}. In this setup, the three computing parties $CP_1$, $CP_2$ and $CP_3$ take an input feature matrix $\mathbf{A}$ representing the node features of $\mathcal{G}$, and collaboratively compute the MPL to generate the output feature matrix. 
In A-SS domain, $CP_p$ holds shares of node features $\mathbf{A}$, source index $S$, and destination index $D$, denoted as $\llbracket \mathbf{A} \rrbracket_p$, $\llbracket S \rrbracket_p$, and $\llbracket D \rrbracket_p$,  respectively. To compute the output node features after executing the MPL, the computing parties execute the read, write, and aggregation protocols (Sections~\ref{sec:cryptmpl_read}, ~\ref{sec:cryptmpl_write}, ~\ref{sec:cryptmpl_agg}).
To protect the graph data, the client shares a noise matrix $\boldsymbol{\xi}$ with the computing parties in A-SS format, enabling each party $C_P$ to mask its share of node features $\llbracket \mathbf{A} \rrbracket_p$ with $\llbracket {\boldsymbol{\xi}}\rrbracket_p$ (Section~\ref{sec:cryptmpl_mask}). The parties then execute the read, write and aggregation protocols on the masked matrix $\llbracket \mathbf{A}_{\boldsymbol{\xi}}\rrbracket_p$, which prevents any leakage of information about the actual shares $\llbracket \mathbf{A} \rrbracket_p$.

Alg.~\ref{alg:Server_MPL_Client_Noise} presents the pseudo-code for CryptMPL, which consists of invoking secure read, write, and aggregate functions (lines 3-5). The details of the protocols behind these operations are presented in Sections~\ref{sec:cryptmpl_read},~\ref{sec:cryptmpl_write}, and~\ref{sec:cryptmpl_agg}, respectively.  
Read and write require each party to communicate with the other parties in a ring-like structure, where the $p$-th party receives data from the $(p-1)$-th and sends data to the $(p+1)$-th party (the $\mathcal{P}$-th party transfers data to the first party). While executing the read and write protocols, CryptMPL uses a novel data masking technique to protect the transferred feature matrices, and the indices of source and destination nodes. To facilitate data masking, the client preprocesses a noise matrix and helps the parties mask their data with noise (Section~\ref{sec:cryptmpl_mask}). The accuracy of computation remains unaffected because the noise is eliminated from the final result (line 7).

\begin{algorithm}[!h] 
\caption{Secure Message Passing Layer, $\mathcal{F}_{CryptMPL}$}
\textbf{Input}: $ \llbracket \mathbf{A} \rrbracket $ (Feature Matrix), $ \llbracket \mathbf{S} \rrbracket $ (Source Indices), $ \llbracket \mathbf{D} \rrbracket $ (Destination Indices), $ \llbracket \boldsymbol{\xi^*} \rrbracket $ (Noise) \\
\textbf{Output}: Output Features $\mathbf{ \llbracket {A^*} \rrbracket }$
    \begin{algorithmic}[1] 
    \STATE $\mathbf{ \llbracket {A_\xi^*} \rrbracket } \gets \mathcal{F}_{InitMatrix}(N, K)$ 
    \FOR {$i \gets 1, \ldots, M$} 
        \STATE $ \llbracket \mathbf{Y} \rrbracket $ $\gets$ $\mathcal{F}_{SR}( \llbracket \mathbf{A} \rrbracket ,  \llbracket \mathbf{S[}i\mathbf{]} \rrbracket)$
        \STATE $ \llbracket \mathbf{G} \rrbracket $ $\gets$ $\mathcal{F}_{SW}( \llbracket \mathbf{Y} \rrbracket ,  \llbracket \mathbf{D[}i\mathbf{]} \rrbracket)$
        \STATE $ \llbracket \mathbf{A_\xi^*} \rrbracket $ $\gets$ $\mathcal{F}_{SA}( \llbracket \mathbf{A_\xi^*} \rrbracket ,  \llbracket \mathbf{G} \rrbracket)$
    \ENDFOR
    \STATE $ \llbracket \mathbf{A^*} \rrbracket $ $\gets$ $ \llbracket \mathbf{A_\xi^*} \rrbracket  -  \llbracket \boldsymbol{\xi^*} \rrbracket $
    \STATE \textbf{return} $\mathbf{ \llbracket {A^*} \rrbracket }$
    \end{algorithmic}
    \label{alg:Server_MPL_Client_Noise}
\end{algorithm}

\begin{figure*}[!h]
\centering
  \subfloat[Secure Read]
  {%
    \includegraphics[width=0.52\textwidth]{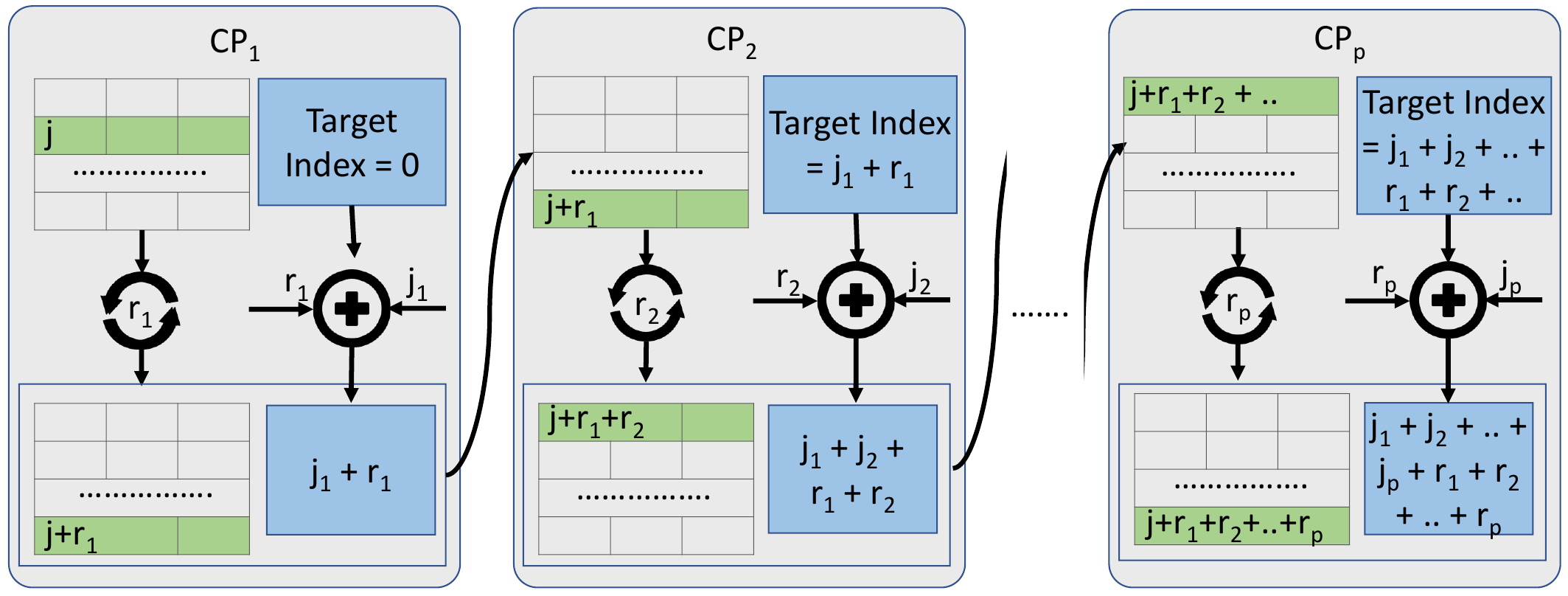}
  }%
  \hspace{7mm}
  \subfloat[Secure Write]
  {%
    \includegraphics[width=0.34\textwidth]{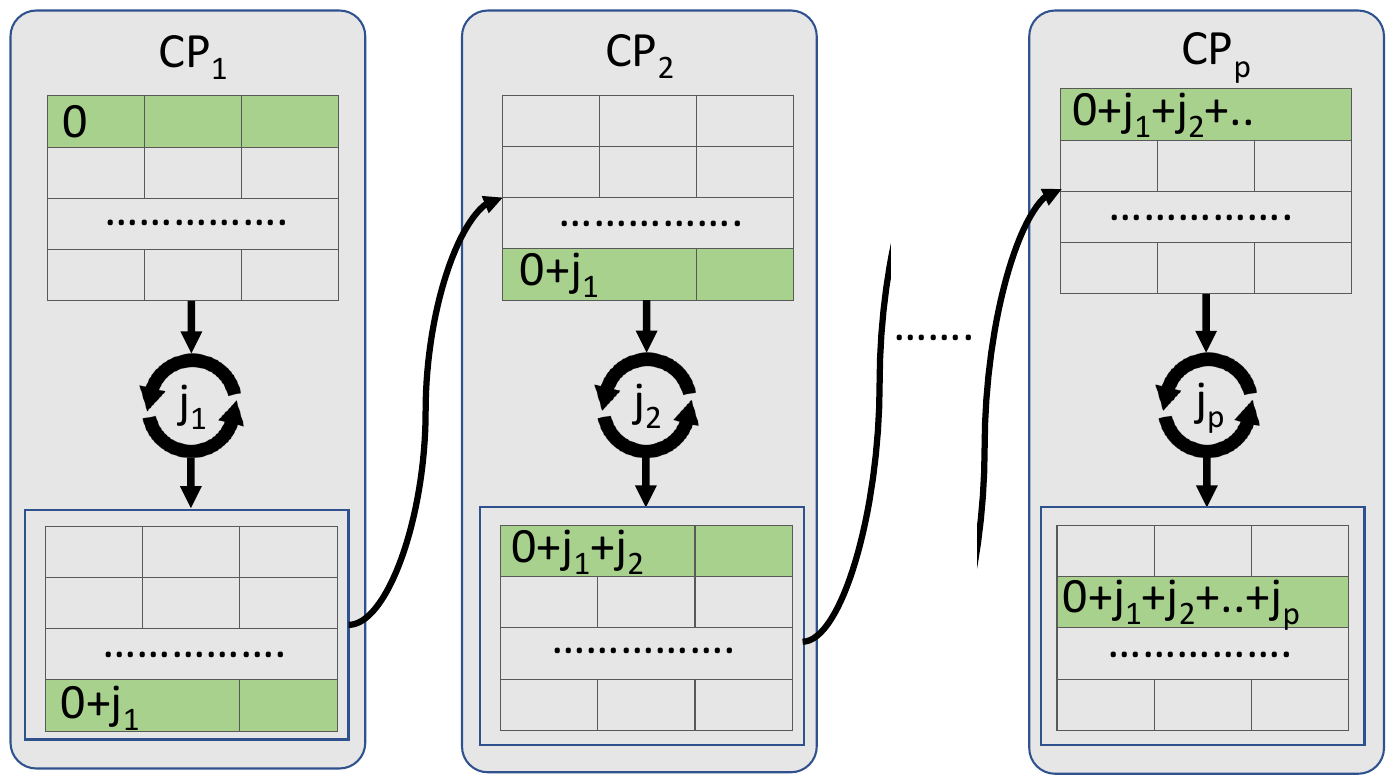}
  }%
\caption{Flow of a share of a matrix for read (a) and write (b) in CryptMPL with $\mathcal{P}$ SMPC parties}
\label{fig:basic_read_write}
\vspace{-5pt}
\end{figure*}

\subsection{Reading the feature vector of a source node} 
\label{sec:cryptmpl_read}

The index in the source nodes' array and the feature vector of a source node for an edge are stored in the A-SS domain. The secure read ($\mathcal{F}_{SR}$ in Alg.~\ref{alg:Server_MPL_Client_Noise}) accesses the feature vector without leaking the index and the features of the source node. 
The main idea of $\mathcal{F}_{SR}$ requires each party to rotate their share of the feature matrix and shift their share of the source index by the same random amount, and then share the updated matrix and index with the next party. The random amount is different at each party. 
After all the parties have rotated the feature matrix and shifted the source index, each party reads the vector at the updated index of the rotated matrix. 
As both the rotation and shift are performed by the same total amount, each party receives a correct share of the source feature vector.

This procedure is illustrated in Fig.~\ref{fig:basic_read_write}(a) and is detailed below. 
For a source node $j \in \mathbf{S}$, its corresponding feature vector is ${\bf A}[j]$. The secret-shared versions of the index and the vector are $ \llbracket j \rrbracket$ and $\llbracket {\bf A}[j] \rrbracket$, respectively. 
In $\mathcal{F}_{SR}$, party $CP_p$ securely retrieves $ \llbracket {\bf A}[j] \rrbracket _p$. 
For example, to retrieve $ \llbracket {\bf A}[j] \rrbracket _1$, the parties execute these steps:

\begin{enumerate}[leftmargin=*,topsep=0pt,itemsep=0ex,partopsep=0ex,parsep=0ex]
\item $CP_1$ initializes two variables: a target index $j' = 0$ and a target matrix $\llbracket {\bf A'} \rrbracket _1 = \llbracket {\bf A} \rrbracket _1$. 
The target index and the target matrix pass through the parties in the ring and are updated by the parties (Steps 2-5). In Step 6, $CP_1$ reads the vector at the updated index of the updated matrix.
\item To protect the share of the source index, $CP_p$ adds a random integer $r_p$ to $\llbracket j \rrbracket _p$,  and updates the target index $j' \leftarrow j' + \llbracket j \rrbracket _p + r_p$.
\item To align the target matrix, $CP_p$ rotates the rows of $\llbracket {\bf A'} \rrbracket _1$ by $r_p$, i.e., $\llbracket {\bf A'} \rrbracket _1 \leftarrow rotate(\llbracket {\bf A'} \rrbracket _1, r_p )$.  
\item $CP_p$ transfers $\llbracket {\bf A'} \rrbracket _1$ and $j'$ to $CP_{p+1}$, which repeats Steps 2 \& 3 to update $\llbracket {\bf A'} \rrbracket _1$ and $j'$. 
\item After the operations at the $\mathcal{P}$-th party, the information is transferred to the first party $CP_1$.
\item $CP_1$ gets $j'= \sum\limits_{p=1}^\mathcal{P} \llbracket j \rrbracket_p +  r_p = j + \sum\limits_{p=1}^\mathcal{P} r_p $. Correspondingly, $\llbracket {\bf A'} \rrbracket _1$ is rotated for $ \sum\limits_{p=1}^\mathcal{P} r_p$ times, i.e, $\llbracket {\bf A}[j] \rrbracket_1 = \llbracket {\bf A'}[j+ \sum\limits_{p=1}^\mathcal{P} r_p] \rrbracket_1$.  Thus, $CP_1$ gets $\llbracket {\bf A'}[j'] \rrbracket_1 = \llbracket {\bf A}[j] \rrbracket _1$.
\end{enumerate}

All parties follow the same procedure in parallel to retrieve the $\mathcal{P}$ shares of $\llbracket {\bf A}[j] \rrbracket$. This procedure protects each party's share of the source index through the random shifting of its value. However, this is insufficient to safeguard the graph data, because: (a) It is possible to reconstruct the feature matrix, as each party gets all the shares of $\mathbf{A}$, and (b) Each party can determine the actual source index by searching the accessed vector in $\mathbf{A}$. 
To solve these problems, each party adds a random noise (a matrix containing random values) to mask the shares of $\mathbf{A}$ before transferring them to the other parties. Thus, $\mathcal{F}_{SR}$ addresses the aforementioned issues, since: (a) Due to the noise, the feature matrix cannot be reconstructed correctly; (b) Since the feature matrix is modified by all parties, the parties cannot determine the source index. Therefore, the parties can securely access the source feature vector. Eliminating the noise from the final result is discussed in Section~\ref{sec:cryptmpl_mask}.

\subsection{Writing messages to the destination node}
\label{sec:cryptmpl_write}

This secure protocol ($\mathcal{F}_{SW}$ in Alg.~\ref{alg:Server_MPL_Client_Noise}, Line 4) creates an intermediate matrix ${\bf G}$ of the same dimensions as the output feature matrix, and writes the feature vector ${\bf Y}$ at the index in ${\bf G}$ corresponding to the destination node's index in the destination nodes' array ${\bf D}$.
Unlike read, which can use rotation operations to preserve index privacy, write must know the destination index to write the vector at the correct position.
As the destination node of each edge is encrypted, to write a feature vector ${\bf Y}$ at the destination index $j \in \mathbf{D}$ of a matrix ${\bf G}$,
the parties need to coordinate. In the secret-shared domain, each party $CP_p$ initializes the share of ${\bf G}$ as $ \llbracket {\bf G} \rrbracket _p = {\bf 0}$ (i.e., ${\bf G}$ with all entries zero). If party $CP_p$ has the shares of ${\bf Y}$ and $j$, i.e., $ \llbracket {\bf Y} \rrbracket _p$ and $ \llbracket j \rrbracket _p$, our goal is to get $ \llbracket {\bf G}[j] \rrbracket_p =  \llbracket {\bf Y} \rrbracket _p$, without leaking the target index and the feature vector. 

The main idea of the write protocol requires each party to write its share of the feature vector ${\bf Y}$ at the $0$-th index of its share of {\bf G} and transfer it to the next party in the ring. Each party rotates the share of the matrix ${\bf G}$ by its share of the destination index. Thus, the feature vector reaches the correct destination index of $\bf{G}$.
For example, Fig.~\ref{fig:basic_read_write}(b) shows the following steps to write $ \llbracket {\bf Y} \rrbracket _1$ at $ \llbracket {\bf G}[j] \rrbracket_1$.

\begin{enumerate}[leftmargin=*,topsep=0pt,itemsep=0ex,partopsep=0ex,parsep=0ex]
\item $CP_1$ writes the vector $ \llbracket \mathbf{Y} \rrbracket _1$ at index 0 of $ \llbracket {\bf G} \rrbracket _1$ as $ \llbracket {\bf G}[0] \rrbracket _1 =  \llbracket {\bf Y} \rrbracket _1$. The matrix $ \llbracket {\bf G} \rrbracket _1$ will pass through the parties in the ring and be updated by other parties in Steps 2-4. In Step 5, $CP_1$ gets the updated matrix $ \llbracket {\bf G} \rrbracket _1$, where $ \llbracket \mathbf{Y} \rrbracket _1$ is written at the correct destination index.
\item $CP_p$ rotates the matrix $ \llbracket {\bf G} \rrbracket _1$ by $ \llbracket j \rrbracket _p$.
\item $CP_p$ transfers  $ \llbracket {\bf G} \rrbracket _1$ to $(p \mod \mathcal{P}) +1$-th party for $ 1\leq p\leq \mathcal{P}$ . $CP_{p+1}$ repeats Steps 2 to update $\llbracket {\bf G} \rrbracket _1$.
\item After the operations at the $\mathcal{P}$-th party, the matrix $\llbracket {\bf G} \rrbracket _1$ is transferred to the first party.
\item $CP_1$ gets $ \llbracket {\bf G} \rrbracket_1 $ which is rotated by $ \sum\limits_{p=1}^\mathcal{P} j_p = j$ times. Due to the overall rotation, $ \llbracket {\bf Y} \rrbracket _1$ is moved to $j$-th index of $ \llbracket \bf G \rrbracket _1$, equivalent to $ \llbracket {\bf G}[j] \rrbracket _1 =  \llbracket {\bf Y} \rrbracket _1$. 
\end{enumerate}

All parties follow the same procedure in parallel to write their shares of $\bf{Y}$ at the destination index of the matrix $\bf{G}$. During write, each party's share of the destination index is protected. However, the actual destination index $j$ is revealed from the final matrix, since the values in the final matrix are zero for all indices other than $j$. To solve this problem, $CP_p$ adds random noise to mask its share of 
$ \llbracket {\bf G} \rrbracket_p $ while sharing it with the other parties. Thus, the final matrix is masked by all parties, and the destination index cannot be determined by observing the values in the matrix. The procedure to remove the noise from the final result is discussed in Section~\ref{sec:cryptmpl_mask}.

\subsection{Updating the feature matrix}
\label{sec:cryptmpl_agg}

Unlike read and write, secure aggregation ($\mathcal{F}_{SA}$ in Alg.~\ref{alg:Server_MPL_Client_Noise}, Line 5) can be implemented using standard SMPC techniques.
The output feature matrix of the same size as the input feature matrix is initialized by each party as $ \llbracket \mathbf{A^*} \rrbracket  =  \llbracket \mathbf{0} \rrbracket $. 
Executing one round of read and write protocols process one edge, where the intermediate result matrix $ \llbracket \mathbf{G} \rrbracket $ contains the feature vector of node $ \llbracket \mathbf{S}[i] \rrbracket $ at index $ \llbracket \mathbf{D}[i] \rrbracket $ after processing the $i$-th edge. $CP_p$ updates $ \llbracket \mathbf{A^*} \rrbracket$ with the result: $ \llbracket \mathbf{A^*} \rrbracket _p =  \llbracket \mathbf{A^*} \rrbracket _p +  \llbracket \mathbf{G} \rrbracket _p$.

\subsection{Putting things together with preprocessing} 
\label{sec:cryptmpl_mask}

The protocols $\mathcal{F}_{SR}$, $\mathcal{F}_{SW}$ and $\mathcal{F}_{SA}$ process all the edges in the graph. During read and write, each party masks the shares of the feature matrices to protect the graph data. Masking a matrix with noise involves adding random values to the original matrix. Here, we describe the preprocessing stage executed at the client side to help the parties generate the noise matrices to mask the original data and eliminate the noise from the output for the correct result of the MPL.

As the client has the graph data structure, it can execute the message-passing on a noise matrix $\boldsymbol{\xi} \in \mathbb{R}^{N \times K}$ to get a feature matrix $\boldsymbol{\xi^*}$ and share both $\boldsymbol{\xi}$ and $\boldsymbol{\xi^*}$ with the SMPC parties in a secret-shared manner. Each party can mask its feature matrix with the share of $\boldsymbol{\xi}$ and calculate the feature matrix $\mathbf{A_\xi^*}$ by executing MPL on the masked feature matrix. Finally, it removes the noise $\boldsymbol{\xi^*}$ from the $\mathbf{A_\xi^*}$ to generate the actual result $\mathbf{A^*}$. The steps of this process are as follows:

\begin{itemize}[leftmargin=*,topsep=0pt,itemsep=0ex,partopsep=0ex,parsep=0ex]
\item The client calculates the effect of noise on each node after executing an MPL round (Eq. \ref{eqn:overall_noise}).
\item Each party executes the MPL to get the output feature matrix on the masked node features (Eq. \ref{eqn:MLP_with_noise}).
\item Each party removes the effect of noise to obtain the correct feature matrix (Eq. \ref{eqn:remove_noise}).
\end{itemize}
{\small
\begin{equation}
\boldsymbol{\xi^*}[j]=\sum_{j \in \mathcal{N}(i)} \boldsymbol{\xi}[j]
\label{eqn:overall_noise}
\end{equation}
\begin{equation}
\mathbf{A_\xi^*}[i]=\sum_{j \in \mathcal{N}(i)} \mathbf{A}[j] + \boldsymbol{\xi}[j]
\label{eqn:MLP_with_noise}
\end{equation}
\begin{equation}
\mathbf{A^*}[i]=\mathbf{A_\xi^*}[i] - \boldsymbol{\xi^*}[i]
\label{eqn:remove_noise}
\end{equation}
}


To mask the feature matrices, $CP_p$ creates a noise matrix $\boldsymbol{\xi}_p$, and adds $\boldsymbol{\xi}_p$ to the share of the feature matrix $ \llbracket \mathbf{A} \rrbracket _p$. However, if the same noise is used to mask $ \llbracket \mathbf{A} \rrbracket _p$ while processing each edge, an attacker can identify the node degree based on the number of times the same value is accessed by a party. To prevent this, $CP_p$ needs to generate different noise matrices $\boldsymbol{\xi}_{pr}$ at each round $r$ of the read and write operations. In read and write, noise is added to the party's own share, and to the matrices received from other parties, such that the matrices cannot be recognized at the end of the process, and the source and destination indices cannot be inferred.

During the initialization stage, the client shares different integer values as seeds to each party, which are used in a pseudo-random function (PRF) to generate all the rotation amounts and noise matrices. At the client side, similar noise matrices are used to compute the effect of noise $\boldsymbol{\xi^*}$. The client shares $\boldsymbol{\xi^*}$ in secret-shared manner with each party. After processing all edges, each party removes the noise $\boldsymbol{\xi^*}$ to retrieve the correct feature matrix. 

\textbf{Processing edges in batches.} To execute an MPL, CryptMPL needs to process all edges in the graph, which involves $M$ rounds of read and write executions, where $M$ is the number of edges. To reduce the number of rounds, and consequently the computation and communication overhead, CryptMPL processes the edges in batches. Our batching technique executes MPL with low overhead while preserving the privacy of the graph structure. The number of edges in a batch is configurable. For example, a batch of 3 nodes in secret-shared format $ \llbracket a_1, a_2, a_3 \rrbracket $ can be represented as relative indices $[0, a_2 - a_1, a_3 - a_1]$ in plain text with respect to $a_1$, which is still represented in secret-shared format as $ \llbracket a_1 \rrbracket $. Both source and destination indices can be represented in this way. 

For batching, the client divides the edges into batches and calculates the relative indices with respect to the first index of a batch. The first indices of all batches from $\mathbf{S}$ and $\mathbf{D}$ are stored as two vectors $ \llbracket \mathbf{S_f} \rrbracket $ and $ \llbracket \mathbf{D_f} \rrbracket $. The relative indices for all batches are concatenated to create two vectors $\mathbf{S_r}$ and $\mathbf{D_r}$ in plain text. The client uploads $\mathbf{ \llbracket X \rrbracket }$, $\mathbf{ \llbracket S_f \rrbracket }$, $\mathbf{ \llbracket D_f \rrbracket }$, $\mathbf{ \llbracket \boldsymbol{\xi}^* \rrbracket }$, $\mathbf{S_r}$, $\mathbf{D_r}$ and the seed to the SMPC parties.

To further reduce the number of communication rounds, each party concatenates masked feature matrices for all batches to create a matrix of size $(R, N, K)$, where the dimension of the feature matrix is $(N, K)$ and the number of batches is $R$. Each sub-matrix of size $(1, N, K)$ can be rotated by a different amount and the concatenated version can be passed to the other parties for read operation. In this way, the read operations for all batches can be executed in a single round. Similarly, write operations can be executed in a single round by concatenating ${\bf G}$ matrices for all batches. 
Finally, let us note that there is a trade-off between performance (fewer batches) and security (more batches). Using the relative order of the indices in each batch, the parties may infer the graph structure. The analysis in Section~\ref{sec:analysis} 
shows that the probability of correct reconstruction of the graph structure is $N^{-2R}$. 

\section{CryptMUL}
\label{sec:cryptmul_design}
Typically, a GNN comprises multiple FTLs, along with MPLs. In A-SS, the primary bottleneck in evaluating FTLs is the multiplication operation. Linear layers need matrix multiplications, and non-linear layers require element-wise multiplications. To generate Beaver Triples (discussed in Section~\ref{sec:Beaver_triple}) for multiplications in A-SS, previous studies~\cite{NEURIPS2021_27545182} have relied on techniques such as HE, oblivious transfer (OT), a trusted third party (TTP), or a combination thereof. However, using HE and OT is costly in terms of both computation and communication. Since CryptGNN is designed for MLaaS, it must scale to support numerous inference requests from each client. Consequently, repeatedly employing HE or OT is impractical. While using a TTP is less resource-intensive, it requires an additional third party that must not collude with the computing parties. Moreover, communication with the TTP is necessary for each multiplication operation.

To execute the multiplication operations in GNN, we design CryptMUL, which offers two benefits: (i) performing multiplications without a trusted server, and (ii) lower overhead due to preprocessing.
Our CryptMUL conducts offline preprocessing to generate auxiliary data, which can be used to easily create a set of Beaver triples~\cite{beaver1992efficient} required for multiplication operations in multiple inference requests from the same client, thereby improving performance while preserving data privacy. Although our protocols use existing HE- or OT-based techniques in the offline phase to generate auxiliary data for a client, our contribution lies in efficiently reusing this data for multiple inference requests from the same client with only one round of communication among the parties.

In this section, we describe the two CryptMUL protocols for: (a) secure matrix-multiplication, and (b) secure element-wise multiplication.
While we discuss these protocols in the context of GNN inference, they are generic and can be applied in similar scenarios for other types of model architectures.

\subsection{Secure matrix-multiplication}
\label{subsec:matrix_mul}

In order to compute the linear layers in GNNs, we must conduct matrix multiplication between the intermediate state matrix $\llbracket \mathbf{X} \rrbracket \in \mathbb{R}^{N \times K}$ and the parameter matrix of the linear layer $\llbracket \mathbf{Y} \rrbracket \in \mathbb{R}^{K \times K'}$ to transform $K$ features into $K'$ features. While the values of $K$ and $K'$ remain constant in GNN inference, the number of nodes in the graph, denoted as $N$, can vary with each inference request for the same GNN model.
To perform matrix multiplication efficiently in A-SS, we employ the Beaver triple technique, as explained in Section~\ref{sec:Beaver_triple}. By utilizing a Beaver triple ($\llbracket \bf{A} \rrbracket$, $\llbracket \bf{B} \rrbracket$, $\llbracket \bf{C} \rrbracket$), we can calculate $\llbracket \bf{Z} \rrbracket = \llbracket \bf{X} \rrbracket \otimes \llbracket \bf{Y} \rrbracket$. However, it is crucial to note that using the same triple to compute the same linear layer for two different inference requests may lead to privacy risks. This is because such a scenario reveals the differences ($\bf{U} = \bf{X}-\bf{A}$) and ($\bf{V} = \bf{Y}-\bf{B}$), and using the same $\bf{A}$ and $\bf{B}$ could disclose the relative changes in $\bf{X}$ and $\bf{Y}$ across different requests. Unfortunately, generating a new Beaver triple for each inference request using state-of-the-art techniques such as HE or OT is impractical due to their high overhead.

The following observations help us to solve this problem: 
\begin{itemize}
[leftmargin=*,topsep=0pt,itemsep=0ex,partopsep=0ex,parsep=0ex]
    \item The intermediate feature matrix $\bf{X}$ is derived from the input feature matrix of the graph. Revealing $\bf{U}$ does not disclose $\bf{X}$, unless $\bf{A}$ becomes known to any party. However, it is infeasible to use the same $\bf{A}$ for different inference requests, as it would reveal the relative changes in $\bf{X}$ across requests.
    \item The trained parameter matrix $\bf{Y}$ remains constant for a GNN model, and the same $\bf{Y}$ is used for all inference requests. Therefore, we can use the same $\bf{B}$ in the Beaver triple for all inference requests. As long as $\bf{B}$ remains unknown to the computing parties, revealing $\bf{V}$ will not disclose $\bf{Y}$.
\end{itemize}

\begin{figure}[!h]
\centering
  \subfloat[]
  {%
    \raisebox{1mm}{\includegraphics[width=0.45\linewidth]{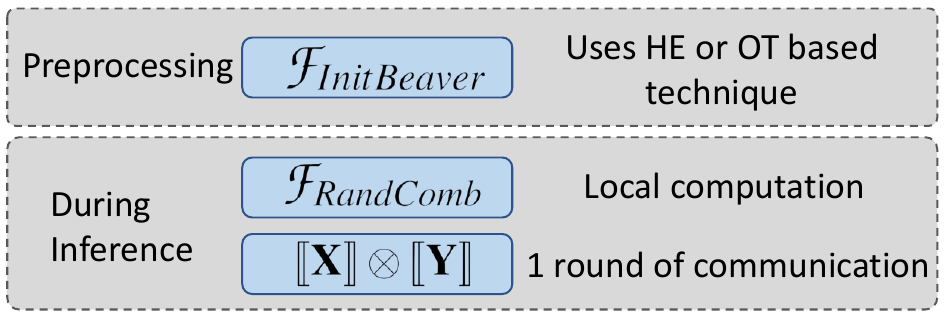}}
  }%
  \subfloat[]
  {%
    \includegraphics[width=0.45\linewidth]{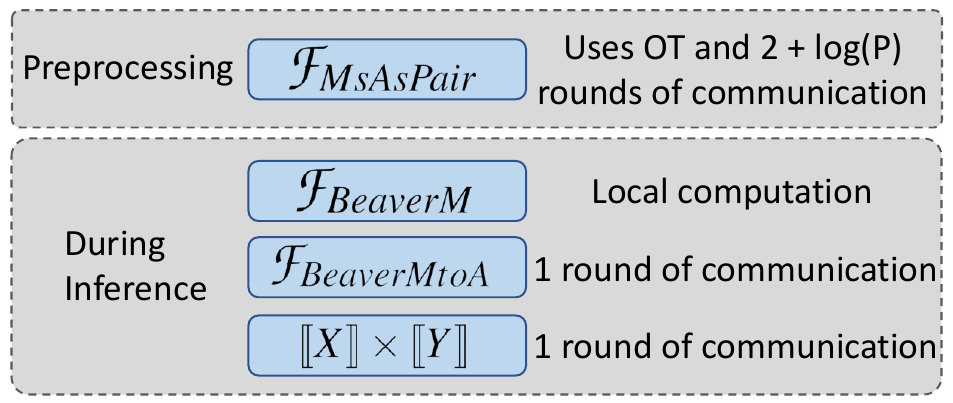}
  }%
\caption{(a)  Secure Matrix Multiplication Protocol (b)  Secure Element-wise Multiplication Protocol}
\label{fig:cryptmul_protocol_stack}
\end{figure}

Based on these observations, we plan to use the same $\bf{B}$ in all Beaver triples. However, $\mathbf{A}$ needs to be changed in different requests, and consequently, $\mathbf{C}$ needs to be adjusted to hold the property of Beaver triples. 
To derive a new set of matrices, denoted as ($\mathbf{A'}, \mathbf{B}, \mathbf{C'})$, from the initial Beaver triple ($\mathbf{A}, \mathbf{B}, \mathbf{C})$, we construct a row in $\mathbf{A'}$ through a linear combination of the rows of $\mathbf{A}$ and use a similar linear combination of rows from $\mathbf{C}$ to compute the corresponding row in $\mathbf{C'}$. Specifically, if we express $\mathbf{A'}[j] = \sum_{i=1}^{N}k_{ji} * \mathbf{A}[i]$, then $\mathbf{C'}$ can be computed as $\mathbf{C'}[j] = \sum_{i=1}^{N}k_{ji} * \mathbf{C}[i]$. Following this approach, we propose a secure protocol $\mathcal{F}_{MatMul}$ to perform matrix multiplication in a GNN's linear layer using the following steps as shown in Fig.~\ref{fig:cryptmul_protocol_stack}(a):

\begin{enumerate}[leftmargin=*,topsep=0pt,itemsep=0ex,partopsep=0ex,parsep=0ex]
\item During the preprocessing stage, $\mathcal{F}_{InitBeaver}$ generates an initial Beaver triple ($\llbracket \bf{A} \rrbracket, \llbracket \bf{B} \rrbracket, \llbracket \bf{C} \rrbracket$) using HE~\cite{mouchet2021multiparty} or OT~\cite{keller2016mascot} for each client. Here, $\mathbf{A} \in \mathbb{R}^{N \times K}$, $\mathbf{B} \in \mathbb{R}^{K \times K'}$ and $\mathbf{C} \in \mathbb{R}^{N \times K'}$. $K$ and $K'$ are fixed and depend on the number of input and output features of the linear layer respectively. $N$ is the maximum number of rows we may need to support and depends on the maximum number of nodes in a graph.

\item During inference, each party uses the same pseudo-random function $\mathcal{F}_{RandComb}$ to pick a random linear combination of the rows to modify $\llbracket \mathbf{A} \rrbracket$ to $\llbracket \mathbf{A'} \rrbracket$ and $\llbracket \mathbf{C} \rrbracket$ to $\llbracket \mathbf{C'} \rrbracket$. All parties execute the same operations locally to generate new $\llbracket \mathbf{A'} \rrbracket$ and $\llbracket \mathbf{C'} \rrbracket$ based on the number of nodes in the graph.

\item Use the new triple ($\llbracket \bf{A'} \rrbracket, \llbracket \bf{B} \rrbracket, \llbracket \bf{C'} \rrbracket$) to compute $\llbracket \mathbf{X} \rrbracket \otimes \llbracket \mathbf{Y} \rrbracket$.
\end{enumerate}

CryptGNN uses $\mathcal{F}_{MatMul}$ to execute a linear layer to compute the output feature states $\bf{Z}$ from the intermediate feature state $\bf{X}$ and the trained parameter matrix $\bf{Y}$ of the linear layer. The pre-processing step is executed once for each client to generate the initial Beaver triples. To further reduce communication costs, we can compute and reveal $\bf{V}$ once and use the same $\bf{V}$ (since $\bf{Y}$ and $\bf{B}$ are fixed) for subsequent inference requests from the same client. 

\subsection{Secure element-wise multiplication}
\label{app:subsec:element_mul}

Several types of FTLs require support for secure element-wise multiplication $\llbracket Z \rrbracket = \llbracket X \rrbracket \times \llbracket Y \rrbracket$, where neither $X$ nor $Y$ are assumed to be constant. In A-SS, we can compute the result of multiplication using a Beaver triple as discussed in Section~\ref{sec:Beaver_triple}. The maximum number of element-wise multiplications required for a single GNN inference request is predetermined by the specific model architecture. 
One approach is to pre-compute this specific number of Beaver triples and employ them during inference. However, as outlined in the preceding subsection, we should not use the same triple for multiple inference requests due to the potential risk of information leakage. Therefore, we introduce $\mathcal{F}_{ElemMul}$ to perform element-wise multiplications within the GNN layers by generating a fresh set of Beaver triples for each inference request. The steps followed in $\mathcal{F}_{ElemMul}$ (as shown in Fig.~\ref{fig:cryptmul_protocol_stack}(b)) are described below: 

\begin{enumerate}[leftmargin=*,topsep=0pt,itemsep=0ex,partopsep=0ex,parsep=0ex]
    \item $\mathcal{F}_{MsAsPair}$: At the pre-processing stage, the parties generate a set of numbers both in A-SS and M-SS format, which will be used in Step 3.
    \item $\mathcal{F}_{BeaverM}$: The parties generate the triples $(\llangle A \rrangle, \llangle B \rrangle, \llangle C \rrangle)$ in the multiplicative format.
    \item $\mathcal{F}_{BeaverMtoA}$: The parties communicate with each other and use the data generated in $\mathcal{F}_{MsAsPair}$ to convert the Beaver triple from M-SS to A-SS as ($\llbracket A \rrbracket$, $\llbracket B \rrbracket$, $\llbracket C \rrbracket$).
    \item The parties use the Beaver triples ($\llbracket A \rrbracket$, $\llbracket B \rrbracket$, $\llbracket C \rrbracket$) to compute $\llbracket X \rrbracket \times \llbracket Y \rrbracket$ following the steps in Section~\ref{sec:Beaver_triple}.
\end{enumerate}

Next, we describe the $\mathcal{F}_{MsAsPair}$, $\mathcal{F}_{BeaverM}$ and $\mathcal{F}_{BeaverMtoA}$ in more detail. 

\noindent\textbf{Generating additive-multiplicative pair: $\mathcal{F}_{MsAsPair}$.} At the pre-processing stage, $\mathcal{F}_{ElemMul}$ generates a list of pairs $\bf{AM}$ for each client, where $i$-th ($i \in \left| \mathbf{AM} \right|)$ element of $\bf{AM}$ is a pair $(\llbracket R_i \rrbracket, \llangle R_i \rrangle)$, i.e., a random value $R_i$ in A-SS and M-SS. The size of the list, $k = \left| \mathbf{AM} \right|$ depends on the maximum number of element-wise multiplications required for a GNN inference. If the total number of multiplications is $m$, then $k = 3 \times m$, since each multiplication operation uses 3 pairs from $\bf{AM}$ in $\mathcal{F}_{BeaverMtoA}$. To protect the relative changes in values between two different inference requests from the same client, it is necessary to use a different set of pairs for multiplication when processing an element at the same index of a GNN layer.
This can be done by shifting the elements of $\bf{AM}$ by 3 for each inference. In this way, $\mathcal{F}_{ElemMul}$ can support $m$ inference requests from the same client.

To generate a number in A-SS and M-SS in $\mathcal{P}$ parties, $\mathcal{F}_{MsAsPair}$ extends the algorithm in~\cite{xiong2020efficient}, which works only for two parties, to make it work for any number of parties. In~\cite{xiong2020efficient}, two parties $CP_1$ and $CP_2$ generates random numbers $X_1$ and $X_2$, and convert ($X_1 + X_2$) to the multiplicative secret-shared M-SS format $\llangle X_1 + X_2 \rrangle$, without revealing $X_1$ and $X_2$ to each other. Following this approach, $\mathcal{F}_{MsAsPair}$ can compute $\llbracket R_i \rrbracket$ and $\llangle R_i \rrangle$ for $i \in [1,.., P-1]$, where $R_i = X_i + X_{i+1}$. Here, the value $X_i$ is known to $i$-th party only. Since, $\llangle R_i \rrangle$ is in multiplicative format, each party can compute $\llangle R \rrangle = \prod_{i=1}^{P-1} \llangle R_i \rrangle$ locally. The parties can compute the additive share of $\llbracket R \rrbracket = \prod_{i=1}^{P-1} \llbracket R_i \rrbracket$ using the Beaver triples generated using a state-of-the-art technique~\cite{keller2016mascot}. Thus, $\mathcal{F}_{MsAsPair}$ can generate a pair ($\llbracket R \rrbracket$, $\llangle R \rrangle$), the A-SS and M-SS version of the same value $R$. 
Algorithm~\ref{alg:gen_AM} presents the steps required to generate a pair $(\llbracket R \rrbracket, \llangle R \rrangle)$) for a random value $R$.
Following this approach, we can generate $k$ pairs to prepare the list $\bf{AM}$ within the same communication round.

\begin{algorithm}[!h] 
\caption{Generate additive-multiplicative pair, $\mathcal{F}_{MsAsPair}$}
\label{alg:gen_AM}
\textbf{Output:} Generate additive share $\llbracket R \rrbracket$ and multiplicative share $\llangle R \rrangle$ of a random value $R$
\begin{algorithmic}[1]
    \FOR {$i \gets 1$ to $P-1$}
        \STATE $CP_i$ and $CP_{i+1}$ generates two random values $X_i$ and $X_{i+1}$. Thus, a $R_i = X_i + X_{i+1}$  is generated in A-SS format $\llbracket R_i \rrbracket$.
        \STATE Two parties communicates to generate multiplicative shares $\llangle R_i \rrangle \gets \llangle X_i + X_{i+1} \rrangle$ using ~\cite{xiong2020efficient}
        \STATE $\llbracket R_i \rrbracket_j \gets 0$ and $\llangle R_i \rrangle_j \gets 1$, for $j \neq i$, $j \neq (i+1)$
    \ENDFOR
    
    \STATE Compute $\llbracket R \rrbracket \gets \prod_{i=1}^{P-1} \llbracket R_i \rrbracket$ using Beaver triples generated using~\cite{keller2016mascot}, thereby each party gets a A-SS of $R$.
    \STATE Each party $p$ computes locally $\llangle R \rrangle_p \gets \prod_{i=1}^{P-1} \llangle R_i \rrangle_p$, thereby computing the multiplicative shares of $R$.
\end{algorithmic}
\end{algorithm}

\noindent\textbf{Generating Beaver triples in multiplicative format: $\mathcal{F}_{BeaverM}$.}
Each party $CP_i$ generates two random variables $A_i$ and $B_i$, and computes $C_i = A_i \times B_i$. In this way, $A_i$, $B_i$ and $C_i$ constitute the $A$, $B$ and $C$ in multiplicative share format. Since $C = \prod_{i=1}^{P} C_i = \prod_{i=1}^{P} A_i \times B_i = \prod_{i=1}^{P} A_i \times \prod_{i=1}^{P} B_i = A \times B$, the parties can generate Beaver triples in the M-SS format without any communication.

\noindent\textbf{Converting Beaver triples in additive format: $\mathcal{F}_{BeaverMtoA}$.}
$\mathcal{F}_{ElemMul}$ converts the Beaver triples from M-SS to A-SS 
using $\mathcal{F}_{BeaverMtoA}$, which internally calls $\mathcal{F}_{MtoA}$ to convert each value in the triples.
To convert a value $U$ from M-SS format $\llangle W \rrangle$ to A-SS format $\llbracket W \rrbracket$, $\mathcal{F}_{MtoA}$ selects a pair ($\llbracket R \rrbracket$, $\llangle R \rrangle$) from pre-computed list $\bf{AM}$, where $R$ is in A-SS and M-SS format as $\llbracket R \rrbracket$ and $\llangle R \rrangle$, respectively. Then, each party computes locally and communicates with each other to reveal the ratio ($\alpha$) of $W$ and $R$. Next, each party uses $\alpha$ and $\llbracket R \rrbracket$ to compute the additive share of $W$ as $\llbracket W \rrbracket$.

$\mathcal{F}_{MtoA}$ follows the below steps to convert a value $U$ from multiplicative format to additive format. 

\begin{enumerate}[topsep=0pt,itemsep=0ex,partopsep=0ex,parsep=0ex]
\item Pick a pair ($\llbracket R \rrbracket$, $\llangle R \rrangle$) from $\bf{AM}$.
\item Apply Extended Euclidean Algorithm~\cite{extended_euclidean} to compute the inverse of $R$ as $\llangle R^{-1} \rrangle$ in M-SS format~\cite{ghodosi2012multi}.
\item Each party $CP_i$ computes locally the product of $\llangle W \rrangle$ and $\llangle R^{-1} \rrangle$ and reveals the ratio $\alpha$.
\item $CP_i$ computes $\llbracket W \rrbracket_i \gets \alpha \times \llbracket R \rrbracket_i$ to get $W$ in A-SS.
\end{enumerate}

$\mathcal{F}_{BeaverMtoA}$ converts each value of Beaver triple from M-SS $\llangle A \rrangle, \llangle B \rrangle, \llangle C \rrangle$ to A-SS format $\llbracket A \rrbracket, \llbracket B \rrbracket, \llbracket C \rrbracket$ respectively using $\mathcal{F}_{MtoA}$.
\section{System Analysis}
\label{sec:analysis}
\subsection{Security Analysis}
\label{sec:security}

This section proves the protocol security throughout the execution of CryptGNN, which preserve the privacy of the client's (DO's) input graph and the model owner's (MO's) model parameters against the threat model $TM$.
CryptGNN follows the standard A-SS approach to store model parameters and uploads graph data to the computing parties. Thus, data at rest (model parameters, feature matrix, source and destination index of each edge) are information-theoretically secure \cite{10.1007/978-3-642-40203-6_1} against adversaries following Axiom 1.

\textbf{Axiom 1.} A value $x$ is information-theoretically secure in A-SS format even if $P-1$ out of $P$ parties collude.

To provide the security analysis in a structured way, we consider three different cases within the threat model $TM$,
(a) $TM_P$: At most $P-1$ parties may collude to learn DO's input graph data or MO's GNN model parameters (but they do not collude with DOs or MO), (b) $TM_M$: The colluding parties in $TM_P$ may collude with MO to gain access to DO's private graph data, (c) $TM_D$: The colluding parties in $TM_P$ may collude with a data owner $DO_{fake}$ to learn MO's GNN model parameters or the input graph the other DOs. To prove the security of our protocols against $TM_P$, $TM_M$ and $TM_D$, we adopt the standard simulation-based security definitions from~\cite{10.1007/978-3-030-57808-4_19}:

\textbf{Definition 1.} Let parties $CP_1, \cdots, CP_P$ engage in a protocol $\pi$ that computes function $\mathcal{F}(in_1, \cdots, in_P) = (out_1, \cdots, out_P)$, where $in_i$ and $out_i$ denote the input and output of party $CP_i$, respectively. Let, $VIEW_{\pi}(CP_i)$ denote the view of participant $CP_i$ during the execution of protocol $\pi$. More precisely, $CP_i$'s view is formed by its input, internal random coin tosses $r_i$, pseudo-random values $pr_i$, as well as messages $m_1, \cdots, m_k$ passed between the parties during protocol execution: $VIEW_{\pi}(CP_i) = (in_i, r_i, pr_i, m_1, \cdots, m_k)$. Let $I$ denote a subset of at most $P-1$ parties that collude in our threat model $TM_P$. $VIEW_{\pi}(I)$ denote the combined view of participants in $I$ during the execution of protocol $\pi$ (i.e., the union of the views of the parties in $I$), and $\mathcal{F}_I(in_1, \cdots, in_P)$ denote the projection of $\mathcal{F}(in_1, \cdots, in_P)$ on the coordinates in $I$ (i.e., $\mathcal{F}_I(in_1, \cdots, in_P)$ consists of the output of function $\mathcal{F}$ of the colluding parties). We say that the protocol $\pi$ is secure against $TM_P$ if for each coalition of size at most $P-1$ there exist a probabilistic polynomial time (PPT) simulator $S_I$ such that $\{S_I(in_I, \mathcal{F}_I(in_1, \cdots, in_P)), \mathcal{F}(in_1, \cdots, in_P)\} \equiv \{VIEW_{\pi}(I), (out_1, \cdots, out_P)\}$, where $in_I = \bigcup_{CP_i \in I}{in_i}$ and $\equiv$ denotes computational or statistical indistinguishability.

\textbf{Definition 2.} In $TM_M$, we consider the model parameters $\Theta$ are also known to the colluding parties $I$, and the protocol $\pi$ is secure against $TM_M$ if there exists a probabilistic polynomial time (PPT) simulator $S_{I}$ such that $\{S_I(in_I, \Theta, \mathcal{F}_I(in_1, \cdots, in_P)), \mathcal{F}(in_1, \cdots, in_P)\}\\ \equiv \{VIEW_{\pi}(I), (out_1, \cdots, out_P)\}$, where $in_I = \bigcup_{CP_i \in I}{in_i}$.

\textbf{Definition 3.} In the case of $TM_D$, a data owner $DO_{fake}$'s private input  ($\mathbf{X}_{f}, \mathbf{S}_{f}, \mathbf{D}_{f}, \mathbf{\xi}_{f}$) is known to the colluding parties, and the protocol $\pi$ is secure against $TM_D$, if there exists a probabilistic polynomial time (PPT) simulator $S_{I}$ such that $\{S_I(in_I, \mathbf{X}_{f}, \mathbf{S}_{f}, \mathbf{D}_{f}, \mathbf{\xi}_{f},\\ \mathcal{F}_I(in_1, \cdots, in_P)), \mathcal{F}(in_1, \cdots, in_P)\} \equiv \{VIEW_{\pi}(I), (out_1, \cdots, out_P)\}$, where $in_I = \bigcup_{CP_i \in I}{in_i}$.

The following theorems ensure protocol security throughout the execution of CryptGNN. 

\noindent\textit{\textbf{Theorem 1.}} The node features are protected against $TM$ in CryptMPL.

\noindent\textit{\textbf{Proof.}} To execute the message-passing, CryptMPL executes secure read, write and aggregation protocols for each edge of the input graph data stored in A-SS format as ($\llbracket \mathbf{X} \rrbracket, \llbracket \mathbf{S} \rrbracket, \llbracket \mathbf{D} \rrbracket)$. To prove this theorem we consider that the simulator $S_{CryptMPL}$ calls the simulators $S_{SR}$, $S_{SW}$ and $S_{SA}$ of $\mathcal{F}_{SR}$, $\mathcal{F}_{SW}$ and $\mathcal{F}_{SA}$ respectively.    

In the secure read protocol with data masking $\mathcal{F}_{SR}$, each party $CP_i$ exchanges the share of feature matrix $\llbracket \mathbf{A} \rrbracket_i$ with other parties. We consider the case where the parties do not rotate the feature matrix to prove this theorem. In this case, a party has all the shares of the feature matrix and can reconstruct the original feature matrix by taking the sum of the shares as, $\mathbf{A} = \sum\limits_{p=1}^P \llbracket \mathbf{A} \rrbracket_p$. To protect the feature matrix, $CP_i$ masks data with noise matrix $\boldsymbol{\xi}_i$ and shares $\llbracket \mathbf{A_{\xi}} \rrbracket_i = \llbracket \mathbf{A} \rrbracket_i + \boldsymbol{\xi}_i$ with other parties. 

As in Definition 1, $I$ denotes the set of at most $P-1$ parties that collude in our threat models. We build a simulator $S_{SR}$, which simulates the view of parties in $I$. In the simulated view, $S_{SR}$ can compute $\mathbf{A_\xi} = \sum\limits_{p=1}^P \llbracket \mathbf{A_{\xi}} \rrbracket_p = \sum\limits_{p=1}^P \llbracket \mathbf{A} \rrbracket_p + \sum\limits_{p=1}^P \boldsymbol{\xi}_p$, as it has all the masked shares $\llbracket \mathbf{A_{\xi}} \rrbracket_i$ for $i \in [1, \cdots, P]$. However, $\mathbf{A_\xi}$ is uniformly random in $I$'s View, since there is at least one mask matrix $\boldsymbol{\xi}_i$ (from the non-colluding party) which is unknown to $I$. Therefore, the distribution over the real $\mathbf{A_\xi}$ received by the colluding parties and over the simulated $\mathbf{A_\xi}$ generated by $S_{SR}$ is identically distributed. 

Similarly, during the secure write, the share of the temporary matrix $\llbracket \mathbf{G} \rrbracket_i$ from $CP_i$ is masked with a mask matrix. In the simulated view of $\mathcal{F}_{SW}$, $I$ can compute $\mathbf{G_\xi} = \sum\limits_{p=1}^P \llbracket \mathbf{G_{\xi}} \rrbracket_p = \sum\limits_{p=1}^P \llbracket \mathbf{G} \rrbracket_p + \sum\limits_{p=1}^P \boldsymbol{\xi}_p$, which is uniformly random in $I$'s View, since there is at least one mask matrix $\boldsymbol{\xi}_i$ (from non-colluding party) which is unknown to $I$. Therefore, the distribution over the real $\mathbf{G_\xi}$ received by the colluding parties and over the simulated $\mathbf{G_\xi}$ generated by the simulator is identically distributed. 

The secure aggregation operation $\mathcal{F}_{SA}$ does not require exchanging data with other parties, as the addition operations on additive secret-shared data can be executed locally. Since the views produced by the simulator $S_{CryptMPL}$ in the read and write protocols are indistinguishable from the parties’ views in the real protocol execution, the views remain indistinguishable after the simulation of $S_{SA}$, which proves that the input and output node features of CryptMPL are secure against $TM_P$, while processing an edge. CryptMPL follows the same procedure to process all the edges. Since the collective view of the protocol execution while processing each edge is computationally indistinguishable from a simulated view, the node features are protected in $\mathcal{F}_{CryptMPL}$. Finally, each party locally subtracts the noise from their share of the result. Since no data is exchanged in this step, the node features remain protected. Processing multiple edges in a batch has no additional impact on the node features, ensuring their security during batching.

The node features are also protected against $TM_M$, since the model parameters $\Theta$ are not involved in any step of $\mathcal{F}_{CryptMPL}$.
In the case of $TM_D$, the input of the data owner $DO_{fake}$ is known to the colluding parties $I$. However, since each client generates its own mask matrix, knowing the $DO_{fake}$'s data does not reveal other client's private input. Thus, for each client, the views of $I$ produced by the simulator $S_{CryptMPL}$ remain indistinguishable from the parties’ views in the real protocol execution, which proves that the input and output node features are secure against $TM_D$.

Since, the node features are secured against $TM_P$, $TM_M$, and $TM_D$, thereby protected against the threat model $TM$. 

\noindent\textit{\textbf{Theorem 2.}} The graph structure is secured against $TM$ in CryptMPL with probability of correct reconstruction at most $N^{-2R}$, where $N$ is the number of nodes in the graph and $R$ is the number of batches.

\noindent\textit{\textbf{Proof.}} To prove this theorem, first, we analyze the protocols followed to process an edge in CryptMPL. During the read operation, $CP_i$ shifts the share of the source node index $\llbracket S \rrbracket_i$ by a random amount $r_i$. Thus, the simulator $S_{SR}$ of $\mathcal{F}_{SR}$ gets an aggregated value as $S' = \sum\limits_{p=1}^P \llbracket S \rrbracket_p + \sum\limits_{p=1}^P r_p$. Since at least one $r_i$ from the $i$-th party is unknown to the view of $S_{SR}$, the source index is uniformly random in the $I$'s view. The destination index for an edge is also protected in the view of the simulator $S_{SW}$ of $\mathcal{F}_{SW}$, since the parties use the share of that index to rotate the intermediate matrix locally and do not share the destination index during the write operation. 

Since both source and destination indices are protected, if the colluding parties $I$ try to estimate the source-destination pair, the probability of correct estimation is $\frac{1}{N \times (N-1)}$, where $N$ is the number of nodes in the graph. Processing multiple edges in CryptMPL does not reveal additional information. Since CryptMPL rotates the matrices by different amounts and uses different noise matrices for each edge, an adversary can not learn anything from the access pattern. Similar to processing an edge in each round, the probability of correct estimation of source-destination pairs is $\frac{1}{N \times (N-1)}$ for a batch in case of batch processing. Thus, to process all edges in $R$ batches in CryptMPL, the probability of correct reconstruction of graph structure is $N^{-2R}$. Therefore, increasing the number of batches ensures stronger security against $TM_P$. 

Similar to the logic described in Theorem 1, the graph structure is also protected against $TM_M$ and $TM_D$, thereby it is protected against $TM$.

\noindent\textit{\textbf{Lemma 1.}} Let $A$ and $B$ be two secrets encrypted using A-SS in a $P$-party SMPC setting, represented as $\llbracket A \rrbracket$ and $\llbracket B \rrbracket$, respectively. 
Let the linear combination of $\llbracket A \rrbracket$ and $\llbracket B \rrbracket$ be $\llbracket C \rrbracket = a \cdot \llbracket A \rrbracket + b \cdot \llbracket B \rrbracket$, where $a$ and $b$ are public coefficients. The shares of $C$ preserve the information-theoretic security of the original secrets against $TM$.

\noindent\textit{\textbf{Proof.}} We analyze the combined view of the participants in $I$ as defined in Defintion 1. We build a simulator $S_{LC}$, which simulates the view of parties in $I$. In the simulated view, $S_{LC}$ can compute $C = a \cdot \sum_{i=1}^P \llbracket A \rrbracket + b \cdot \sum_{i=1}^P \llbracket B \rrbracket$. However, since at least one share of $A$ and $B$ is unknown to $I$, the distribution over the real $C$ received by the colluding parties and the simulated $C$ generated by the simulator is identically distributed.

\noindent\textit{\textbf{Theorem 3.}} In CryptGNN, the DO’s input graph and the MO’s model parameters are secured against $TM$ within the FTLs using the CryptMUL protocols.

\noindent\textit{\textbf{Proof.}} Considering the scenario where each client may upload data for numerous inference requests, to prove the security of DO's input graph and MO's model parameters in CryptGNN, we demonstrate that the CryptMUL protocols employed to generate a fresh set of Beaver triples for matrix multiplication ($\mathcal{F}_{MatMul}$) and element-wise multiplication ($\mathcal{F}_{ElemMul}$) operations in FTLs are secure.

\textit{$\mathcal{F}_{MatMul}$ is secure against $TM$.} To prove $\mathcal{F}_{MatMul}$ is secure, we consider a simulator $S_{MatMul}$ that uses the simulators $S_{InitBeaver}$ and $S_{RandComb}$ of $\mathcal{F}_{InitBeaver}$ and $\mathcal{F}_{RandComb}$ respectively. 

In $\mathcal{S}_{MatMul}$, $S_{InitBeaver}$ generates the initial Beaver triples. We refer to ~\cite{keller2016mascot} for the security proof that shows $\mathcal{F}_{InitBeaver}$ is secure against $TM_P$. Each element of the generated triple, $\mathbf{A}$, $\mathbf{B}$, and $\mathbf{C}$ is secure in additive secret-shared format according to Axiom 1.

To compute $\llbracket \mathbf{X} \rrbracket \otimes \llbracket \mathbf{Y} \rrbracket$, $S_{RandComb}$ modifies $\mathbf{A}$ and $\mathbf{C}$ using the same linear combination of the rows (Step 2 in Section ~\ref{subsec:matrix_mul}). Although, the combination of the rows used to generate $\mathbf{A'}$ and $\mathbf{C'}$ is known to the view of $S_{RandComb}$, since the elements in each row involved in the computation are in the A-SS domain, $\mathbf{A'}$ and $\mathbf{C'}$ remain secure against the threat models (as shown in Lemma 1).

Following Step 3 of $\mathcal{F}_{MatMul}$, the matrices $\mathbf{U} = \mathbf{X} - \mathbf{A'}$ and $\mathbf{V} = \mathbf{Y} - \mathbf{B}$ are revealed to $S_{MatMul}$. Since, $\mathbf{A'}$ and $\mathbf{B}$ are unknown to $S_{MatMul}$, the distribution over elements in the private inputs $\mathbf{X}$ and $\mathbf{Y}$ remain identically distributed. $\mathcal{F}_{MatMul}$ uses a newly generated matrix $\mathbf{A'}$ for each inference request. Therefore, revealing $\mathbf{U}$ will not reveal the relative changes in the private input $\mathbf{A}$ in two different requests.
Finally, $\mathbf{U}$ and $\mathbf{V}$ are used to compute $\llbracket \mathbf{Z} \rrbracket = \llbracket \mathbf{X} \rrbracket \otimes \llbracket \mathbf{Y} \rrbracket$, which does not involve any data sharing between the parties. Therefore, the simulated view is identical to the real view of $I$. This proves that $\mathcal{F}_{MatMul}$ is secure against the threat model $TM_P$.

In the case of $TM_M$, the model parameters $\Theta$ are known, which means the input $\mathbf{Y}$ of the matrix multiplication in the linear layer is known to $I$. Using $\mathbf{V}$ and $\mathbf{Y}$, then $I$ can learn $\mathbf{B}$. However, since $\mathbf{A'}$ and $\mathbf{C'}$ are protected, the distribution over the private input $\mathbf{X}$ and the output $\mathbf{Z}$ received by $I$ and over the simulated matrices generated by the simulator are identically distributed.

In the threat model $TM_D$, using $DO_{fake}$'s private input, the colluding parties $I$ can learn $\mathbf{X}$ and consequently $\mathbf{A'}$ from $\mathbf{U}$ in the inference requests from that data owner. However, $\mathbf{B}$ and $\mathbf{C'}$ remain protected, and the distribution over a linear layer's private input (model parameter) $\mathbf{Y}$ and the output $\mathbf{Z}$ received by $I$ and over the simulated matrices generated by the simulator are identically distributed. Since, $\mathcal{F}_{InitBeaver}$ generates a fresh set of Beaver triples for each client, learning $\mathbf{A'}$ using $DO_{fake}$'s input by $I$ does not help to learn other DO's input. Thus, $\mathcal{F}_{MatMul}$ is secure against $TM_D$.

\textit{$\mathcal{F}_{ElemMul}$ is secure against $TM$.}
At the pre-processing stage, \linebreak[4] $\mathcal{F}_{ElemMul}$ generates additive and multiplicative shares of random values for two parties using oblivious transfer. We refer to~\cite{xiong2020efficient} for the proof of this step. In $\mathcal{F}_{MsAsPair}$, the parties $CP_i$ and $CP_{i+1}$ for $i \in [1, \cdots, P-1]$ generate $P-1$ numbers of $R_i$ values. Then the additive share $\llbracket R_i \rrbracket$ are multiplied using Beaver triples generated using the secure protocol used in ~\cite{keller2016mascot}. The multiplicative shares $\llangle R_i \rrangle$ are used locally to compute the multiplicative shares of $R$. Since at least one $R_i$ value is unknown to $I$, the distribution over the real pair ($\llbracket R \rrbracket$, $\llangle R \rrangle$) received by $I$ and over the simulated ($\llbracket R \rrbracket$, $\llangle R \rrangle$) generated by the simulator is identically distributed.

The simulator $S_{BeaverM}$ of $\mathcal{F}_{BeaverM}$ generates the Beaver triples in multiplicative format, which does not require any communication among the parties. Since the parties in $I$ do not receive any data from other parties, the simulated view is identical to the real view. To covert multiplicative shares to additive shares, the simulator $S_{BeaverMtoA}$ of $\mathcal{F}_{BeaverMtoA}$ uses the additive-multiplicative pairs generated in $\mathcal{F}_{MsAsPair}$. We refer to the proof of protocol $SecMulResh$ described in~\cite{9404811} to prove that $\mathcal{F}_{BeaverMtoA}$ is secure against $TM_P$. Thus, $\mathcal{F}_{ElemMul}$ is secure, since its sub-protocols are proven to be secure. 

To compute element-wise multiplication in different layers of GNN, we precompute the required amount of additive-multiplicative pairs in $\mathcal{F}_{MsAsPair}$. For each inference request, $\mathcal{F}_{BeaverM}$ generates a fresh Beaver triple in multiplicative format and $\mathcal{F}_{BeaverMtoA}$ uses a different pair from $\mathcal{F}_{MsAsPair}$ to compute an element-wise multiplication. Thus, the ratio $\alpha$ recovered in $\mathcal{F}_{BeaverMtoA}$ does not reveal any relative value for two different inference requests.

In the case of $TM_M$, the simulator may learn $B$ in the Beaver triple from $V$, if the model parameter is used as $Y$ in the multiplication step and it is known to $I$. However, the elements $A$ and $C$ are still protected, which are related to the DO's private input and the result of the multiplication. Therefore, the DO's private input and the private output are secure against $TM_M$.

In the threat model $TM_D$, using $DO_{fake}$'s private input, the colluding parties $I$ can learn $A$ of the Beaver triple, but the model parameters and the final result are protected since $B$ and $C$ are unknown. Furthermore, the other DO's data is also protected, since we generate a new set of (additive-multiplicative) shares for each client, which are used to generate the Beaver triples.
\subsection{Overhead Analysis}
The protocols in CryptGNN are able to avoid high overhead, while providing stronger privacy guarantees than existing solutions by leveraging application-specific knowledge.
Unlike general-purpose frameworks such as CrypTen, which rely heavily on costly matrix multiplications for message passing, CryptMPL minimizes this overhead by executing custom secure protocols for read, write, and aggregation operations. In CryptMPL, data privacy is preserved using masking techniques, and efficiency is further enhanced through batch processing. CryptGNN’s secure multiplication protocol, CryptMUL, leverages key characteristics of GNN inference (e.g., known model architecture and fixed number of parameters) to enable secure and efficient execution.

\textbf{Overhead analysis of CryptMPL.} We present the overhead analysis of CryptMPL in terms of: (i) the number of nodes in the graph $N$, (ii) the number of edges $M$, (iii) the number of features of each node $K$, (iv) the number of computing parties $P$, and (v) the number of batches $R$ to process $M$ edges.

\setlength{\heavyrulewidth}{1.5pt}
\setlength{\abovetopsep}{4pt}
\begin{table}[ht!]
\centering
\caption{Performance comparison of secure message-passing}
\resizebox{1.0\linewidth}{!}{
\begin{tabular}{*3c}
\toprule
{}              & CryptMPL  & AdjacencyMatrix\\
\midrule
Computation Cost (Client)        &    $O(N \times K \times P^2 \times R)$      &   $O(N^2 \times K)$  \\
Computation Cost (Each party)   &    $O(N \times K \times P \times R)$   &   $O(N^2 \times K)$   \\
Communication Cost (Client to each CP)        &    $(N \times K + M \times 2 + P) \times L$    &   $(N^2 + 2 \times N \times K) \times L$  \\
Communication Cost (Each CP to others)   &  $(N \times K \times R + M) \times P \times L$    &    $(N^2 + 2 \times N \times K) \times P \times L$  \\

\bottomrule
\label{table:cost_comparison}
\end{tabular}
}
\end{table}

CryptMPL preserves data privacy at the cost of a certain level of computational and communication overhead among servers. It also introduces overhead on the client side, as the client computes a noise matrix at the pre-processing stage and uploads the noise-matrix along with the graph data. 
Table~\ref{table:cost_comparison} presents the overhead of CryptMPL and compares it with that of an adjacency matrix–based approach. The computation cost at the client side is proportional to the size of feature matrix $(N \times K)$, the number of batches ($R$), and the number of parties ($P^2)$, since it requires to compute the effect of noise added by each party on its own data and on the data of the other parties.
To protect the feature matrix and graph structure, CryptMPL adds noise and rotates matrices by a random amount. Overall, there are $2 \times P \times R$ numbers of rotations and $(P+1) \times R + 1$ numbers of addition of matrices of size $(N,K)$ by each server. Similar to the plain text solution, each party needs to add two $(1,K)$ sized vectors $M$ times. Therefore, the computation overhead with respect to the plain text version is $O(N \times K \times P \times R)$. These computations can be performed independently in a multi-threaded way (or transferred to GPUs) to make them faster.
Additionally, to process all $M$ edges in $R$ batches, each server sends a total of $(N \times K \times R + M) \times P \times L$ bits to the other servers, where ($N \times K$) is the size of the feature matrix in a batch and $L$ is the number of bits required to represent a share of a value in A-SS format. As described in Section~\ref{sec:cryptmpl_mask}, CryptMPL processes all batches in a single round by transferring all bits for the $R$ batches at once, thereby reducing propagation and queuing delays~\cite{7445500}.

CryptMPL has significantly less overhead than MPL using an adjacency matrix, which requires multiplying two matrices of size $(N,N)$ and $(N,K)$. 
In the adjacency matrix-based approach, either the client or the trusted server needs to distribute 3 matrices $\mathbf{A} \in \mathbb{R}^{N \times K}, \mathbf{B} \in \mathbb{R}^{N \times N}, \mathbf{C} \in \mathbb{R}^{N \times K}$ as the Beaver triples to $P$ parties to support secure multiplication in A-SS domain following ~\cite{crypto-1991-1013}.

\textbf{Overhead analysis of CryptMUL.}.
In $\mathcal{F}_{MatMul}$, to compute $\llbracket \mathbf{Z} \rrbracket = \llbracket  \mathbf{X} \rrbracket \otimes \llbracket \mathbf{Y} \rrbracket$,  $\mathbf{X} \in \mathbb{R}^{N \times K}, \mathbf{Y} \in \mathbb{R}^{K \times K'}, \mathbf{Z} \in \mathbb{R}^{N \times K'}$ during inference, the parties locally generate a new Beaver triple from the pre-computed Beaver triple. The computation cost in this process is $O(N^2 \times (K + K'))$.
This process does not require any communication among the computing parties, while~\cite{NEURIPS2021_27545182} requires one round of communication between the computing parties and the trusted server to get a new Beaver triple. The cost associated with computing matrix-multiplication using the Beaver triple is the same (one round of communication and local computation) in both ~\cite{NEURIPS2021_27545182} and $\mathcal{F}_{MatMul}$. Thus, the overall cost of matrix-multiplication using CryptMUL is lower compared to~\cite{NEURIPS2021_27545182}, since the communication overhead is reduced through the local computation in $\mathcal{F}_{MatMul}$.

To compute element-wise multiplication, $\llbracket Z \rrbracket = \llbracket X \rrbracket \times \llbracket Y \rrbracket$, $\mathcal{F}_{ElemMul}$ requires each party to locally generate a random Beaver triple $(\llangle A \rrangle, \llangle B \rrangle, \llangle C \rrangle)$. Then, one round of communication among the parties is required to convert the Beaver triple from M-SS to A-SS format.  ~\cite{NEURIPS2021_27545182} also requires a round of communication between the parties and the trusted server to obtain the Beaver triple. The cost associated with computing element-wise multiplication using the Beaver triple is the same (one round of communication and local computation) in both~\cite{NEURIPS2021_27545182} and $\mathcal{F}_{ElemMul}$. Therefore, the overhead of both CryptMUL and ~\cite{NEURIPS2021_27545182} is of the same order.
\section{Evaluation}
\label{sec:evaluation}
\label{subsec:eval_cryptMPL}

We implement a CryptGNN prototype in Python and conduct experiments to compare its performance with baselines. To create arithmetic shares of the private data and to implement FTLs, we use CrypTen~\cite{NEURIPS2021_27545182}.
We implement CryptMPL with client-side data preprocessing and server-side batching.
Additionally, we develop CryptMUL protocols for secure matrix multiplication and element-wise multiplication.
To avoid using a trusted party, we replace the multiplication operations in the FTLs of CrypTen using our CryptMUL. We use $L=64$ bits to represent the values in A-SS format. We perform the experiments on a 3.4GHz Intel Core i7, with the parties running in separate processes. We also use AWS instances to evaluate CryptGNN in a realistic distributed cloud setting.
We conduct each experiment 30 times and report the average execution time.
For evaluation, we use benchmark datasets for graph classification tasks (FAUST~\cite{bogo2014faust}, TUDataset (PROTEINS, ENZYMES)~\cite{DBLP:journals/corr/abs-1910-12091}) and the well-known GIN~\cite{xu2018how} architecture as a GNN model. Let us note that, CryptGNN protocols can be extended to support other complex message-passing GNN architectures (beyond GIN) by incorporating additional operations (e.g., node sampling, concatenation, etc.) using standard SMPC techniques or by designing efficient protocols.

\subsection{Overall CryptGNN Performance}
\label{subsec:GIN}
We use FAUST, PROTEINS, and ENZYMES datasets. For each dataset, 70\% of the graphs are used for training the GIN model; the remaining graphs are considered as the client's private input graphs. 
In all experiments, the batch size is set to group all edges in 20 batches, and CryptGNN processes all the batches in a single round as described in Section~\ref{sec:cryptmpl_mask}.

\noindent\textbf{CryptGNN vs. plain-text performance.} 
We train three GIN models on the three benchmark datasets, and then compare the results of the plain-text versions of the models with the CryptGNN versions. The plain-text version utilizes PyTorch APIs to compute the GNN layers. 
In CryptGNN, we leverage CrypTen’s API, which uses numerical approximations to compute non-linear functions. The necessary multiplication operations are performed using CryptMUL. The fixed-point encoding to represent floating-point values and the approximation techniques may introduce some precision errors in intermediate results. For instance, in a graph with 2000 nodes and 10 features, the mean difference between the plaintext values and A-SS domain values is $5.1 \times 10^{-5}$. This error is acceptable for deep learning tasks. Since our focus is on predicting the classification IDs rather than obtaining the exact float values, the final result remains unaffected.
We achieve the same inference accuracy results for each pair of models (i.e., plain-text vs. CryptGNN). This demonstrates that CryptGNN works correctly from a machine learning point of view. 

\begin{figure}
\centering
    \includegraphics[width=0.37\linewidth]{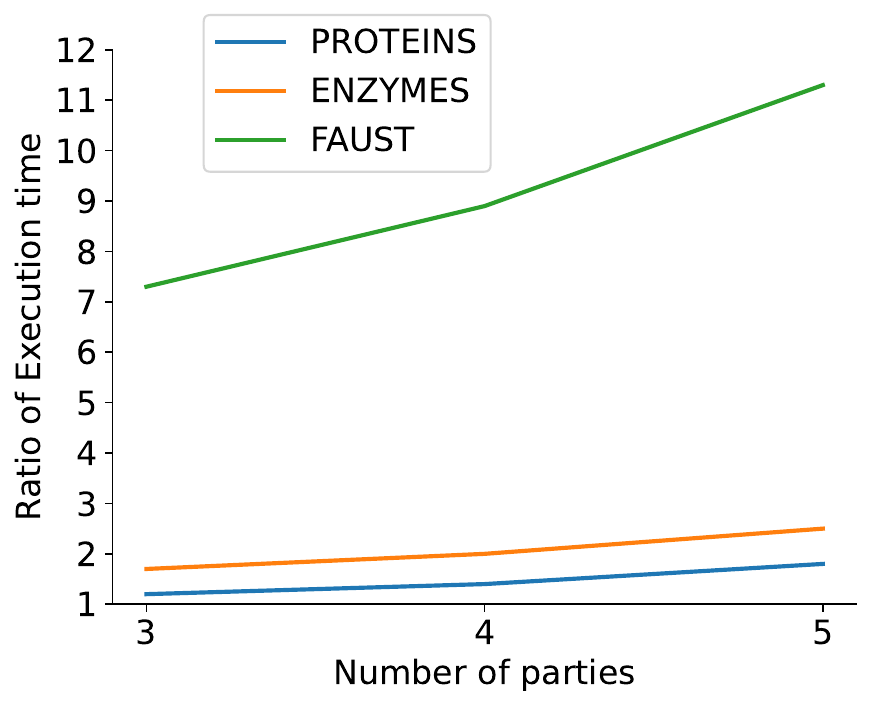}
\caption{Execution time ratio of CrypTen over CryptGNN for GIN, while varying the number of parties}
\vspace{-5mm}
\label{fig:cryptgnn_result}
\end{figure}

\noindent\textbf{Efficiency.}
We compare the performance of CryptGNN with an implementation of the GIN model using CrypTen~\cite{NEURIPS2021_27545182}. 
For CryptGNN, we measure execution time at the server after the data is uploaded by the client. The execution time does not include the server side preprocessing required in CryptMUL, since it is a one-time process for each client that generates the initial Beaver Triples.
To execute the operations required for GNN inference for the Baseline, we use CrypTen's functions for FTLs and implement an adjacency-matrix based solution for MPL.
Unlike CryptGNN, CrypTen uses a trusted server.

Fig.~\ref{fig:cryptgnn_result} shows that CryptGNN is significantly faster than CrypTen, particularly for large-scale graphs in FAUST.
This is because CrypTen incurs higher computation overhead in MPLs and FTLs, which increases with the number of nodes and features in the graph.
The execution time ratio between CrypTen and CryptGNN also increases with the number of parties, since CrypTen requires the parties to communicate with the trusted server for many operations. In terms of absolute inference time, the models using CryptGNN work well in practice. For instance, on the FAUST dataset with 6890 nodes and 41328 edges, the CryptGNN model achieved an average inference time of $22.3s$ in a 3-party setting. Furthermore, in order to evaluate scalability, we used a synthetic graph dataset with an average of 20,000 nodes and 200,000 edges. CryptGNN's average inference time for the graphs in this dataset is approximately $75s$. These results show that CryptGNN can work efficiently in practical situations where security matters more than inference latency, such as in drug discovery and automated code analysis (discussed in Section~\ref{sec:intro}).

To evaluate CryptGNN considering network delay and bandwidth restrictions, we performed an experiment, where we used 3 AWS instances (t2.micro, us-east-1 region) as the computing parties. For the graphs in TUDataset, CryptGNN takes around 2.3 seconds to obtain the inference results for each graph. This indicates that network latency has minimal impact on the overall inference time.

\textbf{End-to-end execution times.} In CryptGNN, the end-to-end execution time can be divided into three main components: (i) offline preprocessing for data masking at the client side, (ii) offline Beaver triple generation at the server side, and (iii) the online phase. The offline preprocessing time at the client side is low compared to the overall execution time. For a benchmark dataset (TUDataset), it takes around 0.1s, with the ratio of offline/online overhead being approximately 1:25. We measure the end-to-end execution time on a large benchmark dataset (FAUST), where client-side computation, Beaver triple generation, and the online phase take approximately 0.9s, 7.5s, and 23.1s, respectively. Since CryptMUL can generate new sets of Beaver triples from an initial set, offline preprocessing is required only once per client. Thus, the overall execution time consists of the time needed for client-side preprocessing and the online phase, which is feasible in real-life scenarios.

\textbf{Communication overhead.} To evaluate the communication overhead during the online phase, we use two datasets: TUDataset, which consists of graphs with an average of 36 nodes, and a synthetic dataset with larger graphs averaging 20,000 nodes. For both datasets, we measure the communication overhead for each inference request using a trained GIN model with parameters encrypted in a 3-party SMPC setting. We report the number of communication rounds, the size of the data (in MB) uploaded by the client, and the total amount of data (in MB) communicated per party using the CryptGNN protocols. Additionally, we compare the communication overhead with CrypTen, which also requires a trusted party during the online phase. 

As shown in Table~\ref{comm_performance}, CryptGNN achieves significantly lower overhead compared to CrypTen on larger datasets. It requires fewer rounds and transfers 74 times less data. The high overhead in CrypTen arises from its adjacency matrix-based implementation, which necessitates large matrix multiplications. Additionally, in CryptGNN, the client uploads substantially less data since there is no need to upload a large adjacency matrix. On smaller datasets, CryptGNN still outperforms CrypTen, reducing overhead by 5\%, while being more secure. For both datasets, the majority of data is transferred in the MPL, which CryptGNN optimizes using CryptMPL protocols. The majority of communication rounds occur in the non-linear layers due to element-wise multiplications needed for the numerical approximation of non-linear functions, which can be further optimized through parallelized computations.

\begin{table}[t!]
\centering
\caption{Communication Performance}
\addtolength{\tabcolsep}{-0.25em}
\resizebox{0.4\textwidth}{!}{
\begin{tabular}{*5c}
\toprule
 &  \multicolumn{2}{c}{TUDataset} & \multicolumn{2}{c}{Synthetic Dataset}\\
\midrule
{}              &  \shortstack{CryptGNN}   & \shortstack{CrypTen}  & \shortstack{CryptGNN} & \shortstack{CrypTen}\\
Client  Comm. (MB) &    0.1    &  0.4      &   12  & 3050\\
Trusted Server Comm. (MB) &    -    &   0.4      &   -  & 3050\\
Each Party Comm. (MB) &    1.24    &  1.31      &   81.9  & 6103\\
Number of  Comm. Rounds  &    120    &   128      &   122  & 130\\
\bottomrule
\label{comm_performance}
\end{tabular}
} 

\end{table}
\subsection{CryptMPL Results}

\begin{figure}
\centering
  \subfloat[]
  {%
    \includegraphics[width=0.35\linewidth]
    {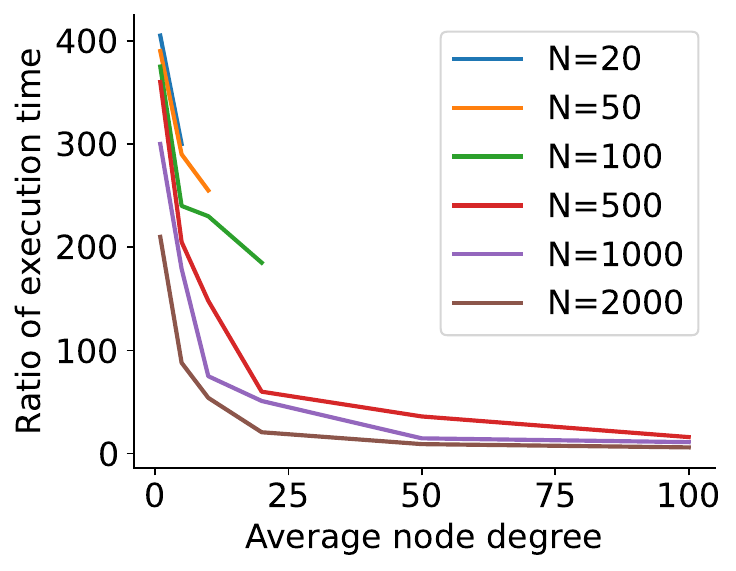}
  }%
  \subfloat[]
  {%
    \includegraphics[width=0.35\linewidth]{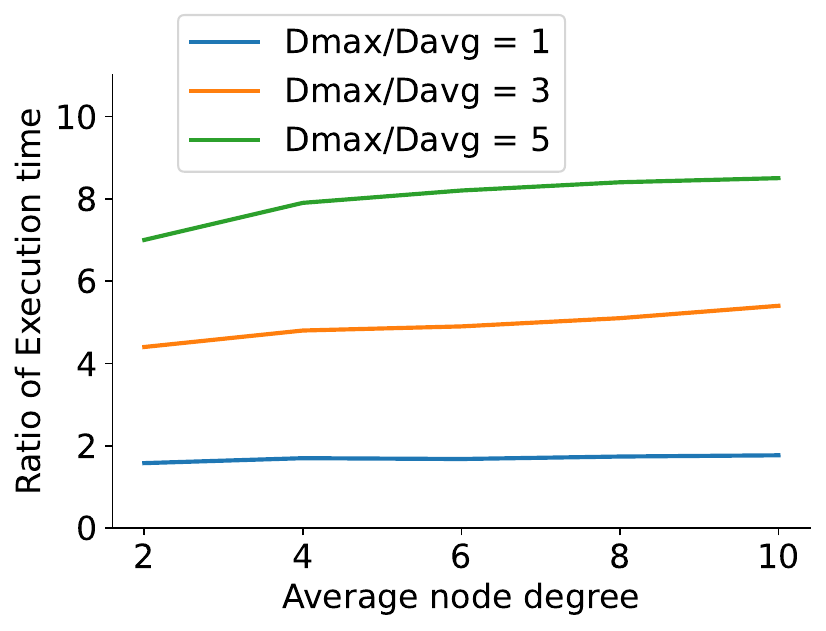}
  }%

\caption{Comparison of CryptMPL with existing techniques: Ratio of execution time — (a) between CryptMPL and Plaintext, and (b) between SecGNN and CryptMPL.}
\label{fig:cryptmpl_compare}
\end{figure}

\textbf{Overhead.} We generate graphs with the numbers of nodes $N$ ranging from 20 to 2000. The average degree $D_{avg}$ is varied from 1 to $max(100, (N-1)/2)$, having a total number of edges $M = N \times D_{avg}$. We choose a batch size $ceil(M/20)$ to process the edges in 20 batches. Fig.~\ref{fig:cryptmpl_compare}(a) shows the ratio of the execution time of CryptMPL and the non-secure MPL for different graphs. The results demonstrate that CryptMPL performs efficiently for medium and large-scale graphs, which are the types of graphs encountered in real-life scenarios. As the plain text computation has a linear relation with the number of edges $M$, the ratio of the execution times decreases as $M$ increases. Despite the high overhead for small graphs, the absolute time for execution remains low
(e.g., 190ms) 
and does not have a major impact on the inference latency.

\noindent\textbf{Comparison with SecGNN.} 
\label{subsec:vs_secgnn}
This experiment compares the efficiency of CryptMPL with that of SecGNN (see Section~\ref{sec:background_related_works}).
We measure the ratio of execution time of the MPL between SecGNN and CryptMPL for graphs with $N=2000$ nodes and $K=10$ features, while varying the node degrees. Fig.~\ref{fig:cryptmpl_compare}(b) shows that CryptMPL is about $1.5\times$ faster than SecGNN. Moreover, the execution time ratio increases linearly as $D_{max}/D_{avg}$ increases, where $D_{max}$ and $D_{avg}$ are the maximum and average node degree in the graph, respectively. This demonstrates that CryptMPL works better for real-life graphs, as $D_{max}$ is usually much higher compared to $D_{avg}$.
In addition, CryptMPL provides better security, as it works with more than 2 parties and does not require a trusted party.

\begin{figure}
\centering
\vspace{-12pt}
  \subfloat[]
  {%
    \includegraphics[width=0.35\linewidth]
    {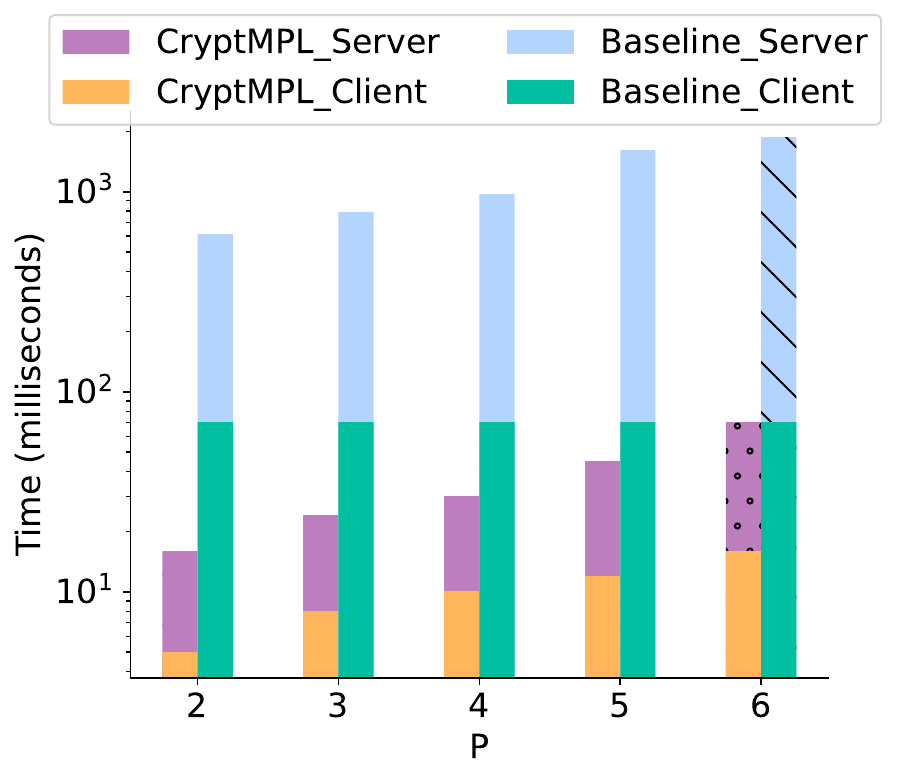}
  }%
  \subfloat[]
  {%
    \includegraphics[width=0.35\linewidth]
    {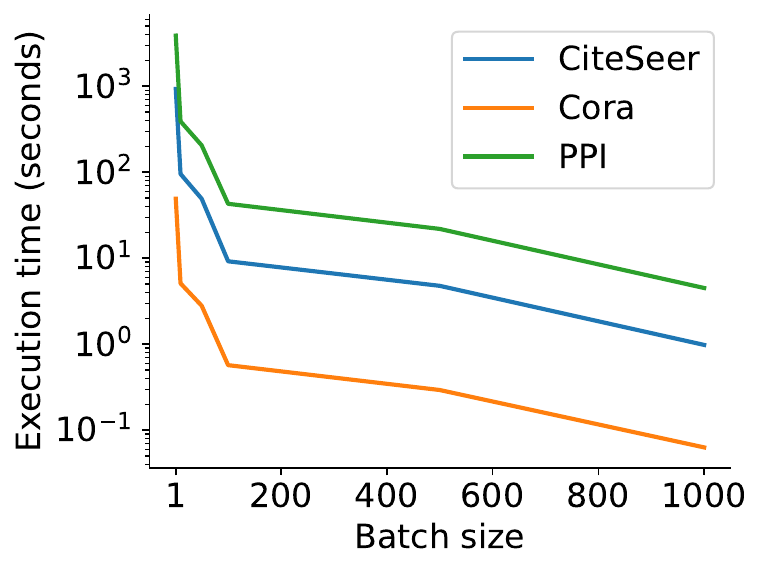}
  }%

\caption{Effect of (a) number of parties and (b) batch size  on CryptMPL (Y-axis is log scaled)}
\label{fig:cryptmpl_effect_P_R}
\end{figure}

\noindent\textbf{Comparison with adjacency matrix-based MPL.} We compare the execution time of CryptMPL with a hypothetical solution based on representing the graph as an adjacency matrix. The experiment uses a PPI dataset and varies the number of parties.
Fig.~\ref{fig:cryptmpl_effect_P_R}(a) shows 
CryptMPL is 25 times faster compared to the adjacency matrix solution when using 6 parties. This demonstrates that CryptMPL's choice of graph representation and its novel SMPC techniques to compute MPL lead to large performance improvements. 

\noindent\textbf{Effect of batching.}
\label{subsec:effect_batch}
We use three datasets (Cora, CiteSeer, and PPI) in a 3-party SMPC setting, each with a large number of nodes and commonly used for node classification tasks.
Fig.~\ref{fig:cryptmpl_effect_P_R}(b) shows that the execution time of CryptMPL decreases as the batch size increases, since CryptMPL requires a low number of rounds $R=\lceil M/B \rceil$ to process the edges, where $M$ and $B$ are the number of edges and the size of each batch, respectively. However, as discussed in Section~\ref{sec:analysis}, the security guarantees improve exponentially with $R$ and CryptMPL can use a relatively large batch size while still guaranteeing a good level of security.
\subsection{CryptMUL Results}
\label{subsec:cryptmul_results}
\begin{figure}
\centering
  \subfloat[]
  {%
    \includegraphics[width=0.33\linewidth]{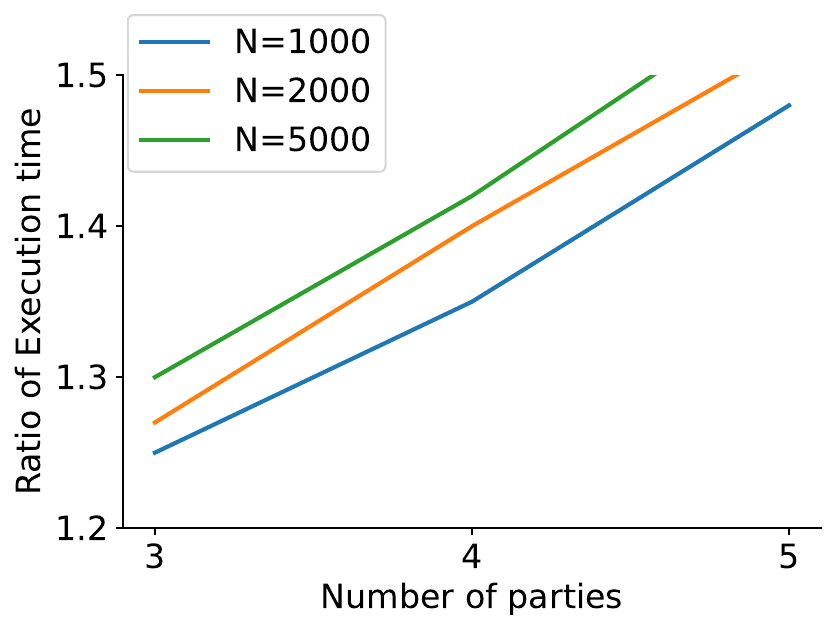}
  }%
  \subfloat[]
  {%
    \includegraphics[width=0.33\linewidth]{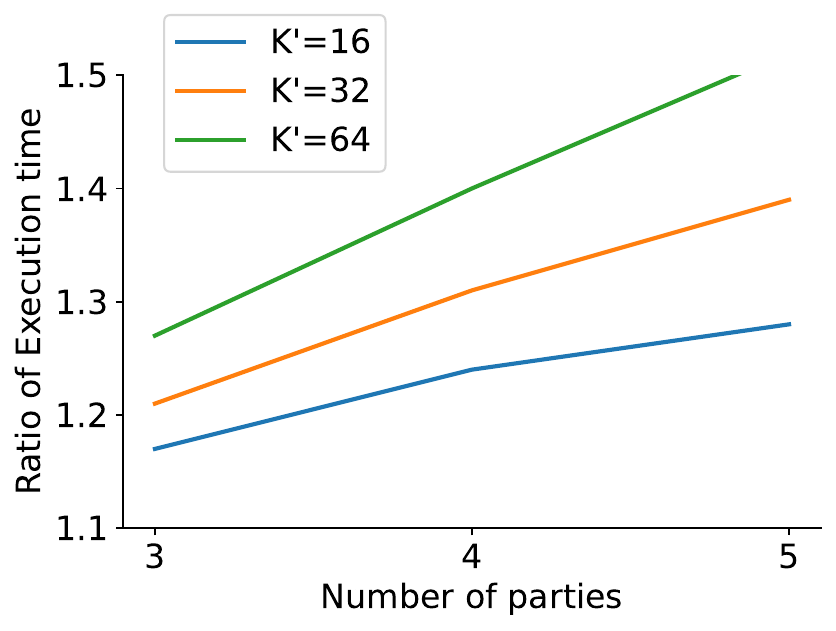}
  }%

\caption{Execution time ratio of a linear layer between CrypTen and CryptMUL by varying: (a) $N$ and $P$ (b) $K'$ and $P$}
\label{fig:cryptmul_matmul}
\end{figure}

In this experiment, we evaluate the performance of a linear layer that computes $\llbracket \mathbf{Z} \rrbracket = \llbracket \mathbf{X} \rrbracket \otimes \llbracket \mathbf{Y} \rrbracket + \llbracket \mathbf{B} \rrbracket$, where $\mathbf{X} \in \mathbb{R}^{N \times K}, \mathbf{Y} \in \mathbb{R}^{K \times K'}, \mathbf{B} \in \mathbb{R}^{N \times K'}$. The linear layer transforms the number of features from $K$ to $K'$ for the same number of nodes $N$. To see the effect of different parameters of the data, we generate synthetic matrices and vary $N \in [1000, 2000, 5000]$ and $K' \in [16, 32, 64]$. We set $K = 3$ in this experiment. We measure the ratio of execution time between linear layers using CrypTen~\cite{NEURIPS2021_27545182} and 
CryptMUL by varying the number of parties $\mathcal{P} \in [3, 4, 5]$. 
Our results demonstrate that the linear layer implemented with CryptMUL outperforms its counterpart using CrypTen. As illustrated in Fig.~\ref{fig:cryptmul_matmul}, the ratio of execution time between CrypTen and CryptMUL exhibits an increasing trend with $\mathcal{P}$, since the communication overhead in CrypTen increases with the number of parties involved. Moreover, this ratio also rises in relation to $N$ and $K'$, since higher values of these parameters increase the computation at the trusted server and the communication between the parties with the trusted server.
\section{Discussion}
\label{sec:discussion}

CryptGNN marks an important advancement toward secure and efficient GNN inference in MLaaS settings. While the current design is effective, there is significant room for improvements.  
This section outlines the current limitations of CryptGNN and identifies several promising directions for future exploration.

\textbf{Supporting Complex Message-Passing Layers.} 
The secure message-passing layer of CryptGNN, referred to as CryptMPL (Section~\ref{sec:cryptmpl_design}), is specifically designed to natively support the widely adopted GIN architecture~\cite{xu2018how} as a GNN model. While this design choice ensures compatibility, it also introduces a limitation in supporting more complex MPLs. However, CryptMPL’s secure read and write protocols can be extended to support such MPLs by incorporating additional operations -- such as node sampling, feature concatenation, and others -- using standard SMPC techniques or by developing efficient, custom protocols.
For example, in GraphSAGE~\cite{10.5555/3294771.3294869}, instead of performing message-passing on the entire graph, nodes are randomly sampled, and message-passing is carried out only through the adjacent edges. In the case of secure inference, this sampling operation must be executed in a way that does not reveal the graph structure. While oblivious sampling operations in SMPC settings may be feasible, they are computationally expensive and significantly increase overhead. An alternative solution is to perform the sampling operation on the client side, where the graph structure is already known. After sampling, the client can upload the list of edges following the CryptMPL protocol, allowing secure message-passing to be executed on the server side.

Similarly, CryptGNN can support Graph Attention Networks (GATs)~\cite{veličković2018graph}, which require the computation of attention coefficients for each edge in the message-passing layer. Since CryptMPL can securely read the feature vector of any source node, it can be extended to concatenate the feature vectors of the two nodes connected by each edge and compute pairwise attention coefficients using standard SMPC techniques. 

\textbf{Supporting Heterogeneous Graphs.} 
In CryptGNN, we focus on state-of-the-art GNN architectures (e.g., GIN, GCN) that are designed for homogeneous graphs. While there are advanced architectures capable of handling heterogeneous graphs, supporting them securely would require additional effort, as it involves protecting the types of nodes and edges to fully preserve the privacy of the graph structure. We plan to explore this in future work.

\textbf{Supporting GNN training.}
In CryptGNN, we design secure protocols to execute the forward pass required for inference in MLaaS, which can benefit many applications, as described in Section~\ref{sec:intro}. CryptGNN could also be extended to support training or fine-tuning, which would require secure protocols for computing the loss, performing backpropagation to calculate gradients, and updating the weight matrices accordingly. While CryptGNN can in principle be extended to support these steps, since they can be decomposed into a fixed number of addition and multiplication operations, doing so would be computationally expensive. Training typically involves multiple epochs, and to preserve data privacy in each epoch, a fresh set of noise matrices (for CryptMPL) and auxiliary data (for CryptMUL) would be required. We plan to explore secure training as part of our future work.

\textbf{Toward Security Against Active Adversaries.}
CryptGNN assumes an honest-but-curious adversarial model, where parties follow the protocol but may attempt to infer private information from observed data. While this model is practical and widely adopted, it does not account for adversaries that may actively deviate from the protocol. In future work, we aim to explore the design of secure and verifiable protocols that can support GNN inference even in the presence of active (malicious) adversaries. Achieving this would require incorporating mechanisms such as zero-knowledge proofs or verifiable computation to ensure correctness and integrity of computations under stronger threat models.
\section{Conclusion}
\label{sec:conclusion}
We presented CryptGNN, a provably secure and effective inference system for GNN in MLaaS scenarios. CryptGNN has two main protocols, CryptMPL and CryptMUL, to support secure MPLs and FTLs in GNN. These novel SMPC protocols preserve the privacy of the model parameters and input graph data while providing the same results as the non-secure inference version. CryptGNN works with an arbitrary number of SMPC parties, and it protects the input data, the intermediate results, and the output, even if $\mathcal{P}-1$ out of $\mathcal{P}$ parties collude. The system analysis and experimental results demonstrate CryptGNN's correctness and low overhead compared to state-of-the-art approaches. 

\bibliographystyle{ACM-Reference-Format}
\balance
\bibliography{main_CCS_2025}

\appendix
\section*{Appendix}
This appendix provides additional technical details to supplement the main paper. In Appendix~\ref{app:sec:cryptmpl}, we present a complete example illustrating the flow of CryptMPL, several straightforward extensions, and the pseudocode for the algorithms introduced in the main paper (Section~\ref{sec:cryptmpl_design}).
Detailed algorithms for CryptMUL are presented in Appendix~\ref{app:sec:cryptmul}. 
While the main paper focuses on the system and security analysis (Section~\ref{sec:analysis}), we provide the correctness analysis of CryptGNN in Appendix~\ref{sec:correctness}.
Finally, Appendix~\ref{appendix-evaluation} includes experimental details and additional evaluation results for CryptMPL and CryptMUL over various graph- and system-specific parameters.

\setcounter{section}{0}
\section{Detailed Description of CryptMPL}
\label{app:sec:cryptmpl}
\setcounter{subsection}{0}

We discuss our secure message-passing layer CryptMPL in Section~\ref{sec:cryptmpl_design}. In this section, we present a simple example illustrating the flow of data in CryptMPL. We also discuss straightforward extensions of CryptMPL and provide the pseudocodes of the algorithms.

\subsection{A Simple Example}
\label{app:subsec:toy_example}
In this example, we consider three computing parties $CP_1$, $CP_2$ and $CP_3$ take an input feature matrix $\mathbf{A}$ and compute the message-passing layer to generate the output feature matrix $\mathbf{A}^*$. Here, the matrix $\mathbf{A}$ represents $K=2$ features for each of the $N=4$ nodes in the graph $\mathcal{G}$. For simplicity, we consider a simple graph and illustrate the protocols for computing the message passing through an edge from node \textcircled{1} to node \textcircled{2}. We consider the indices to be 1-indexed. Therefore, for this edge, the source index $S = 1$ and the destination index $D=2$. In the A-SS domain, $CP_p$ has the shares of node features, source index, and destination index as $\llbracket \mathbf{A} \rrbracket_p$, $\llbracket S \rrbracket_p$, and $\llbracket D \rrbracket_p$  respectively. Here, $\mathbf{A} = \sum_{p=1}^{\mathcal{P}} \llbracket A \rrbracket_p$, $S = (\sum_{p=1}^{\mathcal{P}} \llbracket S_p \rrbracket) \mod N + 1$ and $D = (\sum_{p=1}^{\mathcal{P}} \llbracket D_p \rrbracket) \mod N + 1$.

\begin{figure}[!ht]
    \centering
    \includegraphics[width=0.8\linewidth]{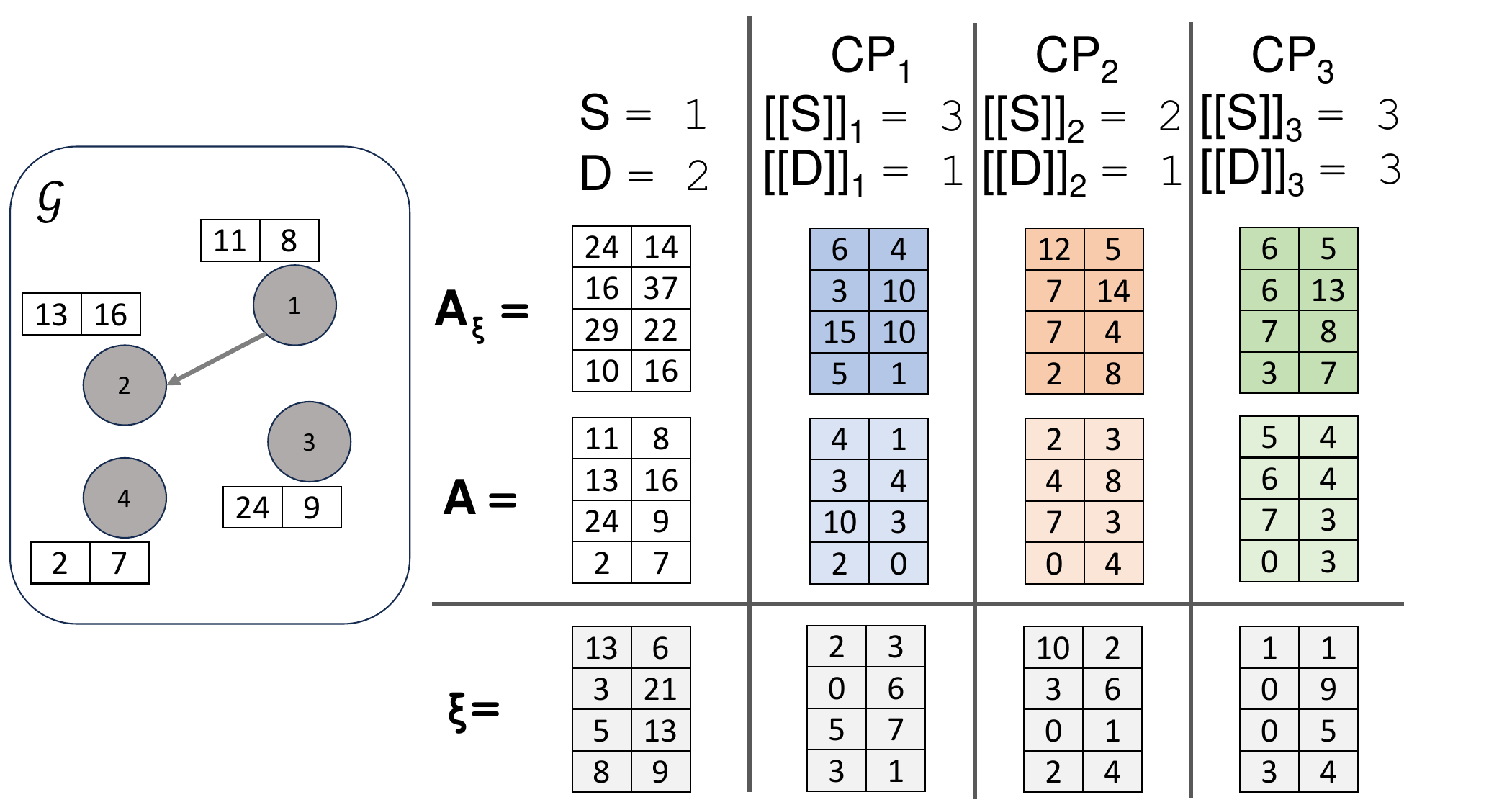}
    \caption{Input feature matrix $\mathbf{A}$, noise matrix $\boldsymbol{\xi}$, and masked feature matrix $\mathbf{A}_{\boldsymbol{\xi}}$, for a graph $\mathcal{G}$ with $N=4$ nodes and $K=2$ features}
    \label{fig:toy_1}
\end{figure}

The owner of the graph, the client, has the information about the graph structure, i.e., it knows the source index $S$ and the destination index $D$. However, the client doesn't have any information about the current feature matrix $\mathbf{A}$. Nonetheless, the client can assist the computing parties in masking their shares and executing the message-passing layer, ensuring that the node features and graph structure ($S$ and $D$) remain protected throughout the execution of CryptMPL.

To preserve the node features, the client shares the seeds $s_p$ with the computing parties $CP_p$. Each party uses its seed to create matrix $\boldsymbol{\xi}_p$ which is used to mask its share $\llbracket \mathbf{A} \rrbracket_p$ as $\llbracket \mathbf{A}_{\boldsymbol{\xi}} \rrbracket_p = \llbracket \mathbf{A} \rrbracket_p + \boldsymbol{\xi}_p$ (shown in Fig.~\ref{fig:toy_1}). The parties follow CryptMPL read protocol $\mathcal{F}_{SR}$ and write protocol $\mathcal{F}_{SW}$ to obtain $\llbracket \mathbf{A}_{\boldsymbol{\xi}}^* \rrbracket$ by executing the message-passing on on $\llbracket \mathbf{A}_{\boldsymbol{\xi}} \rrbracket$. Finally, the computing parties remove the effect of noise from $\llbracket \mathbf{A}_{\boldsymbol{\xi}}^* \rrbracket$ to obtain the correct result $\llbracket \mathbf{A}^* \rrbracket$.

During the read operation (Section~\ref{sec:cryptmpl_read}), using the seed value $s_p$, $CP_p$ generates matrix $\mathbf{X}_{pi}$, which is used to mask the share of the feature matrix from $CP_i$. Additionally, $CP_p$ generates random values $r_{pi}$ and rotates the share of the feature matrix from party $CP_i$ by $r_{pi}$. During the write operation (Section~\ref{sec:cryptmpl_write}), $CP_p$ generates $\mathbf{Y}_{pi}$ which is used to mask the share of the feature matrix from $CP_i$. Additionally, $CP_p$ rotates the share of the feature matrix from party $CP_i$ by $D_p$. We use $\mathbf{X}_{pimn}$ and $\mathbf{Y}_{pimn}$ to represent the value $\mathbf{X}_{pi}[m][n]$ and $\mathbf{Y}_{pi}[m][n]$ respectively. 

\begin{figure}[!h]
    \centering    \includegraphics[width=1\linewidth]{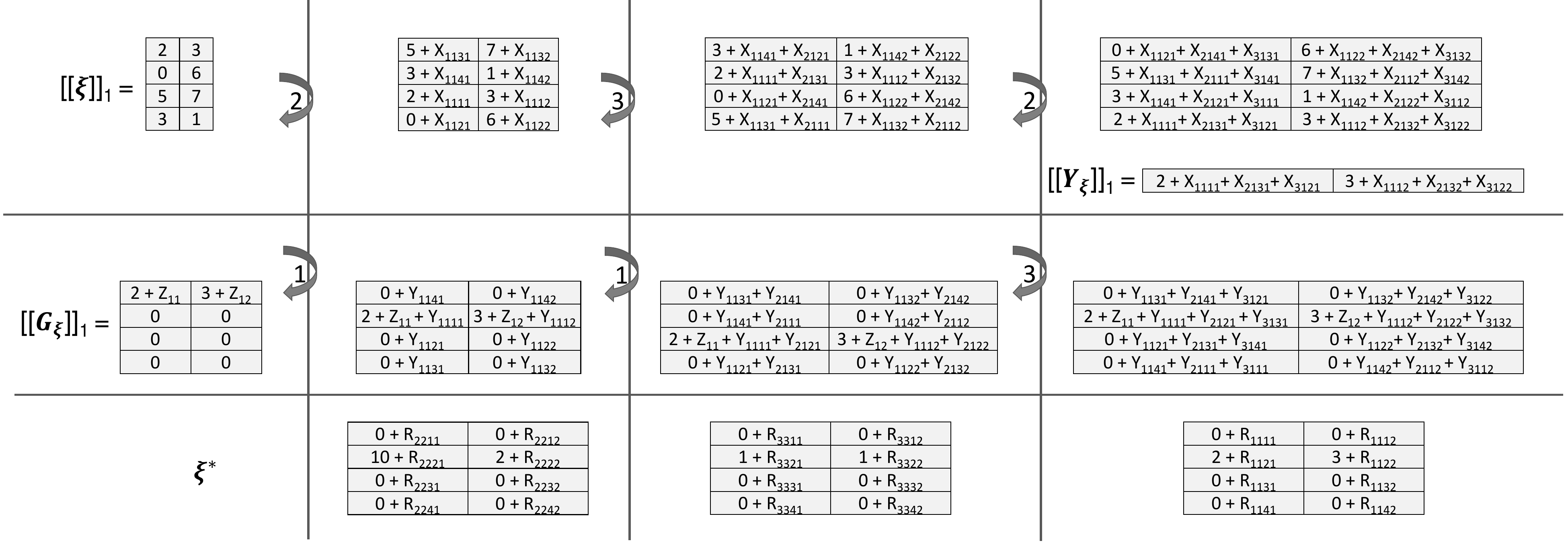}
    \caption{CryptMPL: Client side preprocessing step}
    \label{fig:toy_2}
\end{figure}

\textbf{Client-side preprocessing.}
As described in Section~\ref{sec:cryptmpl_mask}, the client executes the message-passing step on the noise matrix $\boldsymbol{\xi}$ to generate $\boldsymbol{\xi}^*$. $\boldsymbol{\xi}^*$ is shared with the computing parties in A-SS format, so that they can remove the effect of noise from $\llbracket \mathbf{A}_{\boldsymbol{\xi}} \rrbracket_p$. We illustrate the client-side preprocessing step to generate a share $\llbracket {\boldsymbol{\xi}}^* \rrbracket_1$ in Fig.~\ref{fig:toy_2}. During the read operation, the client simulates the effect of rotations $r_{p1}$ and data masking $\mathbf{X}_{p1}$ on $\llbracket {\boldsymbol{\xi}}^* \rrbracket_1$. Figure~\ref{fig:toy_2} shows the final state of $\llbracket {\boldsymbol{\xi}}^* \rrbracket_1$ after passing the share through all parties, assuming $r_{11} = 2, r_{21} = 3 \text{ and } r_{31} = 2$. Client completes the read operation by reading the vector at index $(\sum_{p=1}^{\mathcal{P}} (\llbracket S \rrbracket_p + r_{p1})) \mod N + 1 = ((3+2) + (2+3) + (3+2)) \mod 4 + 1 = 4$ to get the vector $\llbracket \mathbf{Y}_{\boldsymbol{\xi}} \rrbracket_1 = [2 + Z_{11}, 3 + Z_{12}]$, where the cumulative noises, $Z_{11} = \mathbf{X}_{1111} + \mathbf{X}_{2131} + \mathbf{X}_{3121}$, $Z_{12} =  \mathbf{X}_{1112} + \mathbf{X}_{2132} + \mathbf{X}_{3122}$. Similarly, the client computes other shares, $\llbracket \mathbf{Y}_{\boldsymbol{\xi}} \rrbracket_2$ and $\llbracket \mathbf{Y}_{\boldsymbol{\xi}} \rrbracket_3$.

For the write operation, the client creates a feature matrix $\mathbf{G}_{\boldsymbol{\xi}} = \mathbf{0}$ and writes the read result at index 0 as $\llbracket \mathbf{G}_{\boldsymbol{\xi}} [0]\rrbracket_1 = \llbracket \mathbf{Y}_{\boldsymbol{\xi}} \rrbracket_1$. Then, it simulates the effect of rotations $D_p$ and data maskings $\mathbf{Y}_{p1}$ on $\llbracket \mathbf{G}_{\boldsymbol{\xi}}\rrbracket_1$. Finally, it gets the share $\llbracket \mathbf{G}_{\boldsymbol{\xi}}\rrbracket_1$ where $\llbracket \mathbf{Y}_{\boldsymbol{\xi}}\rrbracket_1$ is moved to the target index $\sum_{p=1}^{\mathcal{P}} D_p = D$. Here, all values in $\llbracket \mathbf{G}_{\boldsymbol{\xi}}\rrbracket_1$ are masked with noise added by all parties. Similarly, client gets other shares $\llbracket \mathbf{G}_{\boldsymbol{\xi}}\rrbracket_2$ and $\llbracket \mathbf{G}_{\boldsymbol{\xi}}\rrbracket_3$.

Thus, the output of message passing on $\boldsymbol{\xi}$ is masked with noise as $\llbracket \boldsymbol{\xi} \rrbracket_p^* = \llbracket \mathbf{G} \rrbracket_p + \mathbf{R}_{pp}$, where $\llbracket \mathbf{G} \rrbracket_p$ represents the $p$-th party's share of the final result if noise were not added and $\mathbf{R}_{pp}$ are the accumulated noises. To upload the result $\boldsymbol{\xi}^*$, client creates different shares of $\boldsymbol{\xi}^*$ as $\llbracket \boldsymbol{\xi}^* \rrbracket_p = \llbracket \mathbf{G} \rrbracket_p + \mathbf{T}_{pp}$, where $\sum_{p=1}^{\mathcal{P}}\mathbf{T}_{pp}[m][n] = \sum_{p=1}^{\mathcal{P}}\mathbf{R}_{pp}[m][n]$.

\textbf{CryptMPL protocols in the cloud.} The computing parties $CP_p$ initialize a output feature matrix as $\llbracket \mathbf{A}_{\boldsymbol{\xi}}^* \rrbracket_p = \mathbf{0}$. $CP_p$ perform similar operations on the shares of $\mathbf{A}_{\boldsymbol{\xi}}$ as the clients do on ${\boldsymbol{\xi}}$ during the preprocessing step. 

\begin{figure}
    \centering
    \includegraphics[width=1\linewidth]{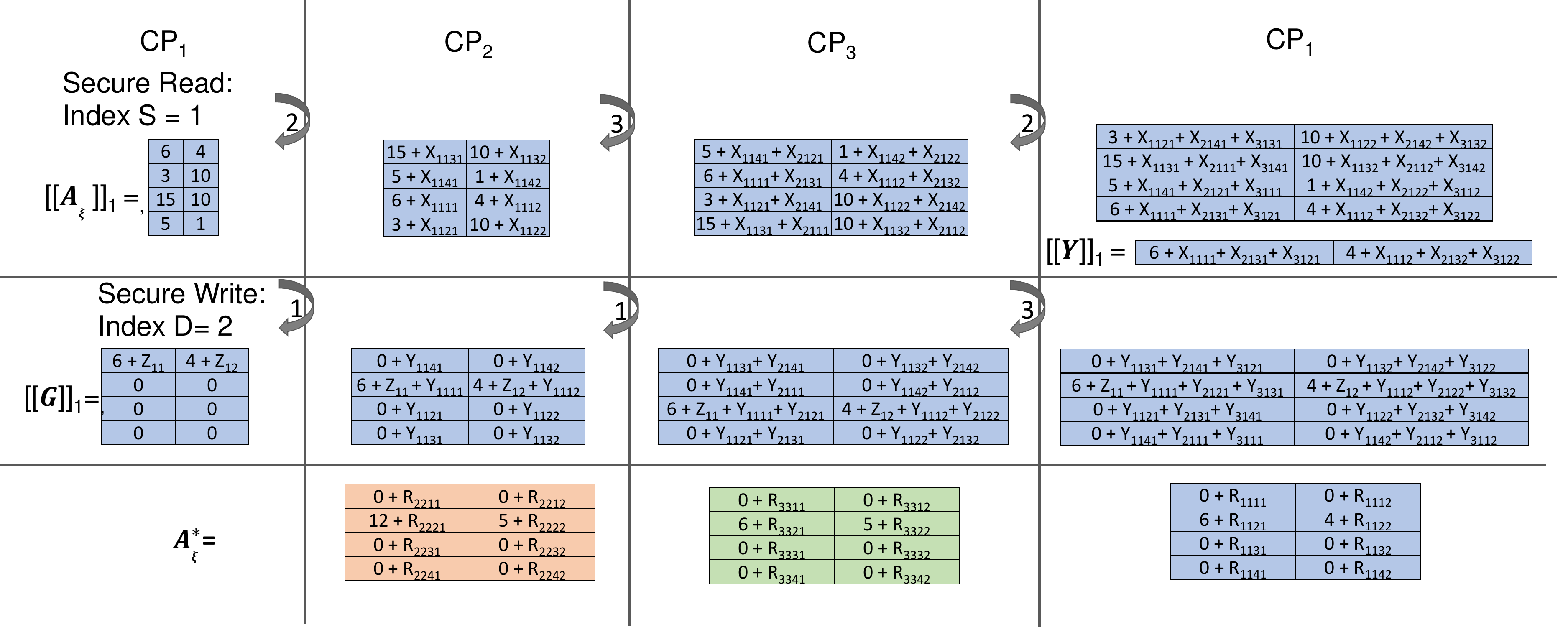}
    \caption{CryptMPL flow at the computing parties}
    \label{fig:toy_3}
\end{figure}

Fig.~\ref{fig:toy_3} shows the operations on $\llbracket \mathbf{A}_{\boldsymbol{\xi}} \rrbracket_1$ to obtain $\llbracket \mathbf{Y} \rrbracket_1$ after read operation and $\llbracket \mathbf{G} \rrbracket_1$ after the write operation. After executing the read operation on $\llbracket \mathbf{A}_{\boldsymbol{\xi}} \rrbracket_1$ for the source index $\llbracket S \rrbracket$, $CP_1$ reads the vector at the updated target index $(\sum_{p=1}^{\mathcal{P}} (\llbracket S \rrbracket_p + r_{p1})) \mod N + 1 = ((3+2) + (2+3) + (3+2)) \mod 4 + 1 = 4$ to get $\llbracket \mathbf{Y} \rrbracket_1$. Since $\llbracket \mathbf{Y} \rrbracket_1$ is obtained by rotation and masking of $\llbracket \mathbf{A}_{\boldsymbol{\xi}} \rrbracket_1$ with noise from all parties, the computing parties cannot learn the original index or values, even if there is collusion among $\mathcal{P} - 1$ parties. Similarly $CP_2$ and $CP_3$ obtain $\llbracket \mathbf{Y} \rrbracket_2$ and $\llbracket \mathbf{Y} \rrbracket_3$ respectively.

During the write operation, each party $CP_p$ sets $\llbracket \mathbf{G} [0]\rrbracket_p = \llbracket \mathbf{Y} \rrbracket_p$. Then, $CP_p$ adds noise $Y_{pi}$ to $i$-th party's share, rotates it by $\llbracket D \rrbracket_p$, and passes it to the next party. As shown in Fig.~\ref{fig:toy_3}, $CP_1$ obtains $\llbracket \mathbf{G} \rrbracket_1$ which is modified by all parties. Due to rotations by $\sum_{p=1}^{\mathcal{P}}D_p \mod N + 1 = (1 + 1 + 3) \mod 4 + 1 = 2$, $\llbracket \mathbf{Y} \rrbracket_1$ reaches at the destination index of $\llbracket \mathbf{G} \rrbracket_1$. Due to the rotations and noises, the computing parties cannot learn the original index or values, even if there is collusion among $\mathcal{P} - 1$ parties. In parallel, $CP_2$ and $CP_3$ obtain $\llbracket \mathbf{G} \rrbracket_2$ and $\llbracket \mathbf{G} \rrbracket_3$ respectively. Then, the matrix $\llbracket \mathbf{G} \rrbracket_p$ is used to update the output feature matrix for the MPL layer as $\llbracket \mathbf{A}_{\boldsymbol{\xi}}^* \rrbracket_p = \llbracket \mathbf{A}_{\boldsymbol{\xi}}^* \rrbracket_p + \llbracket \mathbf{G} \rrbracket_p$. 

\begin{figure}
    \centering
    \includegraphics[width=.9\linewidth]{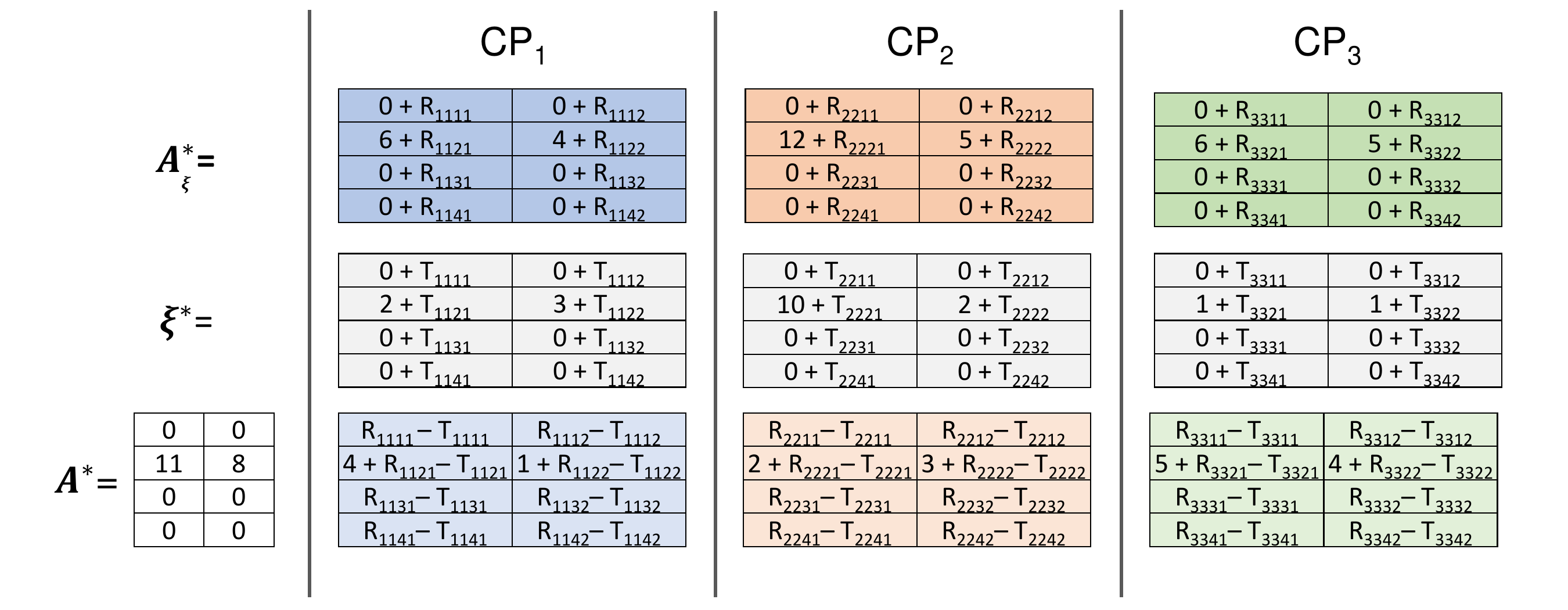}
    \caption{CryptMPL - removal of noise to get the output feature matrix}
    \label{fig:toy_4}
\end{figure}

However, since $\llbracket \mathbf{A}_{\boldsymbol{\xi}}^* \rrbracket$ is computed on the matrix $\llbracket \mathbf{A}_{\boldsymbol{\xi}} \rrbracket$, it is required to remove the noise. As shown in Fig~\ref{fig:toy_4}, $CP_p$ removes the effect of noise by computing $\llbracket \mathbf{A}^* \rrbracket = \llbracket \mathbf{A}_{\boldsymbol{\xi}} \rrbracket - \llbracket \boldsymbol{\xi}^* \rrbracket$. Thus, the shares constitute the correct $\llbracket \mathbf{A}^* \rrbracket$ in A-SS domain as $\llbracket \mathbf{A} \rrbracket^* = \sum_{p=1}^{N} \llbracket \mathbf{A} \rrbracket^*_p$, since $\sum_{p=1}^{\mathcal{P}} (\mathbf{R}_{pp} - \mathbf{T}_{pp}) = \mathbf{0}$. However, the computing parties can not determine the actual values from $\mathcal{P} - 1$ shares of $\llbracket \mathbf{A}^* \rrbracket$.

Following this approach, we can compute the message-passing for each edge of the graph. However, we can process multiple edges in a batch, since secure read and write operation does not reveal the actual source and destination indices of the edges. For each batch, the client and parties use different sets of noises to protect the node features and source-destination indices by observing data in different batches. To obtain the correct result, the client only needs to share the initial noise $\llbracket \boldsymbol{\xi} \rrbracket$, overall noise $\llbracket \boldsymbol{\xi}^* \rrbracket$ and the seeds (to generate rotation amount and noise matrices for each batch) to the computing parties. 

\subsection{Extensions of CryptMPL}
\label{app:subsec:improvements}
For simplicity, the main design of CryptGNN focuses on unweighted and undirected edges in input graphs. Its secure message-passing protocol, CryptMPL, can be extended to support weighted and directed edges, as well as to protect the number of edges in the graph, as discussed below.

\textbf{Supporting directed edges.} An undirected edge (A--B) is represented in CryptGNN as two directed edges (A $\rightarrow$ B and B $\rightarrow$ A) in the edge list.  Therefore, supporting directed edges is straightforward: each edge is simply included as a one-way connection from the source to the destination node.

\textbf{Supporting weighted edges}. To support weighted edges, the client needs to upload the list of edge weights, denoted as $\mathbf{W}$, along with the list of edges, where $\mathbf{W}[i]$ is the weight for the edge from source node $\mathbf{S}[i]$ to destination node $\mathbf{D}[i]$. To protect the weights, $\mathbf{W}$ is represented in A-SS format as $\llbracket \mathbf{W} \rrbracket$.

As discussed in Section~\ref{sec:cryptmpl_read}, CryptMPL reads the node feature of the source node for each $i$-th edge as $\llbracket \mathbf{Y}[i] \rrbracket$. To apply the weights, the computing parties multiply the weight $\llbracket \mathbf{W}[i] \rrbracket$ by the result of the read operation $\llbracket \mathbf{Y}[i] \rrbracket$. CryptMPL leverages CryptMUL’s element-wise multiplication protocol to perform this operation, which requires only one additional round of communication among the computing parties.

In the data-processing step, the client also multiplies the weight of each edge by the corresponding noise vector of the source node in the noise matrix $\boldsymbol{\xi}$ to compute the overall noise $\boldsymbol{\xi}^*$. Taking the weights into account, Eq.~\ref{eqn:overall_noise}, ~\ref{eqn:MLP_with_noise} and ~\ref{eqn:remove_noise} can be modified as follows:
Eq.~\ref{eqn:overall_noise_weight} considers the weights to compute the overall noise, Eq.~\ref{eqn:MLP_with_noise_weight} computes the message-passing on the masked feature matrix and applies the edge weights, and finally Eq.~\ref{eqn:remove_noise_weight} removes the overall noise to obtain the correct output feature matrix. 

{\small
\begin{equation}
\boldsymbol{\xi^*}[j]=\sum_{j \in \mathcal{N}(i)} \mathbf{W}[j] \times \boldsymbol{\xi}[j]
\label{eqn:overall_noise_weight}
    \end{equation}
\begin{equation}
\mathbf{A_\xi^*}[i]=\sum_{j \in \mathcal{N}(i)} \mathbf{W}[j] \times (\mathbf{A}[j] + \boldsymbol{\xi}[j])
\label{eqn:MLP_with_noise_weight}
\end{equation}
\begin{equation}
\mathbf{A^*}[i]=\mathbf{A_\xi^*}[i] - \boldsymbol{\xi^*}[i] = \sum_{j \in \mathcal{N}(i)} \mathbf{W}[j] \times \mathbf{A}[j]
\label{eqn:remove_noise_weight}
\end{equation}
}

\textbf{Protecting the number of edges}. If a client wants to protect the number of edges in the graph, it can insert any number of fake edges $(S_f, D_f)$, $1 \leq S_f, D_f \leq N$, with weight $W_f = 0$. In the A-SS format, the parties cannot determine whether the weight values are zero, and therefore, fake edges cannot be detected. Moreover, these fake edges do not affect the overall result, as their weights are zero. Although fake edges introduce overhead, they can be processed without increasing the number of communication rounds by adjusting the batch size.

\subsection{CryptMPL Algorithms}
\label{app:algorithms}

We consider the following variables to be known to all parties for each graph data uploaded by the client for analysis: (i) the number of nodes $N$, (ii) the number of edges $M$, (iii) the number of features of each node $K$, and (iv) the number of computing parties $\mathcal{P}$. In the pseudo codes, we use the following utility functions:
\begin{itemize}[topsep=0pt,itemsep=0ex,partopsep=0ex,parsep=0ex]
\item $rotate(\mathbf{A}, sz, k)$: rotates an matrix $\mathbf{A}$ of length $sz$ for $k$ times
\item $init\_matrix(a, b)$: returns a matrix of size $(a,b)$ where all the entries are zero 
\item $get\_next\_seed(seed_0, r)$: generates a seeds for round $r$ from an initial seed $seed_0$
\item $PRF(cur\_seed, (a, b))$: generates a pseudo-random matrix of size $(a, b)$ using a seed $cur\_seed$
\item $get\_cur\_party()$: returns the ID of the current party
\item $send\_to(p)$: to send the data to a party with rank $p$
\item $receive\_from(p)$: to receive the data from a party with rank $p$
\end{itemize}

\textbf{Preprocessing at the client side}
This section presents the algorithm $\mathcal{F}_{CN}$ to calculate the overall noise at the preprocessing stage on the client side. In Algorithm \ref{alg:Client_MPL_Noise}, first, the client generates seeds for the $\mathcal{P}$ computing parties (line 2). Next, the client generates noise matrices for all parties using the pseudo-random function and calculates the overall noise $\boldsymbol{\xi^*}$  by considering the effect of noise added in the read stage (line 5-13) and write stage (lines 15-24) during the processing of $M$ edges.

\begin{algorithm}[!h] 
\caption{Preprocessing to create noise, $\mathcal{F}_{CN}$}
\label{alg:Client_MPL_Noise}
\textbf{Input} $\mathbf{S}$ (Source Indices), $\mathbf{D}$ (Destination Indices),  $\mathbf{S_{shares}}$ (Shares of source Indices), $\mathbf{D_{shares}}$ (Shares of destination Indices)\\
\textbf{Output} $\boldsymbol{\xi}^*$ (Overall Noise)
    \begin{algorithmic}[1]
   
    \STATE $\mathbf{seeds}$ $\gets$ $PRF(rand\_value, (\mathcal{P},1))$
    
    \STATE {$\boldsymbol{\xi^*}$ $\gets$ {$init\_matrix(N, K)$}}
    \FOR {$r \gets 1$ to $M$} 
        \STATE $\mathbf{R}$ $\gets$ $init\_matrix(\mathcal{P},1)$ \%Computes the amount of rotation for each party
        
        \FOR {$i \gets$ $1$ to $\mathcal{P}$}  
            \FOR {$j \gets$ $1$ to $\mathcal{P}$}  
                \STATE {$cur\_seed$} $\gets$ $get\_next\_seed(\mathbf{seeds}[j], r)$
                \STATE $\boldsymbol{X_{share}} \gets$ $PRF(cur\_seed, (N,K))$

                \STATE $\boldsymbol{\xi^*}[\mathbf{D}[r]] \gets \boldsymbol{\xi^*}[\mathbf{D}[r]]$ + $\boldsymbol{X_{share}}[\mathbf{S}[r] + \mathbf{R}[j]]$

                \STATE $party\_rank \gets (j - i - 1) \mod \mathcal{P} + 1$
                \STATE $\mathbf{R}[j]$ $\gets$ $\mathbf{R}[j] + \mathbf{S_{shares}}[party\_rank][r]$

            \ENDFOR
        \ENDFOR

        \STATE $\mathbf{R}$ $\gets$ $init\_matrix(\mathcal{P},1)$ \%Computes the amount of rotation for each party

        \FOR {$i \gets$ $\mathcal{P}$ to $1$}  
            \FOR {$j \gets$ $1$ to $\mathcal{P}$}  
                \STATE {$cur\_seed$} $\gets$ $get\_next\_seed(\mathbf{seeds}[j], r)$
                \STATE $\boldsymbol{Y_{share}} \gets$ $PRF(cur\_seed, (N,K))$

                \STATE $party\_rank \gets (j + r) \mod \mathcal{P} + 1$
                \STATE $\mathbf{R}[j]$ $\gets$ $\mathbf{R}[j] + \mathbf{D_{shares}}[party\_rank][r]$

                \STATE $\boldsymbol{Y_{share}} \gets rotate(\boldsymbol{Y_{share}}, N, \mathbf{R}[j])$

                \STATE $\boldsymbol{\xi^*} \gets \boldsymbol{\xi^*}$ + $\boldsymbol{Y_{share}}$
                
            \ENDFOR
        \ENDFOR
    \ENDFOR
    \STATE \textbf{return} $\mathbf{seeds}$, $\boldsymbol{\xi}^*$
\end{algorithmic}
\end{algorithm}

\textbf{Secure Read and Write Protocols with Data Masking}
In the main paper (Section \ref{sec:cryptmpl_read}), we present the protocol to access the feature vector of a source node for an edge in the graph stored in the secret-shared domain. Algorithm \ref{alg:Secure_Access_Noise} shows the following steps to be executed by each party $CP_i$ to read the feature vector of the source node $ \llbracket S \rrbracket $ from the feature matrix $ \llbracket {\bf A} \rrbracket $ while protecting the shares of $ \llbracket {\bf A} \rrbracket $ using noise $\boldsymbol{\xi}$.

\begin{enumerate}[topsep=0pt,itemsep=0ex,partopsep=0ex,parsep=0ex]
\item Generate a random integer $r_i$ (line 7)
\item Initialize a matrix $ \llbracket \mathbf{A'} \rrbracket _i =  \llbracket \mathbf{A} \rrbracket _i$ and add a noise matrix with $\llbracket \mathbf{A'} \rrbracket _i$ (line 9-10)
\item Rotate the matrix $ \llbracket \mathbf{A'} \rrbracket _i$ by $r_i$ to create $ \llbracket \mathbf{A''} \rrbracket _i$ (line 11)
\item Create a new index $R''_i$ with $\llbracket S \rrbracket_i$ and a random integer (line 12)
\item Share matrix $ \llbracket \mathbf{A''} \rrbracket _i$ and $R''_i$ with the next party in the ring, which stores the data as $ \llbracket \mathbf{A'} \rrbracket _i$ and $R'_i$ respectively (line 17-20).
\item Follow steps (2)-(5) for $\mathcal{P} - 1$ times.  Each party adds noise to the others' shares before exchanging the data with other parties (line 23-24)
\end{enumerate}

Finally, each party accesses the feature from $ \llbracket \mathbf{A''} \rrbracket _i$ at index $R''_i$ and stores it as $ \llbracket \mathbf{Y} \rrbracket _i= \llbracket \mathbf{A''}[R''_i] \rrbracket _i$.

\begin{algorithm}[H] 
\caption{Secure Read, $\mathcal{F}_{SR}$}
\label{alg:Secure_Access_Noise}
\textbf{Input:} $ \llbracket \mathbf{A} \rrbracket $ (Feature Matrix), $ \llbracket S \rrbracket $ (Source index), $cur\_round$ (Current round), $seed_p$ (seed for $p$-th party)\\ 
\textbf{Output:} $ \llbracket \mathbf{Y} \rrbracket $ (Feature vector at $\mathbf{A}[S]$)
\begin{algorithmic}[1]
    \STATE {$cur\_party$ $\gets$ $get\_cur\_party()$}

    \STATE {$\mathbf{A'}$ $\gets$ $copy( \llbracket \mathbf{A} \rrbracket _{cur\_party}$)}
    \STATE {$cur\_index\_share$ $\gets$ $ \llbracket S \rrbracket _{cur\_party}$}

    \STATE {$r$ $\gets$ $get\_random\_integer()$}

    \STATE {$cur\_seed$} $\gets$ $get\_next\_seed(seed_p, cur\_round)$
    \STATE $\boldsymbol{\xi}$ $\gets$ $PRF(cur\_seed, (N, K))$,
    {$\mathbf{A'}$ $\gets$ $\mathbf{A'} + \boldsymbol{\xi}$}, 
    {$\mathbf{A''}$ $\gets$ $rotate(\mathbf{A'}, N, r)$}, 
    {$R''$ $\gets$ $cur\_index\_share + r$}
    \STATE {$cur\_send\_data$ $\gets$ $tuple(\mathbf{A''}, R'')$}, {$cur\_receive\_data$ $\gets$ $null$}
    
    \STATE {$pass\_count$ $\gets$ $0$}
    \WHILE {$pass\_count  <  (P-1)$}
       \STATE {$cur\_send\_data.send\_to(next\_party)$}, {$cur\_receive\_data$ $\gets$ $receive\_from(prev\_party)$}
       \STATE {$\mathbf{A'}$ $\gets$ $cur\_receive\_data[0]$}, {$R'$ $\gets$ $cur\_receive\_data[1]$}

        \STATE {$r$ $\gets$ $get\_random\_integer()$}

        \STATE {$cur\_seed$} $\gets$ $get\_next\_seed(seed_p, cur\_round)$
        \STATE $\boldsymbol{\xi}$ $\gets$ $PRF(cur\_seed, (N, K))$,
        {$\mathbf{A'}$ $\gets$ $\mathbf{A'} + \boldsymbol{\xi}$}, 
        {$\mathbf{A''}$ $\gets$ $rotate(\mathbf{A'}, N, r)$}, 
        {$R''$ $\gets$ $ cur\_index\_share + r + R'$}
        \STATE {$cur\_send\_data$ $\gets$ $tuple(\mathbf{A''}, R'')$}

        \STATE {$pass\_count$ $\gets$ $pass\_count + 1$}
       
    \ENDWHILE
    \STATE {$ \llbracket \mathbf{Y} \rrbracket _{cur\_party}$ $\gets$ $\mathbf{A''}[R'' \mod N]$}
    \STATE \textbf{return} $ \llbracket \mathbf{Y} \rrbracket $
\end{algorithmic}
\end{algorithm}

Next, we present Algorithm \ref{alg:Secure_Write_Noise} that shows the following steps to be executed by each party $i$, to write a feature vector $ \llbracket {\bf Y} \rrbracket $ at the target index $ \llbracket D \rrbracket $ of a matrix $ \llbracket {\bf G} \rrbracket $.

\begin{enumerate}[topsep=0pt,itemsep=0ex,partopsep=0ex,parsep=0ex]
\item Initialize a matrix $ \llbracket \mathbf{G} \rrbracket _i$ of size $(N,K)$ with a noise matrix and updates first index of $ \llbracket \mathbf{G} \rrbracket _i$ with $ \llbracket \mathbf{Y} \rrbracket $ (line 5-6)
\item Rotate the matrix $ \llbracket \mathbf{G} \rrbracket _i$ by $ \llbracket D \rrbracket _i$ to create $ \llbracket \mathbf{G'} \rrbracket _i$ (line 7-8)
\item Share the matrix $ \llbracket \mathbf{G'} \rrbracket _i$ to the next party, which stores it as $ \llbracket \mathbf{G} \rrbracket _i$ (line 13-14)
\item Follow steps (2) - (3) for $\mathcal{P} - 1$ times. Each party adds noise to the others' shares before exchanging the data with other parties (line 16-17)
\end{enumerate}

Finally, each party has a share $ \llbracket \mathbf{G'} \rrbracket $, where the destination index is updated with a share $ \llbracket \mathbf{Y} \rrbracket $, and all values of $ \llbracket \mathbf{G'} \rrbracket $ are masked with the noise.

\begin{algorithm}[H] 
\caption{Secure Write, $\mathcal{F}_{SW}$}
\label{alg:Secure_Write_Noise}
\textbf{Input:} $ \llbracket \mathbf{Y} \rrbracket $ (Feature vector), $ \llbracket D \rrbracket $ (Destination index), $cur\_round$ (Current round), $seed_p$ (seed for  p-th party)\\ 
\textbf{Output:} $ \llbracket \mathbf{G} \rrbracket $
\begin{algorithmic}[1]
    \STATE {$cur\_party$ $\gets$ $get\_cur\_party()$}

    \STATE {$cur\_seed$} $\gets$ $get\_next\_seed(seed_p, cur\_round)$
    \STATE $\boldsymbol{\xi}$ $\gets$ $PRF(cur\_seed, (N, K))$
    
    \STATE $\mathbf{G}$ $\gets$ $\boldsymbol{\xi}$, $\mathbf{G[0]}$ $\gets$ $\mathbf{G[0]} +  \llbracket \mathbf{Y} \rrbracket $

    \STATE $cur\_index\_share$ $\gets$ $ \llbracket D \rrbracket _{cur\_party}$
    
    \STATE $\mathbf{G'}$ $\gets$ $rotate(\mathbf{G}, N, cur\_index\_share)$
    \STATE $cur\_send\_data$ $\gets$ $\mathbf{G'}$, $cur\_receive\_data$ $\gets$ $null$
    
    \STATE $pass\_count$ $\gets$ $0$
    \WHILE {$pass\_count  <  (P-1)$}
        \STATE $cur\_send\_data.send\_to(next\_party)$, $cur\_receive\_data$ $\gets$ $receive\_from(prev\_party)$

        \STATE $cur\_seed$ $\gets$ $get\_next\_seed(seed_p, cur\_round)$
        \STATE $\boldsymbol{\xi}$ $\gets$ $PRF(cur\_seed, (N, K))$, 
        $\mathbf{G}$ $\gets$ $cur\_receive\_data + \boldsymbol{\xi}$,
        $\mathbf{G'}$ $\gets$ $rotate(\mathbf{G}, N, cur\_index\_share)$

        \STATE $cur\_send\_data$ $\gets$ $\mathbf{G'}$

        \STATE $pass\_count$ $\gets$ $pass\_count + 1$
       
    \ENDWHILE
    \STATE \textbf{return} $ \llbracket \mathbf{G'} \rrbracket $
\end{algorithmic}
\end{algorithm}

\begin{algorithm}[H] 
\caption{MPL with Batch, $\mathcal{F}_{BatchCryptMPL}$}
\label{alg:Server_MPL_Batch}

\textbf{Input:}$ \llbracket \mathbf{A} \rrbracket $ (Feature matrix), $ \llbracket \mathbf{S_f} \rrbracket $, $ \llbracket \mathbf{D_f} \rrbracket $ (Encrypted source and destination indices), $ \llbracket \boldsymbol{\xi}^* \rrbracket $ (Noise), $\mathbf{S_r}$, $\mathbf{D_r}$ (Relative source and destination indices)\\ 
\textbf{Output:} $\mathbf{ \llbracket {A^*} \rrbracket }$ (Output feature matrix)
\begin{algorithmic}[1]
    \STATE {$M$ $\gets$ {$length(\mathbf{S_r})$}} 
    \STATE {$R$ $\gets$ {$length( \llbracket \mathbf{S_f} \rrbracket )$}} 
    \STATE {$\mathbf{ \llbracket {A_\xi^*} \rrbracket }$ $\gets$ {$init\_matrix(N, K)$}}
    \FOR {$r \gets 0$ to $R-1$}  
        \STATE {$cur\_seed$} $\gets$ $get\_next\_seed(seed, r)$
        \STATE {$\mathbf{N'}$ $\gets$ {$PRF(seed, (N, K))$}}
        \STATE {$ \llbracket srcInd \rrbracket $ $\gets$ $ \llbracket \mathbf{S_f[}r\mathbf{]} \rrbracket $}
        \STATE {$\mathbf{srcIndRel}$ $\gets$ $\mathbf{S_r[}r*b:r*(b+1)\mathbf{]}$} 
        \STATE {$ \llbracket \mathbf{Y} \rrbracket $ $\gets$ $\mathcal{F}_{BatchSR}( \llbracket \mathbf{A} \rrbracket ,  \llbracket srcInd \rrbracket , \mathbf{srcIndRel})$}
        \STATE {$ \llbracket destInd \rrbracket $ $\gets$ $ \llbracket \mathbf{D_f[}r\mathbf{] \rrbracket }$}
        \STATE {$\mathbf{destIndRel}$ $\gets$ $\mathbf{D_r[}r*b:r*(b+1)\mathbf{]}$}
        \STATE {$\mathbf{ \llbracket {G} \rrbracket }$ $\gets$ $\mathcal{F}_{BatchSW}( \llbracket \mathbf{Y} \rrbracket ,  \llbracket destInd \rrbracket , \mathbf{destIndRel})$}
        \STATE {$ \llbracket \mathbf{A_\xi^*} \rrbracket $ $\gets$ $ \llbracket \mathbf{A_\xi^*} \rrbracket  +  \llbracket \mathbf{G} \rrbracket $}
    \ENDFOR
    \STATE {$ \llbracket \mathbf{A^*} \rrbracket $ $\gets$ $ \llbracket \mathbf{A_\xi^*} \rrbracket  -  \llbracket \boldsymbol{\xi}^* \rrbracket $}
    \STATE \textbf{return} $\mathbf{ \llbracket {A^*} \rrbracket }$
\end{algorithmic}
\end{algorithm}

\textbf{Processing edges in batches}
\label{sec:app_cryptmpl_batch}
Algorithm \ref{alg:Server_MPL_Batch} shows how the edges are processed in batches by each party. To protect the feature matrix, each party adds different noise matrices to the feature matrix in each round. For each batch, $\mathcal{F}_{BatchSR}$ reads the feature vector of the encrypted node and uses the relative indices to read the feature vector of other nodes in the batch (line 7-9). The feature vectors $ \llbracket \mathbf{Y} \rrbracket $ accessed via read operation are used to update the feature vectors at the destination nodes of a batch using $\mathcal{F}_{BatchSW}$ function (line 10-12). Specifically, $\mathcal{F}_{BatchSW}$ writes the feature vector of the encrypted node in a batch at the appropriate position of the target matrix and uses relative indices to update the target matrix (line 13). Finally, the overall noise is removed to get the correct feature matrix (line 15).

\section{Detailed Algorithms of CryptMUL}
\label{app:sec:cryptmul}

In this section we present the algorithm $\mathcal{F}_{BeaverMtoA}$ described in Section~\ref{sec:cryptmul_design}. To convert Beaver triples from M-SS to A-SS format $\mathcal{F}_{BeaverMtoA}$ internally uses $\mathcal{F}_{MtoA}$, which invokes the following steps to convert a value $W$ from M-SS to A-SS format.

\begin{enumerate}[topsep=0pt,itemsep=0ex,partopsep=0ex,parsep=0ex]
\item Select a pair ($\llbracket R \rrbracket$, $\llangle R \rrangle$) from $\bf{AM}$ (Alg.~\ref{alg:get_A_from_M} Line 1).
\item Apply Extended Euclidean Algorithm~\cite{extended_euclidean} to compute the inverse of $R$ as $\llangle R^{-1} \rrangle$ in M-SS format~\cite{ghodosi2012multi} (Alg.~\ref{alg:get_A_from_M} Line 2).
\item Each party computes locally the product of $\llangle W \rrangle$ and $\llangle R^{-1} \rrangle$ (Alg.~\ref{alg:get_A_from_M} Line 3) and reveals the ratio $\alpha$ (Alg.~\ref{alg:get_A_from_M} Line 4).
\item Each party computes $\llbracket W \rrbracket_i \gets \alpha \times \llbracket R \rrbracket_i$ to get $W$ in A-SS format (Alg.~\ref{alg:get_A_from_M} Line 5).
\end{enumerate}

\begin{algorithm}[!h] 
\caption{Multiplicative to Additive Shares, $\mathcal{F}_{MtoA}$}
\label{alg:get_A_from_M}
\textbf{Input:} $ \llangle W \rrangle $\\ 
\textbf{Output:} Generate additive shares $\llbracket W \rrbracket$
\begin{algorithmic}[1]
    \STATE Pick a pair ($\llbracket R \rrbracket$, $\llangle R \rrangle$) from $\bf{AM}$ computed in $\mathcal{F}_{MsAsPair}$
    \STATE Each party $p$ computes locally $\llangle R^{-1} \rrangle_p$ using Extended Euclidean Algorithm, thereby computing the multiplicative shares of $R^{-1}$.
    \STATE $CP_i$ computes $\llangle \alpha \rrangle_i \gets \llangle W \rrangle_i \times \llangle R^{-1} \rrangle_i$.
    \STATE All parties collaboratively recover $\alpha$.
    \STATE $CP_i$ computes $\llbracket W \rrbracket_i \gets \alpha \times \llbracket R \rrbracket_i$
\end{algorithmic}
\end{algorithm}

$\mathcal{F}_{BeaverMtoA}$ converts each value of Beaver triple from M-SS format using $\mathcal{F}_{MtoA}$ as shown in Alg.~\ref{alg:generate_Beaver}. 

\begin{algorithm}[!h] 
\caption{Generate Beaver Triples, $\mathcal{F}_{BeaverMtoA}$}
\label{alg:generate_Beaver}
\textbf{Input:}($\llangle A \rrangle, \llangle B \rrangle, \llangle C \rrangle$)\\ 
\textbf{Output:} Beaver Triples in additive format ($\llbracket A \rrbracket, \llbracket B \rrbracket, \llbracket C \rrbracket$)
\begin{algorithmic}[1]
    \STATE $\llbracket A \rrbracket \gets \mathcal{F}_{MtoA}(\llangle A \rrangle)$
    \STATE $\llbracket B \rrbracket \gets \mathcal{F}_{MtoA}(\llangle B \rrangle)$
    \STATE $\llbracket C \rrbracket \gets \mathcal{F}_{MtoA}(\llangle C \rrangle)$  
\end{algorithmic}
\end{algorithm}
\section{Correctness Analysis}
\label{sec:correctness}
In CryptGNN, during the execution of secure protocols in inference, parties mask their shares with random noise to protect the data from adversaries. The protocols also guarantee the correctness of their results by eliminating the effect of noise. In this section, we provide the correctness analysis of the protocols in CryptGNN.

\textbf{Correctness of message-passing layer using CryptMPL.} First, we analyze the correctness of CryptMPL without data masking in a $P$-party SMPC setting. As discussed in the main paper (Section~\ref{sec:cryptmpl_design}), to access the feature vector at the source index of the feature matrix, each share of the matrix is rotated $\sum_{i=1}^{P}r_i$ times, and $\mathcal{F}_{SR}$ accesses the feature vector at index $\sum_{i=1}^{P}(\llbracket S \rrbracket_i + r_i) = S + \sum_{i=1}^{P}r_i$ (equivalent to accessing the value at index $S$ in the original feature matrix). To update the feature at the destination index, $\mathcal{F}_{SW}$ writes the intermediate vector $ \llbracket \mathbf{Y} \rrbracket$ at index 0 of the matrix $ \llbracket \mathbf{G} \rrbracket$, and it is rotated overall by $\sum_{i=1}^{P}D_i = D$. Thus, $\llbracket \mathbf{Y} \rrbracket$ reaches the destination index $D$ of $\llbracket \mathbf{G} \rrbracket $, while the vectors in the other indices are 0. Then, in $\mathcal{F}_{SA}$ each party updates its share of the output feature matrix $\mathbf{A^*}$ with $\mathbf{G}$, which is equivalent to updating the feature vector at index $D$. CryptMPL executes the same operation for all edges, completing the message passing correctly.

Next, to protect the data exchanged with other parties, each party masks the feature matrix and intermediate results with random noise, such that the final feature matrix equals $\mathbf{A^*} + \boldsymbol{\xi}'$, where $\boldsymbol{\xi}'$ is the error due to the added noise. As the same operations are performed at the client side on the input noise matrices, the client can calculate the effect of the overall noise
$\boldsymbol{\xi}^* = \boldsymbol{\xi}'$ and share it with all the SMPC parties. Therefore, the noise can be removed to retrieve the correct feature matrix $\mathbf{A^*}$.

Using batching, CryptMPL reads the feature vector at the first index of a batch, and the feature vectors at the relative indices for the batch. 
Similarly, CryptMPL updates the intermediate matrix $\mathbf{G}$ at the first index of a batch and the indices relative to the first index. Since the first index of a batch can be processed correctly, reads and writes at relative indices give the correct result.

\textbf{Correctness of feature transformation layers using CryptMUL.} To compute FTLs, CryptGNN uses standard techniques as described in Section~\ref{subsec:background_gnn_layers}. In CryptMUL, we propose new techniques to generate Beaver triples for secure matrix multiplication and element-wise multiplications in A-SS format, which are used to implement the secure versions of the FTLs. Other operations in FTLs are straightforward and can be used without requiring any specialized protocol. Here, we demonstrate the correctness of the Beaver triples generated in $\mathcal{F}_{MatMul}$ and $\mathcal{F}_{ElemMul}$ for secure matrix multiplication and element-wise multiplications respectively. 

In $\mathcal{F}_{MatMul}$, $\mathcal{F}_{InitBeaver}$ follows~\cite{keller2016mascot} to generate the initial Beaver triples in A-SS as ($\llbracket \mathbf{A} \rrbracket$, $\llbracket \mathbf{B} \rrbracket$, $\llbracket \mathbf{C} \rrbracket$), where $\mathbf{A} \in \mathbb{R}^{N \times K}, \mathbf{B} \in \mathbb{R}^{K \times K'}, \mathbf{C} \in \mathbb{R}^{N \times K'}$ and $\mathbf{A} \otimes \mathbf{B} = \mathbf{C}$. Next, the parties compute the linear combinations of the rows in the matrix using $\mathcal{F}_{RandComb}$ to generate two new matrices $\mathbf{A'}$ and $\mathbf{C'}$, where $\llbracket \mathbf{A'}[j] \rrbracket = \sum_{i=1}^{N}k_{ji} \times \llbracket \mathbf{A}[i] \rrbracket$ and $\llbracket \mathbf{C'}[j] \rrbracket = \sum_{i=1}^{N}k_{ji} \times \llbracket \mathbf{C}[i] \rrbracket$ for $i \in [1, N], j \in [1, N]$. The random real values $k_{ji}$ are generated by a pseudo-random function (PRF). Since, $\mathbf{B}$ is fixed, modifying $\mathbf{A}$ to $\mathbf{A'}$ and $\mathbf{C}$ to $\mathbf{C'}$ using the same linear combination maintains the correctness for ($\mathbf{A'}, \mathbf{B}, \mathbf{C'}$).

In $\mathcal{F}_{ElemMul}$, $\mathcal{F}_{MsAsPair}$ follows ~\cite{xiong2020efficient} to compute the A-SS and M-SS of a random value $R_i$ for two parties. For other parties, we consider the share of $\llbracket R_i \rrbracket = 0$ and $\llangle R_i \rrangle = 1$, thus the $R_i$ is correct in A-SS and M-SS format for $P$ parties. Following this approach, we generate $P - 1$ random values $R_i, i \in [1, P-1]$ in A-SS and M-SS format. Finally, $\mathcal{F}_{MsAsPair}$ takes the products of $\llbracket R_i \rrbracket = \prod_{i=1}^{P - 1} \llbracket R_i \rrbracket$ using the pre-computed Beaver triples (generated using ~\cite{keller2016mascot}) to compute $R$ in A-SS format. It also calculates the products of $\llangle R_i \rrangle = \prod_{i=1}^{P - 1} \llangle R_i \rrangle$ to get the multiplicative share of  $\llangle R \rrangle$. Thus, $\mathcal{F}_{MsAsPair}$ gets a random value $R$ which is correct in both A-SS and M-SS format. For each element-wise multiplication operation, $\mathcal{F}_{MsAsPair}$ generates 3 random values $R_A, R_B, R_C$ in A-SS and M-SS format following this approach.

As part of $\mathcal{F}_{ElemMul}$, $\mathcal{F}_{BeaverM}$ generates a Beaver triple $(A, B, C)$ in M-SS. Each party $CP_p$ generates random values $\llangle A \rrangle_p$ and $\llangle B \rrangle_p$, and computes $\llangle C \rrangle_p = \llangle A \rrangle_p \times \llangle B \rrangle_p$. In this way, we get the Beaver triples in M-SS as ($\llangle A \rrangle$, $\llangle B \rrangle$, $\llangle C \rrangle)$, since $\llangle C \rrangle = \prod_{i=1}^{P} \llangle C \rrangle_i = \prod_{i=1}^{P} \llangle A \rrangle_i \times \llangle B \rrangle_i = \prod_{i=1}^{P} \llangle A \rrangle_i \times \prod_{i=1}^{P} \llangle B \rrangle_i = \llangle A \rrangle \times \llangle B \rrangle$.

Finally, $\mathcal{F}_{BeaverMtoA}$ converts each element of ($\llangle A \rrangle$, $\llangle B \rrangle$, $\llangle C \rrangle)$ to A-SS format ($\llbracket A \rrbracket$, $\llbracket B \rrbracket$, $\llbracket C \rrbracket)$ using the $R_A, R_B, R_C$ in A-SS and M-SS format following the approach in~\cite{9404811}, which is proved to be correct. Thus, $\mathcal{F}_{ElemMul}$ gets a correct Beaver triple as the required format ($\llbracket A \rrbracket$, $\llbracket B \rrbracket$, $\llbracket C \rrbracket)$, which can be used for the secure element-wise multiplication operation.

\section{Experimental Setup \& Additional Results}
\label{appendix-evaluation}

In this section, we present the statistics of the graph datasets and GNN architecture used in the main paper (Section~\ref{sec:evaluation}) to evaluate the performance of CryptGNN system and its protocols. We also present the additional results on the evaluation of CryptMPL.

\subsection{Graph Datasets}
Table~\ref{table:graph_dataset} summarizes the statistics of the graph datasets we use in our experiments. To evaluate the performance of the CryptGNN system for the graph classification task, we use 3 benchmark datasets: TUDataset (ENZYMES), TUDataset (PROTEINS)~\cite{DBLP:journals/corr/abs-1910-12091}, and FAUST~\cite{bogo2014faust}. Among these three datasets, the FAUST dataset has the largest graphs with 6,890 nodes and 41,328 edges. 
To assess the performance of the secure message-passing layer CryptMPL, we employed three additional datasets: Cora, CiteSeer~\cite{yang2016revisiting}, and PPI~\cite{zitnik2017predicting}. These benchmark datasets are typically well-suited for node classification tasks and have a substantial number of nodes, edges, and features. We use these datasets to assess the performance of our secure MPL on large graphs. We also generate synthetic data to systematically observe the effects of different parameters.

\setlength{\heavyrulewidth}{1.5pt}
\setlength{\abovetopsep}{4pt}
\begin{table}[ht!]
\centering
\caption{Graph dataset statistics}
\resizebox{.7\linewidth}{!}{
\begin{tabular}{*6c}
\toprule
Dataset & \#graphs & \#nodes & \#edges & \#features & \#classes\\
\midrule
ENZYMES & 600 & ~32.6 & ~124.3 & 3 & 6\\
PROTEINS & 1,113 & ~39.1 & ~145.6 & 3 & 2\\
Cora & - & 2,708 & 10,556 & 1,433 & 7\\
CiteSeer & - & 3,327 & 9,104 & 3,703 & 6\\
PPI & 20 & ~2,245.3 & ~61,318.4 & 50 & 121\\
FAUST & 100 & 6,890 & 41,328 & 3 & 10\\

\bottomrule
\label{table:graph_dataset}
\end{tabular}
}
\end{table}

\subsection{Network Architecture}
\begin{figure}[!h]
    \centering
    \includegraphics[width=0.75\linewidth]{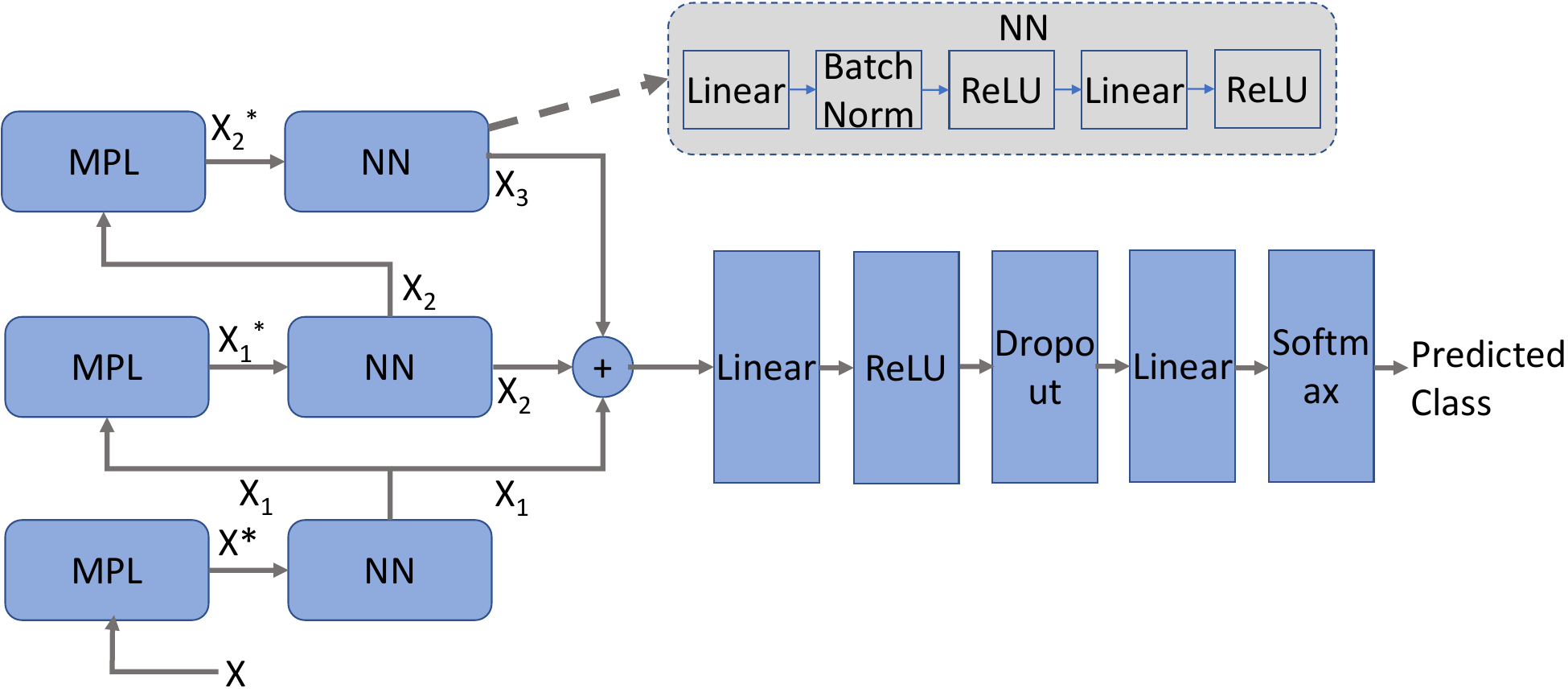}
    \caption{GIN architecture}
    \label{fig:gin}
\end{figure}
Figure~\ref{fig:gin} shows the architecture of the GIN~\cite{xu2018how} model we use to evaluate the performance of secure graph classification tasks. The network comprises 3 message-passing layers stacked one after another. The outputs of the 3 layers are concatenated and passed through a linear layer with the ReLU activation function. Finally, another linear layer with the Softmax activation function predicts the probability of each class for the sample input graph. A dropout layer is used in training, which does not have any effect during inference. The model is trained in private infrastructure and the parameters for Linear and Batch Norm layers are stored in the cloud in secret-shared format.

\begin{figure*}[b]
\centering
\includegraphics[width=.6\linewidth]{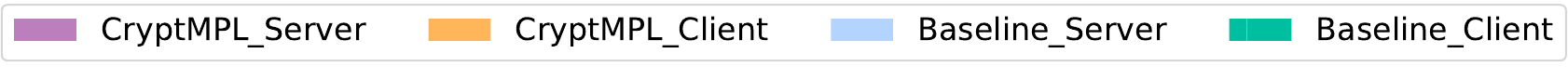}
  \subfloat[Variable $N$]
  {%
    \includegraphics[width=0.25\textwidth]{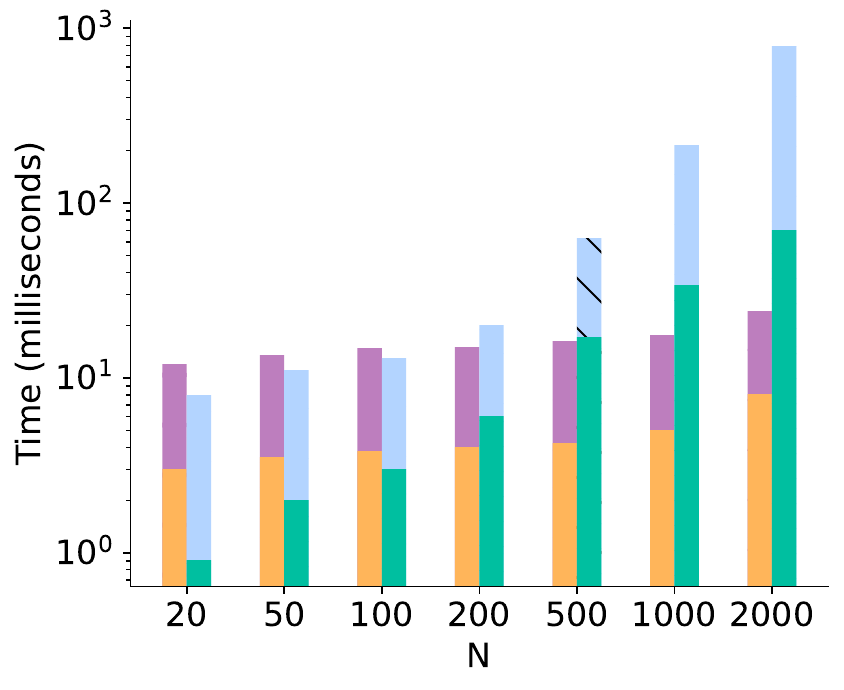}
  }%
  \subfloat[Variable $K$]
  {%
    \includegraphics[width=0.25\textwidth]{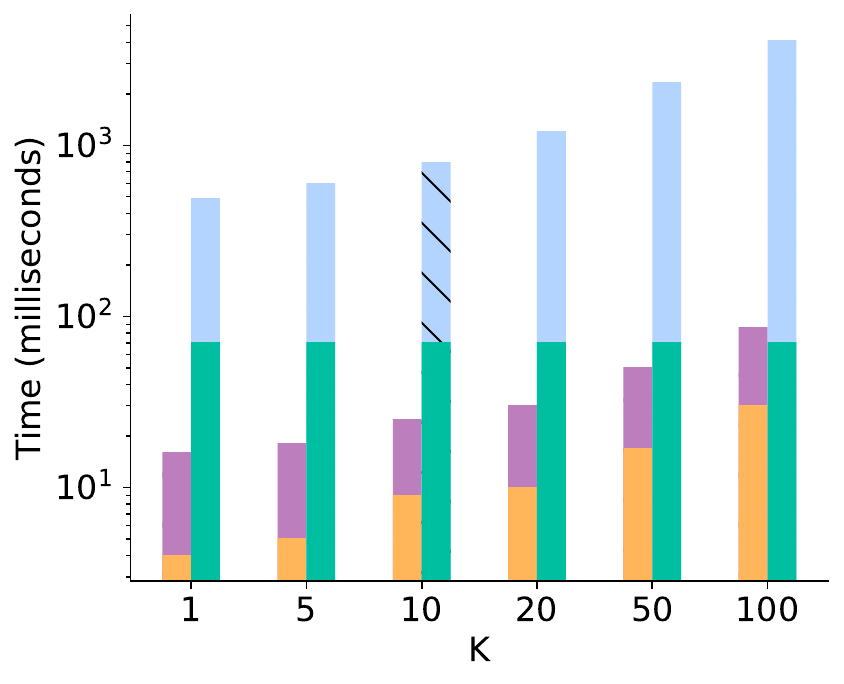}
  }%
  \subfloat[Variable $D_{avg}$]
  {%
    \includegraphics[width=0.25\textwidth]{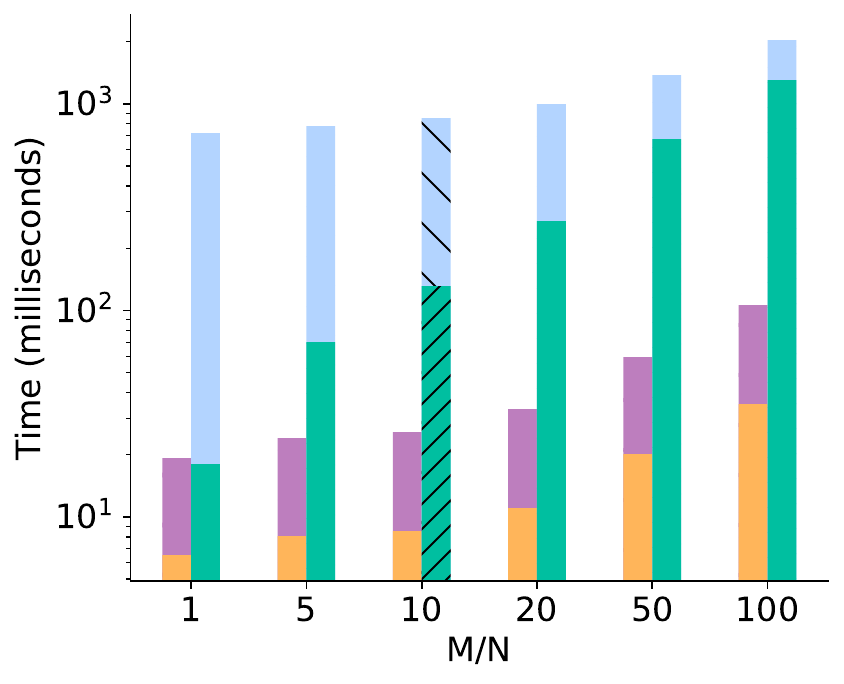}
  }%
\caption{Effect of different parameters of graph data (Y-axis is log-scaled)}
\label{fig:cryptmpl_effect_params}
\end{figure*}

\subsection{CryptMPL Results}
\label{app:subsec:cryptmpl_results}

To evaluate CryptMPL with respect to graph parameters, we use synthetic data and compare the performance with adjacency-matrix based implementation using CrypTen. In each experiment, we vary one parameter. Unless otherwise stated, we use 3 parties in these experiments and the synthetic graph has $N=2000$ nodes, $K=10$ features, $D_{avg}=5$ edges per node. 

\noindent\textbf{Effect of the number of nodes.}
We compare the execution time of CryptMPL with the Baseline using synthetic graphs with different numbers of nodes. 
Figure \ref{fig:cryptmpl_effect_params}(a) shows the execution times at the client and server are not affected much in CryptMPL.
The results show that CryptMPL has superior performance compared to the Baseline for medium and large-scale graphs with many nodes. The reason is that the Baseline approach needs to manage a large adjacency matrix of size $(N,N)$ and performs large matrix multiplications. 

\noindent\textbf{Effect of the number of features.}
We use synthetic graphs with different numbers of features and compare the execution time of CryptMPL with the Baseline.
Figure \ref{fig:cryptmpl_effect_params}(b) shows that CryptMPL performs significantly better than the Baseline. Although the size of the adjacency matrix in the Baseline does not change, increasing the number of features increases the number of multiplication operations, which in turn increases the execution time at a higher rate compared to CryptMPL as shown in Figure \ref{fig:cryptmpl_effect_params}(b). As $N$, $P$ and $M$ are constant, the number of addition and rotation operations remain constant for CryptMPL. However, its execution time increases slightly with $K$, as addition is now executed on larger matrices. 

\noindent\textbf{Effect of the number of edges.}
Figure \ref{fig:cryptmpl_effect_params}(c) shows the execution time of CryptMPL and the Baseline, when we vary the number of edges. 
The results show the execution time increases slightly in the case of CryptMPL, while the rate of change is low compared to the Baseline. For the Baseline, the computation and 
the communication costs remain the same at the server side, as the size of matrices is the same. However, the computation cost at the client side increases to create the adjacency matrix from the list of edges. Increasing the number of edges in a graph increases the computation cost on the client side for CryptMPL, as it needs to process more edges on the noise matrix to compute the overall effect of noise. Similarly, the server needs to do more computation to execute the MPL on the masked feature matrix.

\subsection{Non-linear layers using CryptMUL} 
We compare the execution time of the Sigmoid function, which requires secure multiplications, between the implementations using CrypTen and CryptMUL's protocol $\mathcal{F}_{ElemMul}$. We generate a list of 1000 random values $\bf{X}$ and measure the overall execution time to compute $\llbracket \mathbf{Z} \rrbracket = Sigmoid(\llbracket \mathbf{X} \rrbracket)$ for a different number of parties. We observe that the execution time of CrypTen and CryptMUL is almost the same, since both approaches have similar computation and communication costs. However, the protocol using CryptMUL is more secure since it does not require a trusted server as CrypTen.

\end{document}